\documentclass[%
a4paper,
preprint,
superscriptaddress,
groupedaddress,
runinaddress,
showpacs,
 amsmath,amssymb,
 aps,
prb,
showkeys
]{revtex4-1}
\usepackage{standalone}

\usepackage{graphicx}
\usepackage{dcolumn}
\usepackage{bm}
\usepackage{hyperref}
\usepackage[
scale=0.7, marginratio={1:1, 2:3}, ignoreall,
text={7in,10in},centering,
]{geometry}
\usepackage{amsmath}
\usepackage{braket}

\newcommand{\eqnref}[1]{~(\ref{#1})}%
\newcommand\commutator[2]{\ensuremath{\mathinner{%
    \mathopen[\,#1,#2\,\mathclose]}}}

\newcommand{\bracket}[2]{{\left\langle \vphantom{#1 #2} #1 \,\right|
  \left.\hspace{-0.15em} \vphantom{#1 #2} #2\right\rangle}}

\newcommand*{\braopket}[3]{\ensuremath{\langle{#1}|{#2}|{#3}\rangle}}

\renewcommand{\d}{\downarrow}
\renewcommand{\u}{\uparrow}
\newcommand{\s}{\sigma}
\newcommand\expect[1]{\ensuremath{\left\langle{#1}\right\rangle}}

\newcommand{\ra}{\rightarrow}
\def\bigO#1{\mathcal{O}(#1)}
\DeclareMathOperator{\e}{e} 			
\newcommand{\mbf}[1]{\mathbf{#1}}
\newcommand{\cd}[2]{c^\dagger_{#1,#2}}
\newcommand{\der}{\partial\mspace{2mu}} 
\providecommand{\abs}[1]{\lvert#1\rvert}
\newcommand{\sign}{\text{sign}}
\newcommand{\neci}{\texttt{NECI}}
\newcommand{\mustbe}{\stackrel{!}{=}}

\begin{document}

\title{Compact numerical solutions to the two-dimensional repulsive Hubbard model obtained via non-unitary similarity transformations}

\author{Werner Dobrautz}
 \email{w.dobrautz@fkf.mpg.de}

\author{Hongjun Luo}%
 \email{h.luo@fkf.mpg.de}

\affiliation{%
Max Planck Institute for Solid State Research, Heisenbergstr. 1, 70569 Stuttgart, Germany
}%

\author{Ali Alavi}
\affiliation{%
Max Planck Institute for Solid State Research, Heisenbergstr. 1, 70569 Stuttgart, Germany
}%
\affiliation{
 Dept of Chemistry, University of Cambridge, Lensfield Road, Cambridge CB2 1EW, United Kingdom
}%
\homepage{https://www.fkf.mpg.de/alavi}
\email{a.alavi@fkf.mpg.de}

\date{\today}


\begin{abstract}
Similarity transformation of the Hubbard Hamiltonian using a
Gutzwiller correlator leads to a non-Hermitian effective Hamiltonian,
which can be expressed exactly in momentum-space representation, and
contains three-body interactions. We apply this methodology to study
the two-dimensional Hubbard model with repulsive interactions near
half-filling in the intermediate interaction strength regime
($U/t=4$). We show that at optimal or near optimal strength of the
Gutzwiller correlator, the similarity transformed Hamiltonian has
extremely compact right eigenvectors, which can be sampled to high
accuracy using the Full Configuration Interaction Quantum Monte Carlo
(FCIQMC) method, and its initiator approximation. Near-optimal
correlators can be obtained using a simple projective equation, 
thus obviating the need for a numerical optimisation of the
correlator. The FCIQMC method, as a projective technique, is 
well-suited for such non-Hermitian problems, and its stochastic 
nature can handle the 3-body interactions exactly without undue
increase in computational cost. The highly compact nature of the 
right eigenvectors means that the initiator approximation in FCIQMC 
is not severe, and that large lattices can be simulated, well beyond
the reach of the method applied to the original Hubbard Hamiltonian.
Results are provided in lattice sizes up to 50 sites and compared to
auxiliary-field QMC. New benchmark results are provided in the off half-filling regime, with no severe sign-problem being encountered. In addition, we show that methodology can be used to calculate excited states of the Hubbard model and lay the groundwork for the calculation of observables other than the energy. 

\end{abstract}

\maketitle

\section{\label{sec:intro}Introduction}
The Fermionic two-dimensional Hubbard model\cite{hubbard-model-1,
hubbard-model-2,hubbard-model-3} with repulsive interactions is a 
minimal model of itinerant strongly correlated electrons that is 
believed to exhibit extraordinarily rich physical behaviour. Especially
in the past thirty years, it has been intensively 
studied as a model to understand the physics of  
high-temperature superconductivity observed in layered cuprates\cite{cuprates} . 
Its phase diagram as a function of temperature, interaction strength 
and filling 
includes antiferromagnetism, Mott metal-insulator transition,
unconventional 
superconductivity\cite{high-tc} with d-wave pairing off half-filling,
striped phases, a pseudo gap regime, charge and spin density
waves\cite{scalapino}. Confronted with such a plethora of physical
phenomena, accurate numerical results are indispensable in resolving various competing theoretical scenarios.\\
Unfortunately the numerical study of the 2D Hubbard model has proven
extraordinarily challenging, particularly in the off-half-filling 
regime with intermediate-to-strong interaction strengths $U/t=4-12$.
Major difficulties include severe sign problems for quantum Monte 
Carlo (QMC) methods, whilst the 2D nature of the problem causes
convergence difficulties for density matrix renormalization group
(DMRG)\cite{dmrg-1,dmrg-2,dmrg-3} based methodologies which have
otherwise proven extremely powerful in 1D systems. Nevertheless
extensive numerical studies have been performed with a variety of
methods, such as variational\cite{vmc-1,vmc-2,vmc-3,vmc-4,vmc-5}, 
fixed-node\cite{gf-qmc-1,gf-qmc-2,gf-qmc-3}, constrained-path auxiliary field\cite{afqmc-1,afqmc-2,afqmc-3}, determinental\cite{dqmc} and diagrammatic \cite{Svistunov2015} QMC,
dynamical\cite{dca-1,dca-1, PhysRevLett.104.247001, PhysRevB.88.245110} and variational\cite{vca-1,vca-2} cluster
approximations (DCA/VCA), dynamical mean-field theory
(DMFT)\cite{RevModPhys.68.13, dmft-1,dmft-2, PhysRevLett.110.216405, PhysRevB.82.155101}, density matrix renormalization group
(DMRG) and variational tensor network states \cite{PhysRevB.93.045116, peps}. 
Thermodynamic limit extrapolations
have been carried out with the aim of assessing the accuracy of the
methodologies in various regimes of interaction, filling factor and
temperature \cite{leblanc,zhang,chan}. On the other hand each of these
methods incur systematic errors which are extremely difficult to 
quantify and there is an urgent need to develop methods in which
convergence behaviour can be quantified internally.\\

In this paper, rather than attempting a direct numerical attack on the 
2D Hubbard Hamiltonian with a given technique, we ask if there is an
alternative \emph{exact} reformulation of the problem, the solution of
which is easier to approximate than that of the original problem. If 
this is the case (and this is obviously highly desirable), it should be demonstrable within the framework of a given technique, without 
reference to any other method. The physical basis for any observed
simplification should be transparent. Such an approach turns out to 
be possible, at least for intermediate interactions strengths based,
on a Gutzwiller non-unitary similarity transformation of the Hubbard
Hamiltonian. 


The Gutzwiller Ansatz \cite{hubbard-model-2,Rice1970} and Gutzwiller
approximation \cite{ga, gutzwiller-ansatz-2, Vollhardt1984, Shiba1988, PhysRevLett.62.324} are intensively studied methods to solve 
the Hubbard model. The parameter of the Ansatz is usually optimized
to minimize the energy expectation value by variational Monte Carlo
schemes\cite{rice, Horsch1983} based on a single Fermi-sea reference
state.
It has been long realized that the simple Gutzwiller Ansatz misses
important correlations\cite{fulde-1,vollhardt-1,vollhardt-2}, especially
in the strong interaction regime of the Hubbard model. More general,
Jastrow-like\cite{Jastrow1955}, correlators, including 
density-density\cite{density-density} and 
holon-doublon\cite{holon-doublon-2,holon-doublon-3}, have been proposed
to capture more physical features within the Ansatz. In addition the
Fermi-sea reference function have been extended to HF spin-density
waves\cite{gutzwiller-hf-spin-density} and BCS\cite{Anderson1987} 
wavefunctions\cite{gutzwiller-bcs,gutzwiller-bcs-2,
gutzwiller-bcs-3,gutzwiller-bcs-4,Gros1989,Gros1988,vmc-2, PhysRevB.88.115127, Baeriswyl2018} including
anti-ferromagnetic\cite{bcs-af-1,vmc-5} and charge order\cite{charge-order}. 
Recently there have been developments for a more efficient diagrammatic expansion of the Gutzwiller wavefunction (DE-GWF) \cite{Gebhard2012, PhysRevB.88.115127}, extensions of the GA to quasi-particle excitations \cite{PhysRevB.96.195126}, linear response quantities \cite{PhysRevB.95.075156} and the combination with the Schrieffer-Wolff transformation \cite{PhysRevB.95.161106} to capture Mott physics beyond the Brinkman-Rice scenario. 
\\
An alternative strategy is to use a Gutzwiller correlator  
to perform a non-unitary similarity transformation of the Hubbard
Hamiltonian, whose solution can be well approximated using a Slater
determinant. Such an approach is reminiscent of the quantum chemical
\emph{transcorrelated} method of Boys and 
Handy\cite{transcorrelated-method-1,transcorrelated-method-3}, as well 
as Hirschfelder\cite{transcorrelated-method-0}, in which a 
non-Hermitian Hamiltonian is derived on the basis of a Jastrow
factorisation of the wavefunction.\\
This idea was applied by Tsuneyuki\cite{tsuneyuki} to the Hubbard
model by minimizing the variance of the energy based on projection 
on the HF determinant. Scuseria and coworkers\cite{scuseria-1} and
Chan \emph{et.al.}\cite{non-stochastic} have recently generalized 
to general two-body correlators and more sophisticated reference
states, where the correlator optimization was not performed in a
stochastic VMC manner, but in the spirit of coupled-cluster theory, by
projecting the transformed Hamiltonian in the important subspace
spanned by the correlators. \\ 
These methods have in common that they are based on a single reference
optimization of the correlation parameters and thus the energy
obtained is on a mean-field level. We instead would like to fully
solve the similarity transformed Hamiltonian in a complete momentum
space basis. 
We will use a single reference optimization, based on
projection\cite{scuseria-1,non-stochastic}, to generate a 
similarity-transformed Hamiltonian (non-Hermitian with 3-body
interactions), whose ground-state solution (right-eigenvector) will be
using the projective FCIQMC\cite{initiator-fciqmc} method. \\ 
The remainder of the paper is organized as follows:\\
In Sec.~\ref{sec:tc_method} we recap the derivation of the Gutzwiller
similarity transformed Hubbard Hamiltonian and the projective solution
based on the restricted Hartree-Fock determinant. We also present
analytic and exact diagonalization results, to illustrate the influence
of the transformation on the energies and eigenvectors. In Sec.~\ref{sec:fciqmc_basics} we recap the basics of the FCIQMC method and
necessary  adaptations for its application to the similarity transformed
Hubbard Hamiltonian in a momentum-space basis, named similarity
transformed FCIQMC(ST-FCIQMC). In Sec.~\ref{sec:results} we benchmark the 
ST-FCIQMC method for the exact diagonalizable 18-site Hubbard model,
present ground- and excited-state energies. We observe an 
increased compactness of the right eigenvector of the non-Hermitian
transformed Hamiltonian. We also compare the results obtained with our
method for non-trivial 36- and 50-site lattices, at and off 
half-filling with interaction strengths up to $U/t = 4$. In Sec.~\ref{sec:discussion} we conclude our findings and explain future
applications for observables other than the energy and correct
calculation of left and right excited state eigenvectors. 

\section{\label{sec:tc_method}The similarity transformed Hamiltonian}
We would like to solve for the ground-state energy of the 
two-dimensional, single-band Hubbard 
model\cite{hubbard-model-1,hubbard-model-2,hubbard-model-3} 
with the Hamiltonian in a real-space basis
\begin{equation}
\label{eq:hamil-real}
 \hat H = -t \sum_{\langle ij \rangle,\sigma} a_{i,\sigma}^\dagger a_{j,\sigma} + U \sum_l n_{l,\u} n_{l,\d}.
\end{equation}
$a_{i,\s}^{(\dagger)}$ being the fermionic annihilation(creation)
operator for site $i$ and spin $\s$, $n_{l,\s}$ the number operator,
$t$ the nearest neighbor hopping amplitude and $U \ge 0$ the on-site
Coulomb repulsion. We employ a Gutzwiller-type 
Ansatz\cite{hubbard-model-2,gutzwiller-ansatz-2,
gutzwiller-correlator} for the ground-state wavefunction
\begin{align}
 \label{eq:jastrow-ansatz}
 \ket\Psi &= g^{\hat D} \ket{\Phi} =  \e^{\hat\tau} \ket{\Phi}, \quad \text{with} \\ 
 \hat \tau &= \ln g \hat D = J\sum_l n_{l,\u} n_{l,\d} \quad \text{and} \quad 0 \le g \le 1 \label{eq:gutzwiller-correlator},
\end{align}
where $\hat D$ is the sum of all double occupancies in $\ket{\Phi}$,
which are repressed with $0 \le g \le 1 \ra -\infty < J \le 0$. \\
In the Gutzwiller Ansatz, $\ket{\Phi}$ is usually chosen to be a
single-particle product wavefunction\cite{hubbard-model-2,
correl-functions}, $\ket{\Phi_0}$, such as the Fermi-sea solution of
the non-interacting $U=0$ system, or other similar forms such as
unrestricted Hartree-Fock spin-density 
waves\cite{gutzwiller-hf-spin-density}, or superconducting BCS
wavefunctions\cite{gutzwiller-bcs}. The parameter $J$ is usually
optimized via Variational Monte Carlo(VMC)\cite{Shiba1988},
minimizing the expectation value 
\begin{equation}
\label{eq:vmc}
E_{VMC} = \min_J \frac{\braopket{\Phi_0}{\e^{\hat \tau} \hat H \e^{\hat\tau}}{\Phi_0}}{\braopket{\Phi_0}{\e^{2\hat\tau}}{\Phi_0}}.
\end{equation}
In this work, however, $\ket{\Phi}$ is taken to be a full CI 
expansion in terms of Slater determinants 
\begin{equation}
\ket{\Phi} = \sum_i c_i \ket{D_i}
\end{equation} 
with which we aim to solve an equivalent exact eigenvalue equation
\begin{align}
&\e^{-\hat\tau} \hat H \e^{\hat\tau} \ket \Phi = \bar H \ket{\Phi} =   E\ket\Phi, \quad \text{with} \label{eq:similarity-tranformation} \\
 &\bar H = -t \sum_{<ij>,\sigma} \e^{-\hat\tau}a_{i,\sigma}^\dagger a_{j,\sigma} \e^{\hat\tau} + U \sum_l n_{l,\u} n_{l,\d}, \label{eq:real-space-start}
\end{align}
$\bar H$ denotes a similarity-transformed Hamiltonian. 
Eq.(\ref{eq:similarity-tranformation}) is obtained by 
substituting Eq.~(\ref{eq:jastrow-ansatz}) as an eigenfunction 
Ansatz into Eq.~(\ref{eq:hamil-real}) and multiplying with $e^{-\hat\tau}$ from the left, and due to $\left[n_{i,\sigma},\hat\tau\right] = 0$. The similarity transformation of Eq.~(\ref{eq:hamil-real}) moves the complexity of the correlated Ansatz for the wavefunction $\ket{\Psi}$ into the Hamiltonian, without changing its spectrum. It is a non-unitary transformation, and the resulting Hamiltonian is not Hermitian. Such similarity transformations have been introduced in quantum chemistry \cite{transcorrelated-method-0,transcorrelated-method-1,transcorrelated-method-2}  
in the context of a \emph{Slater}-Jastrow Ansatz, were it is known as the  \emph{transcorrelated}-method of Boys and Handy.
It was first applied to the Hubbard model by Tsuneyuki\cite{tsuneyuki}. The transcorrelated method has been quite widely applied in combination with explicitly correlated methods in quantum chemistry\cite{canonical-transcorr,fciqmc-transcorr,luo-ueg,ali-ueg}, with approximations being employed to terminate the commutator series arising from the evaluation of $\e^{-\hat\tau} \hat H \e^{\hat\tau}$\cite{tenno1,tenno2}. The explicit similarity transformation of the Hubbard Hamiltonian(\ref{eq:hamil-real}) with a Gutzwiller(\ref{eq:jastrow-ansatz})\cite{tsuneyuki,scuseria-2} or more general correlator\cite{scuseria-1,non-stochastic}, which can be obtained without approximations due to a terminating commutator series, has been solved on a mean-field level\cite{tsuneyuki}. In the present work, we will not restrict ourselves to a mean-field solution, but solve for the exact ground state of $\bar H$ with the FCIQMC method\cite{original-fciqmc, initiator-fciqmc}. 

\subsection{\label{subsec:transcorr-derivation} Recap of the derivation of $\bar H$}

Tsuneyuki\cite{tsuneyuki} and Scuseria \emph{et al.}\cite{scuseria-1} have provided a derivation of the similarity transformed Hubbard Hamiltonian, based on the Gutzwiller and more general two-body correlators, respectively. Their derivations result in a  
Hamiltonian expressed in real-space. Here we go one step further and obtain an exact \emph{momentum space} representation of the similarity transformed Hamiltonian, which is advantageous in the numerical study of the intermediate correlation regime. In this representation, the total momentum is an exact quantum number, resulting in a block diagonalised
Hamiltonian. This is computationally useful in projective schemes, especially where there are near-degeneracies in the exact spectrum close to the ground-state, which can lead to very long projection times and be problematic to resolve.  
Additionally, it turns out that even in the intermediate strength regime, 
the ground-state right eigenvector is dominated by a single Fermi determinant for the half-filled system. This is in stark contrast with the ground-state eigenvector of the original Hubbard Hamiltonian, which is highly multi-configurational in this regime. 

As seen in Eq.~(\ref{eq:real-space-start}) we need to compute the following transformation
\begin{equation}
\label{eq:taylor-expansion}
 \hat F(x) = \e^{-x\hat\tau} a_{i,\sigma}^\dagger a_{j,\sigma} \e^{x\hat\tau}
\end{equation}
which can be done by introducing a formal variable $x$ and performing a Taylor expansion (cf. the Baker-Campbell-Hausdorff expansion). The derivatives of (\ref{eq:taylor-expansion}) can be calculated as 
\begin{align}
\label{eq:derivatives}
 \hat F'(0) &= \commutator{a_{i,\sigma}^\dagger a_{j,\sigma}}{\hat \tau} = J \sum_l \commutator{a_{i,\sigma}^\dagger a_{j,\sigma}}{a_{l,\sigma}^\dagger a_{l,\sigma}} n_{l,\bar\sigma} \nonumber \\
  &= J a_{i,\sigma}^\dagger a_{j,\sigma} \left(n_{j,\bar\sigma} - n_{i,\bar\sigma} \right), \nonumber \\
 \hat F''(0) &= \commutator{\commutator{a_{i,\sigma}^\dagger a_{j,\sigma}}{\hat\tau}}{\hat\tau}  = 
 J\commutator{a_{i,\sigma}^\dagger a_{j,\sigma}(n_{j,\bar\sigma} - n_{i,\bar\sigma})}{\hat\tau} \nonumber \\
 &= J^2 a_{i,\sigma} a_{j,\sigma} \left(n_{j,\bar\sigma} - n_{i,\bar\sigma}\right)^2, \nonumber \\ 
 \hat F^{(n)}(0) &= \commutator{\commutator{a_{i,\sigma}^\dagger a_{j,\sigma}}{ \hat\tau}}{\dots,\hat\tau}\nonumber \\
 &  = J^n a_{i,\sigma} a_{j,\sigma} \left(n_{j,\bar\sigma} - n_{i,\bar\sigma}\right)^n.
\end{align}
With this closed form (\ref{eq:derivatives}) the Taylor expansion can be summed up as $\hat F(1) = a_{i,\sigma}^\dagger a_{j,\sigma}\e^{J(n_{j,\bar\sigma} - n_{i,\bar\sigma})}$  and Eq.~(\ref{eq:similarity-tranformation}) takes the final form of\cite{scuseria-1,scuseria-2,tsuneyuki}
\begin{equation}
\label{eq:hamil-exponential}
 \bar H = -t\sum_{<ij>,\sigma} a_{i,\sigma}^\dagger a_{j,\sigma} \e^{J(n_{j,\bar\sigma} - n_{i,\bar\sigma})} + U\sum_l n_{l,\u} n_{l,\d} 
\end{equation}
Due to the idempotency of the (Fermionic) number operators, $n_{i,\sigma}^2 = n_{i,\sigma}$, we have for $m \ge 1$:
\begin{align}
\label{eq:idempotency}
 \left(n_{j,\sigma} - n_{i,\sigma}\right)^{2m-1} = n_{j,\sigma} - n_{i,\sigma}, \quad \text{and} \nonumber \\ \left(n_{j,\sigma} - n_{i,\sigma}\right)^{2m} = n_{j,\sigma} + n_{i,\sigma} - 2 n_{i,\sigma} n_{j,\sigma}.
\end{align}
With Eq.~(\ref{eq:idempotency}) the exponential factor in Eq.~(\ref{eq:hamil-exponential}) can be calculated as 
\begin{align}
\label{eq:exp-factor}
\e^{J(n_{i,\sigma} - n_{j,\sigma})} =& 1 + \sum_{m=1}^\infty \frac{J^{2m-1}}{(2m-1)!}\left(n_{j,\sigma} - n_{i,\sigma}\right) 
+ \sum_{m=1}^\infty \frac{J^{2m}}{(2m)!}\left(n_{j,\sigma} + n_{i,\sigma} - 2n_{i,\sigma} n_{j,\sigma} \right)\nonumber \\
 =& 1 + \sinh(J)\left(n_{j,\sigma} - n_{i,\sigma}\right) + \left(\cosh(J) -1\right)\left(n_{j,\sigma} + n_{i,\sigma} - 2n_{i,\sigma} n_{j,\sigma}\right) \nonumber \\
 =& 1 + \left(\e^J - 1\right) n_{j,\sigma} + \left(\e^{-J}-1\right) n_{i,\sigma} - 2\left(\cosh(J) -1\right)n_{i,\sigma}n_{j,\sigma}. 
\end{align}
With Eq.~(\ref{eq:exp-factor}) we can write the final similarity transformed Hamiltonian as 
\begin{equation}
 \label{eq:final-H-realspace}
 \bar H =  \,\hat H - t \sum_{<ij>,\sigma} a_{i,\sigma}^\dagger a_{j,\sigma} \Big[\left(\e^J-1\right)n_{j,\bar\sigma} + \left(\e^{-J} -1\right)n_{i,\bar\sigma} 
 - 2\left(\cosh(J)-1\right) n_{i,\bar\sigma} n_{j,\bar\sigma}\Big].
\end{equation}
Formulated in a real-space basis the additional factor in Eq.~(\ref{eq:final-H-realspace}) is simply a nearest-neighbor density dependent renormalization of the hopping amplitude. For large interaction $U/t \gg 1$, as already pointed out by Fulde \emph{et al.}\cite{fulde-1}, the simple Ansatz (\ref{eq:jastrow-ansatz}) shows the incorrect asymptotic energy behavior, $E \sim -t^2/(U\ln{U})$ instead of $E \sim -t^2/U$\cite{vollhardt-1,vollhardt-2}, proportional to the magnetic coupling of the Heisenberg model for $U/t \gg 1$, due to the missing correlation between nearest-neighbor doubly occupied and empty sites. The Gutzwiller Ansatz does however provide a good energy estimate in the low to intermediate $U/t$ regime. For these values of $U/t$ the momentum space formulation is a better suited choice of basis, due to a dominant Fermi-sea determinant and thus a single reference character of the ground-state wavefunction. Thus we transform Eq.~(\ref{eq:final-H-realspace}) with
\begin{equation}
 a_{\mbf l,\s}^\dagger = \frac{1}{\sqrt{M}} \sum_\mbf k \e^{-i \mbf{kl}} c_{\mbf k,\s}^\dagger, 
\end{equation}
with $M$ being the size of the system and $c_{\mbf k,\s}^{(\dagger)}$ the annihilation(creation) operator of a state with momentum $\mbf k$ and spin $\s$ into a momentum space representation. The terms of Eq.~(\ref{eq:final-H-realspace}) become
\begin{eqnarray}
 \sum_{<ij>,\s}  a_{i,\s}^\dagger a_{j,\s} n_{j,\bar\s} 
 &= &\frac{1}{M}\sum_{\mbf{pqk},\s}\epsilon_{\mbf{p-k}} \cd{\mbf{p-k}}{\s}\cd{\mbf{q+k}}{\bar\s} 
 c_{\mbf q,\s} c_{\mbf p,\s}, \\
 \sum_{<ij>,\s} a_{i,\s}^\dagger a_{j,\s} n_{i,\bar\s} 
 &=& \frac{1}{M} \sum_{\mbf{pqk},\s} \epsilon_\mbf p \cd{\mbf{p-k}}{\s} \cd{\mbf{q+k}}{\bar\s} 
 c_{\mbf q,\bar\s} c_{\mbf p,\s}  \\
 \sum_{<ij>,\s} a_{i,\s}^\dagger a_{j,\s} n_{i,\bar\s} n_{j,\bar\s} 
 &=  &\frac{1}{M^2} \sum_{\mbf{pqskk'},\s} \epsilon_{\mbf{p-k+k'}} 
  \cd{\mbf{p-k}}{\s} \cd{\mbf{q+k'}}{\bar\s} \cd{\mbf{s+k-k'}}{\bar\s} 
 c_{\mbf s,\bar\s} c_{\mbf q,\bar\s} c_{\mbf p,\s},
\end{eqnarray}
with $\epsilon_\mbf k$ being the dispersion relation of the lattice. The original Hubbard Hamiltonian in k-space is
\begin{equation}
\label{eq:original-k-space}
 \hat H = -t\sum_{\mbf k,\s} \epsilon_\mbf k n_{\mbf k,\s} + \frac{U}{2M} \sum_{\mbf{pqk},\s} 
 \cd{\mbf{p-k}}{\s}\cd{\mbf{q+k}}{\bar\s} c_{\mbf q,\bar\s} c_{\mbf p,\s}.
\end{equation}
while the similarity transformed Hamiltonian in k-space is a function of the correlation parameter $J$
\begin{widetext}
\begin{align}
 \bar H(J) =& -t\sum_{\mbf k,\s} \epsilon_\mbf k n_{\mbf k,\s} + \frac{1}{M} \sum_{\mbf{pqk},\s} 
 \omega(J,\mbf{p,k}) \cd{\mbf{p-k}}{\s} \cd{\mbf{q+k}}{\bar\s} c_{\mbf q,\bar\s} c_{\mbf p,\s} \nonumber \\
 &+ 2t \frac{\cosh(J) - 1}{M^2} \sum_{\mbf{pqskk'},\s} \epsilon_\mbf{p-k+k'}             
 \cd{\mbf{p-k}}{\s} \cd{\mbf{q+k'}}{\bar\s} \cd{\mbf{s+k-k'}}{\bar\s} c_{\mbf s,\bar\s} c_{\mbf q,\bar\s} c_{\mbf p,\s}, \label{eq:final-H-k-space} \\
\omega(J,\mbf{p,k}) &= \frac{U}{2} - t \left[\left(\e^J - 1\right)\epsilon_\mbf{p-k} + 
 \left(\e^{-J} - 1\right) \epsilon_\mbf p \right].  
\end{align}
\end{widetext}
Comparing to the original Hubbard Hamiltonian in k-space (\ref{eq:original-k-space}), $\bar H$ (\ref{eq:final-H-k-space}) has a modified 2-body term and contains an additional 3-body interaction, which for $\mbf k = 0$ gives rise to parallel-spin double excitations. These are not present in the original Hamiltonian. As mentioned above, in contrast to other explicitly correlated approaches \cite{tenno-ohtsuka} this is an \emph{exact} similarity transformation of the original Hamiltonian 
and does not depend on any approximations. Hence the spectrum of this Hamiltonian is the same as that of (\ref{eq:hamil-real}). Unlike the canonical transcorrelation Ansatz of Yanai and Shiozaki\cite{canonical-transcorr}
which employ a unitary similarity transformation, the resulting Hamiltonian (\ref{eq:final-H-k-space}) is not 
Hermitian (the non-Hermiticity arising in the two-body terms), 
and hence its spectrum is not 
bounded from below. Variational minimization approaches are not applicable. The left and right eigenvectors differ, and form a biorthogonal basis $\bracket{\Psi_i^L}{\Psi_j^R} = 0$ for $i\ne j$. Tsuneyuki circumvented the lack of a lower bound of $\bar H$ by instead minimizing the variance of $\bar H$
\begin{equation}
\label{eq:min-var}
\min \braopket{\Phi_{HF}}{\left(\bar H - \expect{\bar H}\right)^2}{\Phi_{HF}},
\end{equation}
which is zero for the exact wavefunction and positive otherwise, to determine the optimal $J_{var}$. \\
Projective methods such as the Power method \cite{Golub}, or a
stochastic variant such as FCIQMC \cite{original-fciqmc}, can 
converge the right/left eigenvectors by multiple 
application of a suitable propagator, without recourse to 
a variational optimisation procedure, and this is the technique 
we shall use here.  
Since the matrix elements of (\ref{eq:final-H-k-space}) can be calculated analytically and on-the-fly, the additional cost of the 3-body term is no hindrance in our calculations and we do not need to apply additional approximations, unlike other explicitly correlated approaches \cite{yanai-chan,yanai-chan-2}. 
While complicating the calculation of observables other than the energy, due to the need to have both the left and right eigenvector of the now non-Hermitian Hamiltonian (\ref{eq:final-H-k-space}), the difference between the left and right eigenvectors actually proves to be beneficial for the sampling of the ground-state wavefunction in the FCIQMC method. This will be numerically demonstrated below in \ref{subsec:ED-results}.
As a side note, the use of more elaborate correlators, like 
density-density\cite{density-density} or 
holon-doublon\cite{fazekas-1,holon-doublon-2,holon-doublon-3} is 
no hindrance in the real-space formulation of the Hubbard model 
and is currently being investigated\cite{to-be-published-1}, but 
in the momentum-space basis would lead to even higher order
interactions and have not been further explored. 

\subsection{\label{subsec:analytic_results}Analytic results for the Hubbard model}
As a starting point we optimize the strength of the correlation
factor, controlled by the single parameter $J$ from the Ansatz
(\ref{eq:jastrow-ansatz}), by projecting the single determinant
eigenvalue equation $\left(\bar H(J) - E\right)\ket{\Phi_{HF}} = 0$ 
to the single basis of the correlation
factor\cite{scuseria-1,scuseria-2,non-stochastic}
\begin{equation}
\label{eq:optimize-j}
 \expect{\left(\hat \tau - \expect{\hat\tau}\right)^\dagger \bar H}_{HF} = \expect{\hat\tau^\dagger \bar H}_c = 0, 
\end{equation}
where $\expect{\dots}_c$ denotes a cumulant expression, where only linked diagrams are evaluated. HF denotes the state with all orbitals with $\abs {\mbf k} \le k_F$ being doubly occupied and $k_F$ being the Fermi surface. Eq.(\ref{eq:optimize-j}) is similar to a Coupled-Cluster equation. We simply report the results here (further information on the solution of Eq.~(\ref{eq:optimize-j}) can be found in Appendix \ref{app:optimize-j}). For an infinite system at half-filling, and only considering the two-body contribution of Eq.~(\ref{eq:optimize-j}), we can express the optimal $J$ which fulfils Eq.~(\ref{eq:optimize-j}), and the corresponding total energy per site, as (see \ref{app0-analytic-j})
\begin{align}
J_{opt}^{TDL} &= \sinh^{-1}\left(-\frac{5U\pi^6}{288t\left(16+\pi^4\right)}\right),\label{eq:2-body-infinite-j}\\
 E_{J}^{TDL} &= -t\frac{64}{4\pi^2} + \frac{U}{4} -t {J_{opt}^{TDL}}^2\left(\frac{16}{4\pi^2} + \frac{64}{\pi^6}\right).\label{eq:2-body-infinite-e}
\end{align}
The results of Eq.~(\ref{eq:optimize-j}-\ref{eq:2-body-infinite-e}) compared with AFQMC results\cite{zhang} on a $16\times 16$ half-filled square lattice are shown in Table \ref{tab:ana-transcorr-results}, for various values of $U/t$. The superscript (TDL) denotes thermodynamic limit results from Eq.~(\ref{eq:2-body-infinite-j}-\ref{eq:2-body-infinite-e}) for both the energy and $J$ parameter, and an absent superscript refers to the solution of Eq.~(\ref{eq:optimize-j}) for the actual finite lattices. At half-filling AFQMC does not suffer from a sign problem\cite{afqmc-sign} and is numerically exact. One can see that the results obtained from Eq.~(\ref{eq:optimize-j}-\ref{eq:2-body-infinite-e}) capture most of the correlation energy for low $U/t$ as expected. For larger 
$U/t$, due to the missing correlation between empty and doubly occupied sites in the Ansatz (\ref{eq:jastrow-ansatz}), the energies progressively deteriorate compared to the reference results. The optimal value $J_{opt}$ is also displayed in Table \ref{tab:ana-transcorr-results}. We use these values of $J$ obtained by solving Eq.~(\ref{eq:optimize-j}) as a starting point for our FCIQMC calculations to capture the remaining missing correlation energy. \\
\begingroup
\squeezetable
\begin{table*}
\centering
\caption{Ground-state energy per site for the half-filled $16\times16$ square lattice with periodic (PBC) and mixed (anti-)periodic (ABPC) boundary conditions along the (y-)x-axis. Thermodynamic limit extrapolations (TDL) for various values of $U/t$ obtained with AFQMC\protect\cite{zhang} are denoted as $E_{Ref}^{(TDL)}$. Results obtained by evaluating Eq.~(\ref{eq:optimize-j}) and Eq.~(\ref{eq:2-body-infinite-j}-\ref{eq:2-body-infinite-e}) are noted as $E_{J}^{(TDL)}$. The optimal value of $J$ is also shown. All energies are in units of $t$.}
\label{tab:ana-transcorr-results}
\begin{tabular}{cccccccccccc}
\toprule
 & \multicolumn{2}{c}{$U/t = 2$}  & \phantom{ab} & \multicolumn{2}{c}{$U/t = 4$} & \phantom{ab}& \multicolumn{2}{c}{$ U/t = 6$} & \phantom{ab} & \multicolumn{2}{c}{$U/t = 8$} \\ 
 & PBC & APBC && PBC & APBC && PBC & APBC && PBC & APBC\\
 \hline
$E_{ref} $ & -1.174203(23) & -1.177977(20) & & -0.86051(16) & -0.86055(16) & & -0.65699(12) & -0.65707(20) & & -0.52434(12) & -0.52441(12)\\[1pt]
$E_{J}$ & -1.151280\phantom{0()0} &  -1.166370\phantom{0()0} & & -0.76354\phantom{0()0} & -0.77769\phantom{0()0} & & -0.42855\phantom{0()0} & -0.44160\phantom{0()0} & & -0.12848\phantom{0()0} & -0.14051\phantom{0()0} \\[1pt]
$E_{J}/E_{ref} \%$ & 98.0\phantom{000}&99.0\phantom{000} & & 88.7\phantom{000} &90.4\phantom{000}  & & 65.3\phantom{000} &67.2\phantom{000}  & &24.5 \phantom{000} & 26.8\phantom{000} \\[2pt]
\hline 
$J_{opt}$ &-0.29233\phantom{00()0} &-0.28957\phantom{00()0}& & -0.56284\phantom{0()0} &-0.55787\phantom{0()0} & & -0.80107\phantom{0()0} &-0.79460\phantom{0()0} & & -1.00701\phantom{0()0} & -0.99956\phantom{0()0} \\[1pt]
\hline
$E_{ref}^{TDL}$ & \multicolumn{2}{c}{-1.1760(2)} & & \multicolumn{2}{c}{-0.8603(2)} & & \multicolumn{2}{c}{-0.6567(3)} & & \multicolumn{2}{c}{-0.5243(2)} \\[1pt]
$E_{J}^{TDL}$ & \multicolumn{2}{c}{-1.1609\phantom{()0}} & & \multicolumn{2}{c}{-0.7686\phantom{(0)}} & & \multicolumn{2}{c}{-0.4203\phantom{(0)}} & & \multicolumn{2}{c}{-0.0943\phantom{(0)}}\\[1pt]
$E_{J}^{TDL}/E_{ref} \% $ & \multicolumn{2}{c}{98.7} & & \multicolumn{2}{c}{89.4} & & \multicolumn{2}{c}{64.0} & &  \multicolumn{2}{c}{18.0}\\
\hline
$J_{opt}^{TDL}$ & \multicolumn{2}{c}{-0.29025\phantom{()}} & & \multicolumn{2}{c}{-0.55911\phantom{()}} & & \multicolumn{2}{c}{-0.79621\phantom{()}} & &  \multicolumn{2}{c}{-1.00142\phantom{()}}\\[1pt]

\botrule
\end{tabular}
\end{table*}
\endgroup
To compare this most basic combination of a on-site Gutzwiller correlator and a single restricted Hartree-Fock determinant as a reference in Eq.~(\ref{eq:optimize-j}) we show in Table \ref{tab:opt-j-comparison}  the percentage of the energy obtained with this method to more elaborate correlators and reference states, for different system sizes $M$, number of electrons $n_{el}$ and interaction strengths $U/t$. $E_{(S)UGST}$ in \ref{tab:opt-j-comparison} denotes a on-site Gutzwiller correlator with a (symmetry-projected) unrestricted Hartree-Fock reference state\cite{scuseria-2}. At half filling and $U/t \le 4$ we can capture more than $80\%$ of the energy obtained with a more elaborate reference determinant. Off half-filling the recovered energy is above $90\%$ up to $U/t = 4$. For a more dilute filling of $\expect n = 0.8$, for $M = 100$ and $U/t=2$, the energies agree to better than $99\%$. Although the absolute error in energy increases off half-filling, as already mentioned in Ref. [\onlinecite{scuseria-1,scuseria-2}] the relative error actually decreases\cite{rice,vmc-2,rice-2}, as can be seen in the comparison with the AFQMC reference results\cite{6x6_nel24,zhang,scuseria-1,leblanc}, $E_{ref}$ in Table \ref{tab:opt-j-comparison}. As expected, for $U/t > 4$ the results from Eq.~(\ref{eq:optimize-j}) drastically deteriorate compared to $E_{(S)UGST}$.\\
$E_{R/U J}$ in Table \ref{tab:opt-j-comparison} refer to energies obtained with restricted/unrestricted Hartree-Fock reference states with a general two-body correlator\cite{scuseria-1}, which includes all possible density-density correlations in addition to the on-site Gutzwiller factor. The comparison with $E_{RJ}$ shows that, as already found in Ref.~[\onlinecite{scuseria-1}], the Gutzwiller factor is by far the most important term in a general two-body correlator for low to intermediate values of $U/t \le 4$, with an agreement of over $98\%$ with $E_{RJ}$. Off half-filling, as can be seen in the $N = 36, n_{el} = 24$ and $U/t = 8$ case, the relative error remains small even for large interaction. The comparison with the available AFQMC reference results\cite{6x6_nel24,zhang,scuseria-1,leblanc}, $E_{ref}$, shows that the solution of Eq.~(\ref{eq:optimize-j}) with a on-site Gutzwiller correlator and a restricted Hartree-Fock reference, retrieves above $80\%$ of the energy for $U/t \le 4$.
This gives us confidence that the optimal $J$ obtained by this method is appropriate in the context of the Gutzwiller similarity transformed Hamiltonian, which we further solve with the FCIQMC method.
\begingroup
\begin{table}
\caption{\label{tab:opt-j-comparison}\small Fraction of the total energy obtained with the Gutzwiller Ansatz(\ref{eq:jastrow-ansatz}) based on the Hartree-Fock determinant(\ref{eq:optimize-j}) compared with a Gutzwiller correlator with a unrestricted Hartree-Fock reference $E_{UGST}$ and subsequent symmetry projection $E_{SUGST}$\cite{scuseria-2} and a general two-body correlator with a Hartree-Fock reference $E_{RJ}$ and unrestricted Hartree-Fock reference $E_{UJ}$\protect\cite{scuseria-1} and numerically exact AFQMC reference results\protect\cite{6x6_nel24,zhang,scuseria-1,leblanc} for different number of sites $M$, number of electrons $n_{el}$ and interaction strengths $U/t$.}
\begin{tabular}{cccccccc}
\toprule
   $M$ & $n_{el}$ & $U/t$ & $\% E_{UGST}$ & $\% E_{SUGST} $ & $\% E_{RJ}$ & $\% E_{UJ} $ & $\% E_{ref}$ \\
 \hline
  16 &   14 &  2 &   97.34 &  97.03 &  99.69 &  97.30 &  96.79 \\
  16 &   14 &  4 &   92.81 &  91.70 &  99.02 &  93.07 &  90.75 \\
  16 &   14 &  8 &   72.68 &  70.16 &  92.28 &  73.84 &  66.60 \\
  16 &   16 &  2 &   80.85 &  80.75 &  99.82 &  93.77 &  93.16 \\
  16 &   16 &  4 &   81.84 &  80.18 &  98.57 &  82.61 &  80.24 \\
  16 &   16 &  8 &   21.37 &  20.19 &  47.54 &  21.81 &  20.08 \\
 \hline
  36 &   24 &  4 &         &        &  99.67 &        &  98.26 \\
  36 &   24 &  8 &         &        &  98.72 &        &  93.53 \\
  64 &   28 &  4 &         &        &  99.74 &        &  99.19 \\
  64 &   44 &  4 &         &        &  99.77 &        &  98.38 \\
 \hline
 100 &   80 &  2 &  100.00\phantom{0} &  99.98 &        &        &  99.84 \\
 100 &   80 &  4 &   99.85 &  99.61 &        &        &  97.43 \\
 100 &  100 &  2 &   97.69 &  97.56 &        &        &  97.27 \\
 100 &  100 &  4 &   88.50 &  88.08 &        &        &  87.39 \\
 100 &  100 &  6 &   65.70 &  65.19 &        &        &  64.04 \\
 100 &  100 &  8 &   25.01 &  24.76 &        &        &  23.89 \\
 \botrule
\end{tabular}
\end{table}
\endgroup

\subsection{Exact diagonalization study.}
\label{subsec:ED-results}
Due to the non Hermitian nature of $\bar H$ (\ref{eq:final-H-k-space}) the left and right ground-state eigenvectors $\ket{\Phi_0^{L/R}}$ differ and depending on the strength of the correlation parameter $J$ they can have a very different form. The most important characteristic for the projective FCIQMC method is the \emph{compactness} of the sampled wavefunction. As a measure of this compactness we chose the $L^2$ norm of the exact $\ket{\Phi_0^{L/R}}$ contained in the leading HF-determinant $\ket{\Phi_{HF}}$ and double excitations $\ket{\Phi_{ij}^{ab}} = c_a^\dagger c_b^\dagger c_i c_j \ket{\Phi_{HF}}$ (spin-index omitted) thereof, i.e. the sum over the squares of the coefficients of these determinants: $L^2_{(0,2)}=c_0^2 + \sum_{i<j,a<b} c_{ijab}^2$. 

\begin{figure}
\centering
\includegraphics[scale=1]{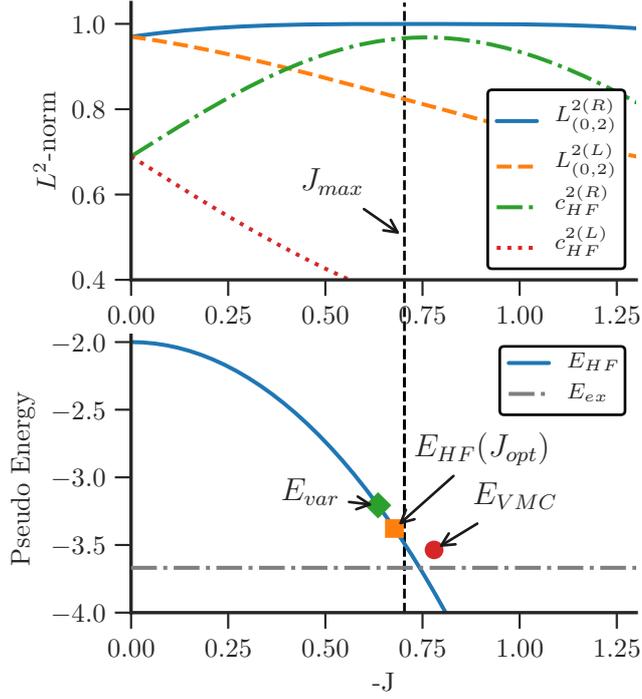}
\caption{\label{fig:l2-norm-ED} (Color online) Top: $L^2$ norm of the HF state $c_{HF}^{2(L/R)}$ and within the HF determinant and double excitations, $L_{(0,2)}^{(L/R)}$, for a half-filled 6-site Hubbard chain with periodic boundary conditions at $U/t = 4$ and $\mbf k = 0$ for the left $\ket{\Phi_0^L}$ and right $\ket{\Phi_0^R}$ ground-state eigenvector of $\bar H$ as a function of $-J$. Bottom: The Hartree-Fock energy $E_{HF}(J) = \braopket{\Phi_{HF}}{\bar H(J)}{\Phi_{HF}}$ as a function of $-J$. The dash-dotted line indicates the exact ground-state energy $E_{ex}$ for this system, and since $\bar H$ is not Hermitian $E_{HF}$ can drop below the exact energy. Also indicated are the results of minimizing the variance of the energy of the similarity transformed Hamiltonian $E_{var}$ by Tsuneyuki \protect\cite{tsuneyuki}, the result of solving Eq.~(\ref{eq:optimize-j}) $E_{HF}(J_{opt})$ and the result from a VMC optimization $E_{VMC}$ \protect\cite{pablo-priv-comm}. The vertical dashed line indicates $J_{max}$ where $L_{(0,2)}^2$ of $\ket{\Phi_0^R}$ is maximal. All energies are in units of $t$ and the two panel share the x-axis.}

\end{figure}

As a simple example, in the top panel of Fig.~\ref{fig:l2-norm-ED} we show the coefficient of the Hartree-Fock determinant $c_{HF}^2$ and $L^2_{(0,2)}$ of $\ket{\Psi_0}$ for the 1D half-filled 6-site Hubbard model with periodic boundary conditions at $U/t = 4$ and $\mbf k = 0$, as a function of the correlation parameter $J$. $J = 0$ corresponds to the original Hamiltonian (\ref{eq:original-k-space}). In the bottom panel of Fig.~\ref{fig:l2-norm-ED} the Hartree-Fock energy $E_{HF}$ and results of minimizing the variance of the energy $E_{var}$ by Tsuneyuki \cite{tsuneyuki}, $E_{HF}(J_{opt})$ with $J_{opt}$ obtained by solving Eq.~(\ref{eq:optimize-j}) and Variational Monte Carlo(VMC) results \cite{pablo-priv-comm} $E_{VMC}$ are shown. Due to the fact that $\bar H$ is not Hermitian any more, and hence not bounded by below, $E_{HF}$ can drop below the exact ground-state energy $E_{ex}$, also displayed in Fig.~\ref{fig:l2-norm-ED}, so following Tsuneyuki\cite{tsuneyuki} we termed the energy axis ``pseudo energy". There is a huge increase in $c_{HF}^2$ and the $L^2_{(0,2)}$ norm of the $\ket{\Phi_0^R}$ until an optimal value of $J_{max}$, close to the $J_{opt}$ obtained by solving Eq.~(\ref{eq:optimize-j}), see Tab.~\ref{tab:opt-j-explicit}, where $L^2_{(0,2)} \approx 1 $, followed by subsequent decrease. The result obtained by minimizing the energy variance $E_{var}$ is higher in energy and farther from $J_{max}$ than $J_{opt}$. And, although $E_{VMC}$ is closer to $E_{ex}$, the optimized correlation parameter obtained by VMC is also farther from $J_{max}$ than $J_{opt}$. 
At the same time $c_{HF}^2$ and $L^2_{(0,2)}$ of $\ket{\Phi_0^L}$ shows a monotonic decrease with increasing $-J$. This shows that the amount of relevant information contained within the HF determinant and double excitations thereof can be drastically increased in the right eigenvector, whilst decreased in the left one. For the calculation of the energy, where only the right eigenvector is necessary, a more efficient sampling with the stochastic FCIQMC method should be possible. 

\section{The FCIQMC method}
\label{sec:fciqmc_basics}
The FCIQMC\cite{original-fciqmc,initiator-fciqmc} method is a projector Monte Carlo method, based on the integrated imaginary-time Schr\"odinger equation
\begin{equation}
\frac{\der\ket\Psi}{\der t} = -\hat H \ket\Psi \xrightarrow[]{\int dt} \ket{\Psi_0} \propto \lim_{t \ra \infty} \e^{-t \hat H} \ket {\Psi(t=0)},
\label{eq:imaginary-schrodinger}
\end{equation}
where $t$ is an imaginary-time parameter and $\ket {\Psi(t=0)}$ is an arbitrary initial wave-function with non-zero overlap with $\ket {\Psi_0}$. One obtains the ground-state energy and wave-function by repeatedly applying a first-order difference approximated projector of (\ref{eq:imaginary-schrodinger}) to the initial state
\begin{equation}
\label{eq:first-order-proj}
\ket{\Psi_0} = \lim_{n\ra\infty} \left[\hat{\mbf 1} - \Delta t \left(\hat H - E_S \hat{\mbf 1}\right) \right]^n \ket{\Psi(0)}, 
\end{equation}
for $\Delta t < E_W^{-1}$\cite{spectral-width}, with $E_W = E_{max} - E_0$ being the spectral width of $\hat H$. If the energy shift $E_S=E_0$, convergence to a non-diverging and non-zero solution can occur. In practice the shift is dynamically adapted to keep the walker number, explained below, constant, which corresponds to keeping the $L^1$ norm of the sampled wave-function constant. If the sampled wave-function is a stationary solution to the projector, adapting $E_S(t)$ to keep the $L^1$ norm constant guarantees $E_S(t) \ra E_0$.\\ $\ket{\Psi(t)}$ is expanded in an orthonormal basis of $N_d$ Slater determinants, $\ket{\Psi(t)} = \sum_i^{N_d} c_i(t) \ket{D_i}$ and the working equation for the coefficients $c_i$ is
\begin{equation}
\label{eq:working-equation}
c_i(t + \Delta t) = \left[1 - \Delta t \left(H_{ii} - E_S(t)\right)\right] c_i(t) - \Delta t \sum_{j\neq i}^{N_d} H_{ij} c_j(t).
\end{equation}
Eq.(\ref{eq:working-equation}) governs the dynamics of a population of $N_w$ signed walkers, which stochastically sample the ground-state wave-function $\ket{\Psi_0}$. Since the number of states, $N_d$, grows combinatorially with system size, only a stochastic ``snap-shot" of $\ket{\Psi_0}$ is stored every iteration, where only states occupied by at least one walker are retained. The diagonal term of Eq.~(\ref{eq:working-equation}), $ 1 - \Delta t \left(H_{ii}-E_S\right)$, increases or decreases the number of walkers on state $i$. The shift energy $E_S(t)$ is dynamically adapted after the chosen number of walkers $N_w$ is reached to keep it constant over time. The off-diagonal term, $-\Delta t H_{ij}$, creates new walkers from an occupied determinant $i$ to a connected state $j$. The sum is is sampled stochastically by only performing one of these ``spawning'' events with a probability 
\begin{equation}
\label{eq:p-spawn}
p_{spawn} = \Delta t \vert H_{ij}\vert/p(j|i)
\end{equation}
and the sign of the new walker is: $-\sign(H_{ij})$. At the end of each iteration, walkers with opposite sign on the same determinant, which is a reflection of the fermionic sign problem, are removed from the simulation. For sufficiently many walkers the sign problem can be controlled for many systems. In the intermediate to high interaction regime of the Hubbard model, the number of necessary walkers is proportional to the Hilbert space size, making this ``original" FCIQMC method impractical. The initiator approximation i-FCIQMC\cite{initiator-fciqmc} overcomes this exponential bottleneck at the cost of introducing an initiator bias. It does so by 
allowing only walkers on determinants above a certain population threshold $n_{init}$ to spawn onto empty determinants (thereby dynamically truncating the Hamiltonian matrix elements between low-population determinants and empty ones). This is the source of the initiator error, which can be systematically reduced by increasing the walker population. Nevertheless, convergence can be slow, especially if the ground state wavefunction is highly spread out over the Hilbert space, as is often the case for strongly correlated systems.  On the other hand, convergence can be rapidly obtained if the ground-state eigenvector is relatively compact, and does not require any prior knowledge of this fact, nor of the nature of the compactness. In fact, it is precisely for this reason that the similarity transformations can be of use in the i-FCIQMC method.\\  
In addition to the shift energy $E_S(t)$, a projected energy 
\begin{equation}
\label{eq:projected-energy}
E_P(t) = \frac{\braopket{D_{ref}}{\hat H}{\Psi(t)}}{\bracket{D_{ref}}{\Psi(t)}},
\end{equation}
with $\ket{D_{ref}}$ being the most occupied determinant in a simulation, is an estimate for the ground-state energy, if $\ket{\Psi(t)} \approx \ket{\Psi_0}$. An improved estimate (with a smaller variance) can also be obtained by projection onto a multi-determinant trial wave-function $\bra{\Phi_{trial}}$, 
\begin{equation}
\label{eq:projected-trial}
E_{trial}(t) = \frac{\braopket{\Phi_{trial}}{\hat H}{\Psi(t)}}{\bracket{\Phi_{trial}}{\Psi(t)}},
\end{equation}
where $\bra{\Phi_{trial}}$ is obtained as the eigenvector of a small sub-space diagonalised similarity-transformed Hamiltonian. This is particularly useful in open shell problems, where there are  several dominant determinants in the ground-state wave-function, and as a result $E_{trial}(t)$ can exhibit notably smaller fluctuations than $E_P(t)$.

\subsection{The ST-FCIQMC approach}
\label{subsec:tc_fciqmc}
In variational approaches the lack of a lower bound of the energy due to the non-Hermiticity of the similarity transformed Hamiltonian poses a severe problem. As a projective technique, however, the FCIQMC method has no inherent problem sampling the right ground-state eigenvector, obtaining the corresponding eigenvalue by repetitive application of the projector (\ref{eq:first-order-proj}). Additionally, the increased compactness of $\ket{\Phi_0^R}$ observed in Section\ref{subsec:ED-results}, due to the suppressed double occupations via the Gutzwiller Ansatz, tremendously benefits the sampling dynamics of i-FCIQMC. 
On the other hand, the implementation of the additional 3-body term in (\ref{eq:final-H-k-space}) necessitate major technical changes to the FCIQMC algorithm. We changed the \neci\cite{neci} code to enable triple excitations. Due to momentum conservation and the specific spin-relations ($\s\s\bar\s$) of the involved orbitals and efficiently analytically calculable 3-body integrals of (\ref{eq:final-H-k-space}), these could be  implemented without a major decrease of the performance of the algorithm.
In fact the contractions of the 3-body term in (\ref{eq:final-H-k-space}), namely $\mbf k = 0 \, \veebar \, \mbf k' = 0 \, \veebar \, \mbf{k = k'} \, \veebar\, \mbf{q+k' = s}$ lead to an $\bigO M$ additional cost of the 2-body matrix element, which have the largest detrimental effect on the performance.
(There is an $\bigO{M^2}$  scaling for the diagonal matrix elements, coming from the $\mbf{k = k' = 0}$ contraction, but this has a negligible overall effect, since we store this quantity for each occupied determinant and is thus not computed often). The additional cost for 2-body integrals is similar to the calculation of 1-body integrals in conventional \emph{ab-initio} quantum chemistry calculations and unavoidably hampers the performance, but is manageable. 
Surprisingly, the performance penalty, due to the additional three-body interactions, decreases with increasing strength of the correlation parameter $J$. 
This is due to the following fact:
the performance of the FCIQMC method depends heavily on the ``worst-case'' $\abs{H_{ij}}/p(i\vert j)$ ratio, where $p(i\vert j)$ is the probability to spawn a new walker on determinant $\ket{D_i}$ from $\ket{D_j}$ and $\abs{H_{ij}}$ is the absolute value of the corresponding matrix element $\braopket{D_i}{\hat H}{D_j}$. The time-step $\Delta t$ of the FCIQMC simulation is on-the-fly adapted to ensure the ``worst-case'' product remains close to unity, $\Delta t \abs{H_{ij}}/p(i\vert j) \approx 1$.
Due to the $M^2$ increased number of non-zero matrix elements in the similarity transformed Hamiltonian (\ref{eq:final-H-k-space}), the time-step $\Delta t$ \emph{formally} scales as $\bigO{M^{-5}}$\textemdash momentum conservation decreases the scaling by a factor of $M$\textemdash, instead of $\bigO{M^{-3}}$ for the original Hubbard Hamiltonian (\ref{eq:original-k-space}). However, although numerous, the 3-body terms are easier to calculate and sampled less often, due to their relatively small magnitude and the actually important scaling measure is the necessary number of walkers to achieve a desired accuracy; which is tremendously reduced for the similarity transformed Hamiltonian.

\begin{figure}
\centering
\includegraphics[scale=1]{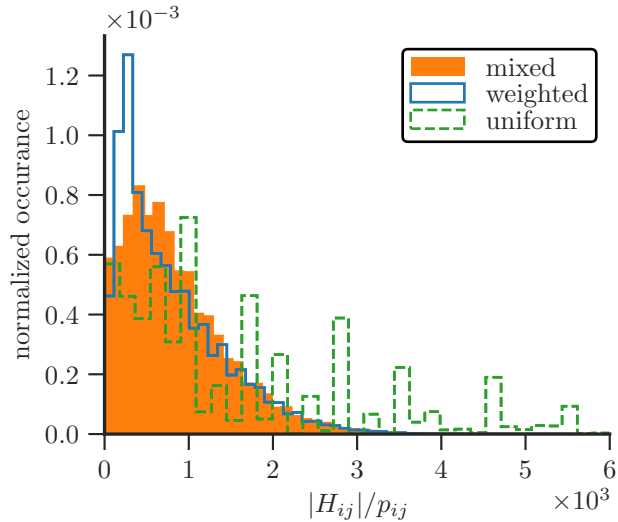}
\caption{\label{fig:histogram} (Color online) Histogram of $\abs{\bar H_{ij}}/p_{ij}$ ratios for the half-filled, 50-site, $U/t = 4$ Hubbard model with periodic boundary conditions for uniform, weighted and mixed generation probabilities.}
\end{figure}

An optimal sampling in FCIQMC would be achieved, if for every pair $(i,j): p(i|j) \approx \abs{H_{ij}}$ and thus $\Delta t \approx \min(1,E_W^{-1})$. Since $\bar H$ is not Hermitian, the off-diagonal matrix elements are not uniform, as in the original Hamiltonian (\ref{eq:original-k-space}). We therefore need to ensure an efficient sampling by a more sophisticated choice of $p(i\vert j)$. Additionally we can separate $p(i\vert j)$ into a probability to perform a double(triple) excitation $p_D$($1 - p_D$) since there are still no single excitations in (\ref{eq:final-H-k-space}), due to momentum conservation. This split into doubles or triples, gives us the flexibility, in addition to $\Delta t$, to also adapt $p_D$ during run-time to bring $\abs{H_{ij}}/p(i\vert j)$ closer to unity. We observed that with increasing correlation parameter $J$ the dynamically adapted probability to create triple excitations increased and thus reducing the detrimental additional cost to calculate 2-body matrix elements. \\ 
When we perform the spawning step in FCIQMC we first decide if we are perform a double excitation with probability $p_D$, or a triple excitation with probability $1-p_D$. Then we pick two or three electrons $mn(l)$ from the starting determinants ($\ket{D_j}$) uniformly, with probability $p_{elec}$. For a double excitation, due to momentum conservation, we only need to pick one unoccupied orbital, since the second is fixed to fulfil $\mbf k_m + \mbf k_n = \mbf k_a + \mbf k_b$. To guarantee $p(i|j) \sim \abs{H_{ij}}$ we loop over the unoccupied orbitals $a$ in $\ket{D_j}$ and create a cumulative probability list with the corresponding matrix elements $\abs{H_{ij}(mn,ab)}$ and thus pick the specific excitation with $p(i|j) \sim \abs{H_{ij}}$. The cost of the is $\bigO{M^2}$, due to the loop over the unoccupied orbitals $\sim M$ and the $\bigO M$ cost of the double excitation matrix element calculation. For triple excitations the procedure is similar, except we pick 3 electrons $m_{\s},n_\s,l_{\bar \s}$ with $p_{elec}$, then we pick orbital $a_{\bar \s}$ of the minority spin uniformly with $p_a = 1/n_{holes}$ and pick orbital $b_\s$ weighted from a cumulative probability list proportional to $\abs{H_{ij}}$; the third orbital $c_\s$ is again determined by momentum conservation $\mbf k_m + \mbf k_n + \mbf k_l = \mbf k_a + \mbf k_b + \mbf k_c.$ As opposed to double excitations, this is only of cost $\bigO M$, due to the loop over unoccupied orbitals in $\ket{D_j}$ do determine $b_\s$. We term this procedure as \emph{weighted} excitation generation algorithm. \\
An alternative and simpler algorithm is to pick the unoccupied orbitals in a uniform way. This decreases the cost per iteration, but  also leads to a worse worst-case $H_{ij}/p_{ij}$ ratio leading to a 
decreased time-step $\Delta t$. Fig. \ref{fig:histogram} shows the histogram of the $\abs{\bar H_{ij}}/p_{ij}$ ratios for the \emph{weighted} procedure, described above, the \emph{uniform} choice of empty orbitals and a \emph{mixed} method for the half-filled 50-site Hubbard model at $U/t = 4$. In the mixed method, the $\bigO{M^2}$ scaling double excitations in the weighted scheme, are done in a uniform manner, while the $\bigO M$ scaling triple excitations are still weighted according to their matrix element $\abs{\bar H_{ij}}$. Longer tails in a distribution indicate the need for a lower time-step to ensure $\Delta t \abs{\bar H_{ij}}/p_{ij} \approx 1$. It is apparent that the \emph{mixed} scheme possesses the optimal combination of favorable $\abs{\bar H_{ij}}/p_{ij}$ ratios similar to the \emph{weighted} method, with manageable additional cost per iteration, shown in Table~\ref{tab:pgen-ratios}. Table \ref{tab:pgen-ratios} shows the relative difference of the time-step $\Delta t$, time per iteration $t_{iter}$, number of aborted excitations $n_{abort}$ and acceptance rate $n_{accept}$ of the different methods compared to the original $J = 0$ uniform sampled half-filled, 50-site Hubbard model with $U/t = 4$. While there is a 7-fold increase of the time per iteration of the \emph{mixed} scheme compared to the original uniform, the time-step is almost a third larger and the accepted rate of spawning events a third higher. $n_{abort}$ indicates those spawning attempts which originally are proposed in the uniform scheme, but are finally rejected, due to zero matrix elements or are Fermi blocked. This quantity is also decreased by more than a half in the mixed method compared to the uniform original scheme. $n_{accept}$ indicates the number of accepted proposed spawning events and is directly related to the $p_{spawn}$ (\ref{eq:p-spawn}). The choice of the excitation generator is therefore not straightforward and depends on the interaction strength and $J$: the \emph{uniform} scheme performs better than expected at small $U/t$, whilst the mixed scheme performs better at large $U/t$. 


\begingroup
\begin{table}
\caption{\label{tab:pgen-ratios}\small Ratio of the time-step $\Delta t$, time per iteration $t_{iter}$, aborted $n_{abort}$ and accepted excitations $n_{accept}$ of the different excitation generation probabilities compared to the original ($J = 0$) method with uniform choice of open orbitals as a reference for the half-filled, 50 site, $U/t = 4$ Hubbard model with periodic boundary conditions.}
\begin{tabular}{cccccc}
\toprule
$J$ & method & $\% \Delta t $ & $\% t_{iter}$ & $\% n_{abort}$ & $\% n_{accept}$ \\
\hline
 $0$ & weighted 	& 100.00 &   \phantom{0}240.12 &   \phantom{0}0.00 &  100.00 \\
 $\neq 0$& uniform 	& \phantom{0}21.02 &   \phantom{0}169.33 &  93.72 &   \phantom{0}77.31 \\
 $\neq 0$& mixed 	& \phantom{0}35.55 &   \phantom{0}719.22 &  40.14 &  130.64 \\
 $\neq 0$& weighted 	& \phantom{0}45.01 &  1506.72 &   \phantom{0}0.00 &  165.29 \\
\botrule
\end{tabular}
\end{table}

\section{Results}
\label{sec:results}
We assessed the performance of initiator ST-FCIQMC (i-ST-FCIQMC) for different Hubbard lattices, 
as a function of the Gutzwiller correlation factor $J$. As a starting guide for $J$, we use  
$J_{opt}$ obtained by solving Eq.~(\ref{eq:optimize-j}) for the specific lattice size $M$, number of electrons $n_{el}$ and interaction strength $U/t$, and calculate the ground-state and excited states energies with i-ST-FCIQMC. In particular, we were interested in the rate of convergence of the energy with respect walker 
number, or in other words, how quickly the initiator error disappeared with increasing walker number. 
The optimal values of $J$ for each studied system can be found in Table~\ref{tab:opt-j-explicit} in the appendix \ref{app:optimize-j}. All energies are given per site and in units of the hopping parameter $t$ and the lines in the figures \ref{fig:18in18_error} to \ref{fig:excited-curve} are guides to the eye.

\subsection{18-site Hubbard model}
\label{subsec:18-site}
We first study the 18-site Hubbard model on a square lattice with tilted boundary conditions (see Fig.~\ref{fig:lattice}), which can be exactly diagonalised: at half-filling and zero total momentum $(\mbf k = 0$) it has a Hilbert space of $\sim 10^8$ determinants. All the exact reference results were obtained by a Lanczos diagonalization\cite{olle-priv-comm}. 
For this system ST-FCIQMC could be run either in ``full'' mode or with the initiator approximation, i-ST-FCIQMC. 
This enables us to assess to two separate questions, namely the performance of i-ST-FCIQMC
with regards to initiator error on the one hand, and compactness of the wave-functions resulting from the 
similarity transformation (without the complicating effects of the initiator approximation), on the other.

Fig. \ref{fig:18in18_error} shows the error (on a double-logarithmic scale) of the energy per site in the initiator calculation, as a function of walker number. The left panel shows results for the $U/t = 2$  system. As one can see there is a steep decrease in the error and even with only  $10^4$ walkers, for a correlation parameter of $J = -1/4$ (close to the $J_{opt}$) 
the error is below $10^{-4}$. At $10^6$ walkers it is well below $10^{-6}$, almost two orders of magnitude lower than the original (i.e. $J=0$) Hamiltonian at this value of $N_w$. This also confirms the assumption that the chosen Ansatz for the correlation function (\ref{eq:jastrow-ansatz}) is particularly useful in the low $U/t$ regime. 

Results for an intermediate strength, $U/t = 4$, are shown in the right panel of Fig.~\ref{fig:18in18_error}. Compared to the $U/t = 2$, more walkers are needed to achieve a similar level of accuracy. The two sources for this behaviour are:\\
Firstly, i-FCIQMC calculations on the momentum-space Hubbard model are expected to become more difficult with increasing interaction strength $U/t$, due to the enhanced multi-configurational character of the ground-state wave-function. Secondly, the chosen correlation Ansatz (\ref{eq:jastrow-ansatz}) is proven to be more efficient in the low $U/t$ regime\cite{fulde-1}. Nevertheless, the results shown in Fig.~\ref{fig:18in18_error} show a steep decrease in the double logarithmic plot of the error with increasing walker number. The decrease is steeper for $J=-1/2$, close to the analytic result obtained with $J_{opt} = -0.5234470$. For $J=-1/2$, at walker numbers up to $5\cdot 10^7$ we are, to within error bars, at the exact result. At a walker number of $10^7$ there is a two order of magnitude difference in the error of the $J=-1/2$ and $J=0$ result.\\
\begin{figure}
\centering
\includegraphics[scale=1]{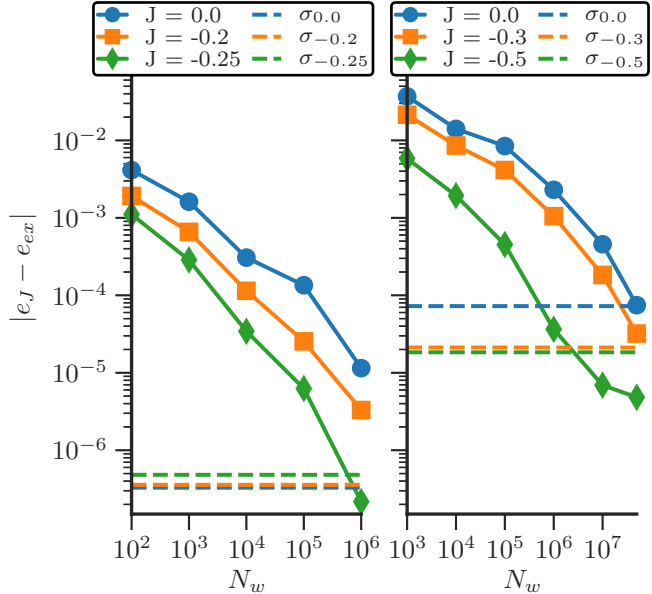}
\caption{\label{fig:18in18_error} (Color online) The error of the energy per site $\abs{e_{J} - e_{ex}}$ for the half-filled 18-site Hubbard model for the original $J = 0$ and different strengths of the correlation parameter $J$ at $\mbf k = 0$ for $U/t = 2$ (left) and $U/t = 4$ (right) versus the walker number $N_w$. The dashed lines indicate the statistical errors of the $N_w = 10^6$ results with $n_{init} = 1.2$ for $U/t = 2$ and of the $N_w = 5\cdot 10^7$ with $n_{init} = 2.0$ for $U/t = 4$. The exact reference results were obtained by Lanczos diagonalization\protect\cite{olle-priv-comm}.}
\end{figure}

To confirm the more compact form of the right ground-state eigenvector, mentioned in Sec.~\ref{subsec:ED-results}, we performed two analyses. First was the study of the $L^2$ norm captured within the HF determinant $c_{HF}^{2(L/R)}$ and additionally double excitations, $L^2_{(0,2)}$, for the ST-FCIQMC wave-function. In Fig.~\ref{fig:l2_norm_18_U4} $L^2_{(0,2)}$ of the left and right ground-state eigenvector of the half-filled 18-site, $U/t = 4$ Hubbard model as a function of $-J$ is shown. The results were obtained by running full non-initiator ST-FCIQMC calculations to avoid any influence of the initiator error. The left eigenvector was obtained by running with positive $J$, which corresponds to a conjugation of $\bar H$. 
\begin{equation}
\label{eq:h-dagger}
\bar H(J)^\dagger = \left(\e^{-\hat\tau} \hat H \e^{\hat \tau}\right)^\dagger = \e^{\hat\tau} \hat H \e^{-\hat \tau} = \bar H(-J),
\end{equation}
since $\hat H^\dagger = \hat H$ and $\hat \tau^\dagger = \hat \tau$. And 
\begin{equation}
\label{eq:left-ev}
\bar H^\dagger \ket{\Phi_L} = E \ket{\Phi_L}, \quad \text{with} \quad \ket{\Phi_L} = \e^{-\hat\tau}\ket\Psi.
\end{equation} 
Similar to the exact results for the 6-site model in Fig.~\ref{fig:l2-norm-ED}, the right eigenvector shows a huge compactification compared to the original $J=0$ result, going from $0.65$ to over $0.9$ for $L_{(0,2)}^2$.
The ``optimal'' value of $J = J_{max} = -0.57444831$, where $L^2_{(0,2)}$ is maximal, is close to the analytical obtained $J_{opt} = -0.5234470$, indicating that we can simply use $J_{opt}$ without further numerical 
optimization of $J$, and still be close to optimal conditions.  
\begin{figure}
\centering
\includegraphics[scale=1]{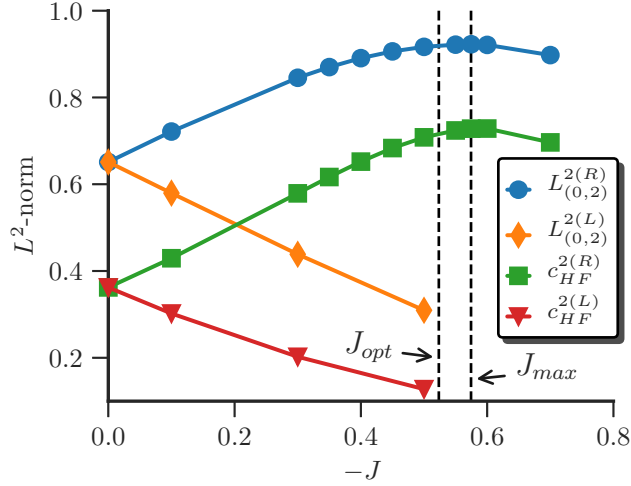}
\caption{\label{fig:l2_norm_18_U4} (Color online) $L^2$ norm captured within the HF determinant $c_{HF}^{2(L/R)}$ and additionally double excitations, $L^2_{(0,2)}$, for the half-filled 18-site Hubbard model at $U/t = 4$ for the left and right ground-state eigenvector of the non-Hermitian similarity transformed Hamiltonian (\ref{eq:final-H-k-space}) as a function of $-J$. The results were obtained by a non-initiator ST-FCIQMC calculation. $\ket{\Phi_L}$ was sampled by running the simulation with positive $J$, which corresponds to conjugating $\bar H$. $J_{max}$ indicates the position of the maximum of $L^2_{(0,2)}$ and $J_{opt}$ is the result of solving Eq.~(\ref{eq:optimize-j}).}

\end{figure}
Fig. \ref{fig:l2_norm_per_exlevel} shows the $L^2$ norm contained in each excitation level relative to the HF determinant for the half-filled, 18-site, $U/t = 4$ Hubbard model for different values of $J$. For $J = -1/2$ there is a huge increase in the $L^2$ norm of the HF determinant, indicated by an excitation level of $0$, while it drops of very quickly for higher excitation levels and remains one order of magnitude lower than the $J = 0$ result above an excitation level of $5$.\\
\begin{figure}
\centering
\includegraphics[scale=1]{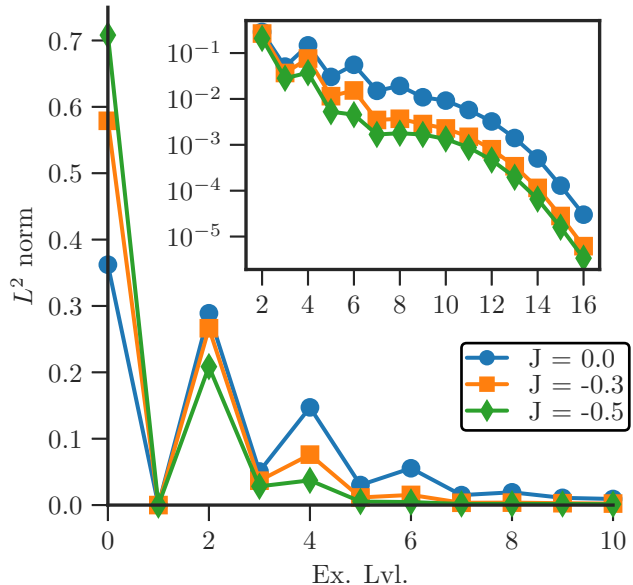}
\caption{\label{fig:l2_norm_per_exlevel}
 (Color online) The $L^2$ norm contained in each excitation level relative to the HF determinant, indicated by excitation level $0$, for the half-filled, 18-site, $U/t = 4$, $\mbf k = 0$ Hubbard model for different values of $J$. The inset shows the tail of the same data on a logarithmic scale.}
\end{figure}
Our second analysis on the compactness of $\ket{\Phi_0^R}$ consisted of running truncated CI\cite{truncated-ci} calculations, analogous to the CISD, CISDTQ, etc. methods of quantum chemistry. Here we truncate the Hilbert space by only allowing excitation up to a chosen value $n_{trunc}$ relative to the HF determinant. Fig. \ref{fig:truncated-ci} shows the error of the energy per site as a function of $n_{trunc}$ for different $J$. For $J = -1/2$ we are below $10^{-4}$ accuracy already at only quintuple excitation, which is two orders of magnitude lower than the original $J=0$ result at this truncation level. The error bars in the inset of Fig.~\ref{fig:truncated-ci} are from the $n_{trunc} = 8$ simulations for each value of $J$, which do not differ much from $n_{trunc} = 5$ to $n_{trunc} = 8$ for each simulation. Already at sextuple excitations we are well within error bars of the exact result for $J = -1/2$, with an error that is two orders of magnitude smaller than the $J = 0$ result. 
\begin{figure}
\centering
\includegraphics[scale=1]{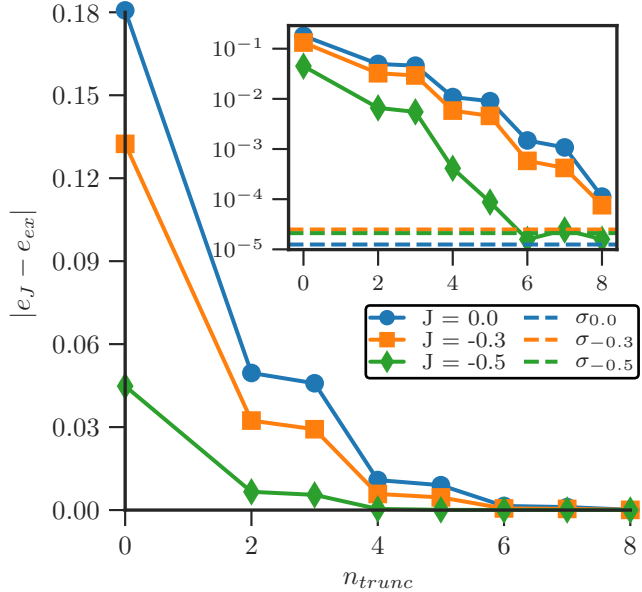}
\caption{\label{fig:truncated-ci} (Color online) Error of the energy per site versus the excitation level truncation $n_{trunc}$ in the half-filled, 18-site, $U/t = 4$, $\mbf k = 0$ Hubbard model for different values of $J$. The inset shows the absolute error on a logarithmic scale. The dashed lines in the inset indicate the statistical error for the $n_{trunc} = 8$ results for each value of $J$.}
\end{figure}

\subsubsection{\label{subsubsec:14-electrons} Off half-filling 14 $e^-$ in 18-sites}
We have also investigated the applicability of the i-ST-FCIQMC method to the off-half-filling case, and also to excited states calculations. To this end we calculated the ground, first and second excited states of the 14 $e^-$ in 18-sites, $U/t = 4$, $\mbf k = 0$ system. Such a system can be prepared by removing 4 electrons (2 $\alpha$ and 2 $\beta$ spins)  from the corners of the Fermi-sea determinant, and using this as a starting point for an i-ST-FCIQMC simulation. Excited states are obtained by running multiple independent runs in parallel and applying a Gram-Schmidt orthogonalization to a chosen number of excited states\cite{excited-fciqmc} 
\begin{equation}
\label{eq:excited-states-projector}
\ket{\Phi_i(t + \Delta t)} = \hat P_i(t+\Delta t) \left[\mbf 1 - \Delta t\left(\hat H - E_{S,i}\right)\right]\ket{\Phi_i(t)}, 
\end{equation}
with $\hat P_i(t)$ being the orthogonal projector
\begin{equation}
\label{eq:orthogonal-projector}
\hat P_i(t) = \mbf 1 - \sum_{j<i} \frac{\ket{\Phi_j(t)}\bra{\Phi_j(t)}}{\bracket{\Phi_j(t)}{\Phi_j(t)}} \quad \text{with} \quad E_j < E_i.
\end{equation}
However, since the set of right eigenvectors $\ket{\Phi_i^R}$ of a non-Hermitian operator
are not guaranteed to be orthogonal, we cannot rely on the projected energy estimate (\ref{eq:projected-energy}) as an estimate for the excited state energy. By orthogonalising each eigenvector $\bracket{\Phi_i^E}{\Phi_j^R} \mustbe 0$ for $i \neq j$ ($i$ and $j$ indicate the excited states), the sampled excited states will in general not be identical to the exact right eigenvectors of $\bar H$. On the other hand, since the spectrum of $\bar H$ does not change due to the similarity transformation~(\ref{eq:similarity-tranformation}), the shift energy $E_S$ in (\ref{eq:excited-states-projector}), dynamically adapted to keep the walker number constant, remains a proper estimate for the excited states energy. This interesting fact is developed further in appendix \ref{app:6-site}. Additionally, if the excited state belongs to a different spatial or total-spin symmetry sector the overlap to the ground-state is zero, so our excited state approach within the FCIQMC formalism, via orthogonalisation, correctly samples these orthogonal excited states. \\
Fig. \ref{fig:excited-curve} shows the energy per site error of the ground-, first and second excited state of the 14 $e^-$ in 18-site, $U/t = 4$, $\mbf k = 0$ system, compared to exact Lanczos reference results\cite{olle-priv-comm} for different values of $J$ versus the walker number $N_w$, obtained via the shift energy $E_{S,i}$. All states show a similar behaviour of the energy per site error. The closer $J$ gets to the optimal value of $J_{opt} = -0.557941$ for $U/t = 4$, which is determined for $E_0$, one observes that more than an order of magnitude fewer walkers are necessary to achieve the same 
accuracy as the $J=0$ case. This is true for all the states considered. For $E_1$, the energy difference of the $N_w = 10^7$ and $J = -1/2$ calculation is already within the statistical error of $10^{-5}$, hence the non-monotonic behaviour. The size of the absolute error of these states is comparable to the absolute error of the half-filled, 18-site, $U/t = 4$ system, shown in the right panel of Fig.~\ref{fig:18in18_error}. Since, without a chemical potential, the total ground-state energy per site of the $n_{el} = 14$ system, $e_0^{(14)} = -1.136437$, is lower than the half-filled one, $e_0^{(18)} = -0.958466$, the relative error is in fact smaller off half-filling. As already mentioned above and shown in Table~\ref{tab:ana-transcorr-results} and \ref{tab:opt-j-comparison}, the projective solution based on the restricted Hartree-Fock determinant (\ref{eq:optimize-j}) also yields smaller relative errors off half-filling. These results give us confidence to also apply the i-ST-FCIQMC method to systems off half-filling and for excited states energy calculations. 

\begin{figure*}
\centering
\includegraphics[scale=1]{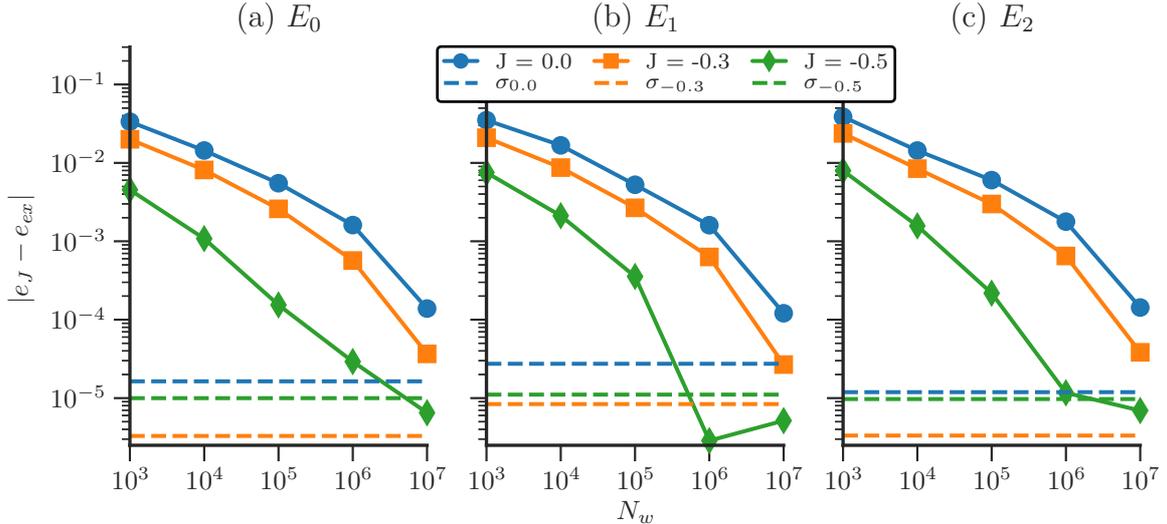}
\caption{\label{fig:excited-curve} (Color online) Energy per site error compared to exact Lanczos results\protect\cite{olle-priv-comm} for the 14 $e^-$ in 18-site, $U/t = 4$, $\mbf k = 0$ Hubbard model for the (a) ground-, (b) first and (c) second excited state as a function of walker number $N_w$. All three panels share the $x$ and $y$ axes. The dashed lines indicate the statistical errors of the $N_w = 10^7$ for each value of $J$.}
\end{figure*}
\subsubsection{Symmetry Analysis}
\label{subsubsec:symmetry-analysis}
As mentioned above, the set of right eigenvectors of a non-Hermitian operator are in general not orthogonal, except when they belong to different irreducible representations and/or total spin symmetry sectors. Here we investigate the interesting influence of the similarity transformation on the symmetry properties of the truncated low-energy subspace of the 14$e^-$ in 18-sites system with total $\mbf k = 0$. There are 8 important low energy determinants with the 5 lowest energy $\mbf k$ points double occupied and 4 $e^-$ distributed among the 4 degenerate orbitals of the corner of the square $\mbf k_1 = (-1,-1), \mbf k_2 = (1,-1), \mbf k_3 = (-1,1)$ and $\mbf k_4 = (1,1)$ to preserve the total $\mbf k = 0$ symmetry. This is illustrated in Fig.~\ref{fig:lattice}, where filled red circles indicate the doubly occupied k-points and half-filled green circles the singly occupied ones (color online). The point group of the square lattice is $D_{4h}$. There are 2 closed shell determinants in this set, with opposite k-points doubly occupied and 6 open-shell determinants with all 4 corners of the BZ singly occupied. Without a correlation parameter all these 8 determinants are degenerate in energy, while with $J\neq 0$ this degeneracy is lifted. To study the low energy properties of this system we diagonalized $\bar H$ in this sub-space. Table \ref{tab:symmetry-analysis} shows the results. We found that with $J = 0$ the ground state of this subspace has a different spatial and spin symmetry, $^5B_{1g}$, than the ground state of the full system, which belongs to $^1A_{1g}$. At approximately $J\approx -0.71$ there is a crossover and the subspace ground state changes to $^1A_{1g}$ symmetry. The first excited state in the subspace is then the $^5B_{1g}$, which is also the symmetry of the first excited state of the full system and the 2nd excited state is of $^1B_{2g}$ symmetry, the same as 2nd excited state of the not truncated system. Therefore the similarity transformation not only ensures a more compact form of the ground- and excited state wavefunctions,  
but also \emph{correctly orders the states obtained from subspace diagonalizations}. The implication is that, in the off half-filling Hubbard model , the structure of ground state has very important contributions arising from high-lying determinants, so much so that they are necessary to get 
a qualitatively correct ground-state wavefunction (i.e. one with the correct symmetry and spin). 
With the similarity transformed Hamiltonian, however, this is not the case. Even small sub-space diagonalizations yield a ground-state wavefunction with the same symmetry and spin as the exact one. In other words, the similarity transformation effectively downfolds information from higher lying regions of the Hilbert space to modify the matrix elements between the low-lying determinants. Since the structure of the ground-state eigenvector already has the correct symmetry (and therefore signs) in small subspaces, the rate of convergence of the solution with respect to the addition of further determinants is much more rapid. We believe this is a crucial property which leads to the observed greatly improved convergence rate of the i-ST-FCIQMC method in the off half-filling regime.

\begingroup
\begin{table}
\caption{\label{tab:symmetry-analysis}\small Irreducible representations and spin symmetry of the ground-state $E_0$ and first two excited states $E_1$, $E_2$ of the $U/t=4$, $\mbf k = 0$, 14 $e^-$ in the 18-site Hubbard model for the full and sub-space(subsp.) solutions for different values of $J$. For a large enough correlation parameter $J$ the ground-state of the low-energy subspace resembles the correct symmetry structure as for the full solution.}
\begin{tabular}{cccccc}
\toprule 
 & $E_0$ & \phantom{ab} & $E_1$ & \phantom{ab} & $E_2$ \\ 
 \hline
Full & $^1A_{1g}$& \phantom{ab} & $^5B_{1g}$ & \phantom{ab} & $^1B_{2g}$\\ 
$J=0$ subsp. & $^5B_{1g}$& \phantom{ab} & $^1A_{1g}$& \phantom{ab} & $^1B_{2g}$\\ 
$J=-0.72$ subsp. & $^1A_{1g}$& \phantom{ab} & $^5B_{1g}$& \phantom{ab} & $^1B_{2g}$\\ 
\botrule
\end{tabular}
\end{table}
\endgroup

\begin{figure}
\centering
\includegraphics[scale=1]{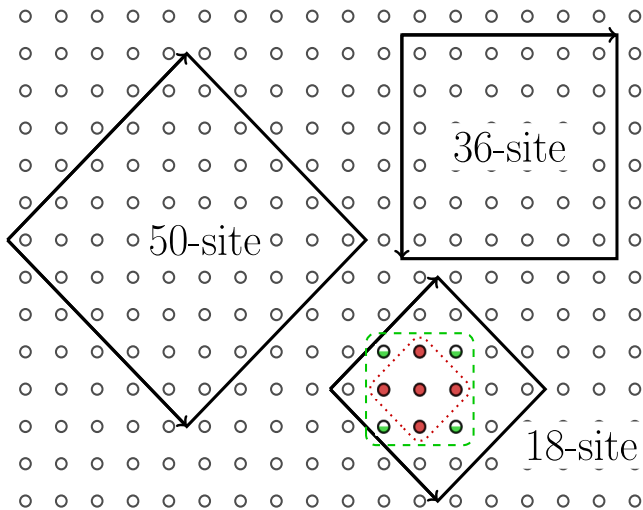}
\caption{\label{fig:lattice}(Color online) The three different square lattices studied in this paper. The 18- and 50-site lattice have tilted periodic boundary conditions with lattice vectors $\mbf R_1 = (3,3), \mbf R_2 = (3,-3)$ and $\mbf R_1 = (5,5), \mbf R_2 = (5,-5)$ respectively. The 36-site lattice is studied with periodic and mixed, periodic along the x-axis and anti-periodic boundary conditions along the y-axis. The filled red circles in the 18-site lattice indicate the doubly occupied states and the half-filled green circles the singly occupied states in the sub-space study in \ref{subsubsec:symmetry-analysis}.}
\end{figure}

\subsection{Results for the 36- and 50-site Hubbard model}
\label{subsec:50-site}
To put the i-ST-FCIQMC method to a stern test, we applied it to two much larger systems, namely 36-site and 50-site lattices, which are well beyond the capabilities of exact diagonalization. In the case of the 36-site ($6\times6$) lattice, we considered two boundary conditions, 
namely fully periodic (PBC) and a mixed periodic-anti-periodic (along the x- and y-axes respectively), the latter being used in some studies 
to avoid degeneracy of the non-interacting solution\cite{vmc-2}. We considered two fillings, namely half-filling and 24$e^-$, at $U/t = 2$ and $U/t = 4$.
The optimal $J_{opt}$ was determined by solving Eq.~(\ref{eq:optimize-j}) and is listed in Table~\ref{tab:opt-j-explicit} in the appendix \ref{app:optimize-j}. For the $6\times 6$ by lattice we compared our results to AFQMC calculations\cite{zhang}, which are numerically exact at half-filling\cite{afqmc-sign}. The results are shown in Table~\ref{tab:36-50-results}. While the original i-FCIQMC method shows a large error even at walker numbers up to $N_w = 5\cdot 10^8$ the i-ST-FCIQMC method agrees with the AFQMC reference to within one $\sigma$ error bars in all but one case (PBC $U/t=4$ half-filled), where the agreement is within $2\sigma$. Even in that case the energies agree to better than $99.8\%$. The small discrepancy could be due to this system being strongly open-shell, making equilibration more challenging.  

The 50-site Hubbard lattice corresponds to a $5\sqrt 5\times 5\sqrt 5$ tilted square, which has been widely investigated using the AFQMC method.  We considered half-filling and various off half-filling, $n_{el} = 26, 42, 44, 46$ and $48$ cases for $U/t = 1,2,3$ and $4$ and calculated the ground-state energy. The optimal $J$ are listed in Table~\ref{tab:opt-j-explicit} in the appendix \ref{app:optimize-j}. This system size, especially with increasing $U/t$ and off half-filling, was previously unreachable with the FCIQMC method. We compare our half-filling results to AFQMC\cite{afqmc-sign,afqmc-white,afqmc-sorella} reference values, which do not have a sign problem at half-filling\cite{afqmc-sign}. The remaining sources of error are extrapolation to zero temperature and finite steps, both of which are expected to be very small. Off half-filling, exact AFQMC results are not available, and we compare against constrained-path AFQMC(CP-AFQMC)\cite{cp-afqmc-1,cp-afqmc-2} and linearized-AFQMC(L-AFQMC)\cite{sorella-lafqmc}. \\
Table \ref{tab:36-50-results} shows the results for various fillings and $U/t$ values the reference calculations, the original i-FCIQMC and the i-ST-FCIQMC method. We converged our results for this system size up to a walker number of $N_w = 10^9$. We can see that the original i-FCIQMC method performs well for the weakly correlated half-filled $U/t = 1$ system, but fails to reproduce the reference results at $U/t = 2$ for this system size, and the discrepancy worsens with increasing interaction. The i-ST-FCIQMC method on the other hand agrees within error bars with the reported reference calculation up to $U/t = 3$ at half-filling. Similar to the half-filled 36-site lattice, the i-ST-FCIQMC results are slightly below the AFQMC reference results at $U/t = 4$, which could be a finite temperature effect of the AFQMC reference results. 

We investigated the half-filled 50-site $U/t=4$ system further by performing the convergence of a  truncated CI expansions, 
similar to the 18-site lattice. The results are shown in Fig. \ref{fig:50-exlevel}. The convergence with excitation level truncation shows that convergence occurs from above, and at 6-fold excitations we are converged to statistical accuracy to the fully unconstrained simulation. The energy at 6-fold truncation is indeed slightly below the AFQMC result, although the discrepancy is small (approximately 0.1$\%$). It is intriguing that the CI expansion of the 50-site lattice is converged at 6-fold excitations, which is the same as observed for the 18-site lattice.
This suggests that linear solutions to the similarity-transformed Hamiltonian may be size-consistent to a greater degree than similar truncations to the original untransformed Hamiltonian. 
This question however is left for a future study. 

For $U/t = 4$ off half-filling the i-ST-FCIQMC energies are consistently slightly above the reference AFQMC results.
However the approximations in the off half-filling AFQMC calculations can lead to energies below the exact ones. For example\cite{6x6_nel24}, 
CP-AFQMC on a $4 \times 4$ lattice with 14 $e^-$ and $U/t = 4$ gives an energy of $-0.9863(1)$ compared to an exact energy of $-0.9840$, i.e. an overshoot $~0.2\%$. Similar overshoots are observed at other fillings.
In the off half-filling regime in the 50-site system at $U/t = 4$, CP-AFQMC overshoots our i-ST-FCIQMC results by similar amounts. 
Therefore our results are in line with ED results for smaller lattices, and thus represent a new set of benchmarks for the off half-filling 50-site Hubbard model.

\begingroup
\begin{table*}
\caption{\label{tab:36-50-results}\small Zero temperature, $\mbf k = 0$ ground-state energy results for the 36-site and 50-site Hubbard model for various interaction strengths $U/t$, number of electrons $n_{el}$ and periodic (PBC) and mixed (anti-)periodic boundary conditions along the (y-)x-axis, obtained with the initiator FCIQMC and the i-ST-FCIQMC method compared with available (CP-)AFQMC and linearised-AFQMC reference results\protect\cite{zhang,6x6_nel24,sorella-priv-comm,sorella-lafqmc,rep-qmc}. The differences to the AFQMC reference energies are displayed as $\Delta E$. The correlation parameter $J$ was chosen close to the optimal $J_{opt}$ obtained by solving Eq.~(\ref{eq:optimize-j}) listed in Table~\ref{tab:opt-j-explicit} of App. \ref{app:optimize-j} for the specific $U/t$ value. An initiator threshold of $n_{init} = 2.0$ was chosen and convergence of the energy up to a walker number of $N_w = 10^9$ was checked.}
\begin{tabular}{ccccccccc}
\toprule
M & $U/t$ & $n_{el}$ & BC & $E_{ref}$ & i-FCIQMC & $\Delta E_{J=0}$ & iST-FCIQMC & $\Delta E_{J}$ \\
\hline
36 & 4 & 24 & APBC&  				  &		-1.155828(40) 	&  					& 	-1.159285(24)\phantom{0} 			&  			\\[1pt]
36 & 4 & 24 & PBC & 	-1.18525(4)\phantom{0} 	  & -1.182003(57) 	&	\phantom{0}0.003247(97)	& 	-1.1852109(52)	& 	 \phantom{00}0.000039(45)	\\[1pt]
36 & 2 & 36 & APBC& 	\phantom{0}-1.208306(56) & 	\phantom{0}-1.2080756(39)	&	\phantom{0}0.000230(60)	&	-1.2082581(17) 	& 	 \phantom{00}0.000048(58) 	\\[1pt]
36 & 2 & 36 & PBC & 	-1.15158(14)	  & -1.149734(95) 	& 	0.00185(24)		&	-1.151544(18)\phantom{0}	& 	 \phantom{0}0.00004(16) 	\\[1pt]
36 & 4 & 36 & APBC& 	-0.87306(56)  & 	-0.847580(84) 	& 	\phantom{0}0.025480(64)	& 	-0.872612(50)\phantom{0} 	& 	 \phantom{0}0.00045(61) 	\\[1pt]
36 & 4 & 36 & PBC & 	-0.85736(25)  & 	-0.82807(87)\phantom{0} 	& 	0.0293(11)\phantom{0} 		&	-0.85625(30)\phantom{00} 	& 	\phantom{0}0.00111(55) 	\\[1pt]
\hline
50 & 1 & 50 & PBC & 	-1.43718(11)  & 	\phantom{0}-1.4371801(18)	&	0.00000(11) 	& 	\phantom{0}-1.43724130(44)	& 	-0.00006(11) 	\\[1pt]
50 & 2 & 50 & PBC & 	-1.22278(17)  & 	-1.220590(16) 	& 	0.00219(19) 	& 	-1.2228426(80) 	& 	-0.00006(18)		\\[1pt]
50 & 3 & 50 & PBC & 	-1.03460(30)  & 	-1.023064(35) 	& 	0.01154(34) 	& 	-1.034788(18)\phantom{0} 	& 	-0.00019(32)		\\[1pt]
50 & 4 & 50 & PBC & 	\phantom{0}-0.879660(20) & 	-0.83401(15)\phantom{0}  	& 	0.04565(17) 	& 	-0.880657(60)\phantom{0} 	& 	\phantom{0}-0.000997(80)	\\[1pt] 
50 & 4 & 48 & PBC & 	-0.93720(15)  & 	-0.89610(12)\phantom{0} 	& 	0.04110(27) 	& 	-0.93642(40)\phantom{00} 	& 	 \phantom{0}0.00078(55) 	\\  
50 & 4 & 46 & PBC & 	\phantom{00}-0.9911420(86)&		-0.95550(15)\phantom{0} 	& 	0.03564(24) 	& 	-0.990564(89)\phantom{0} 	& 	 \phantom{0}0.00058(18) 	\\[1pt] 
50 & 4 & 44 & PBC & 		\phantom{0}-1.037883(59) &		-1.006483(38) 	& 	\phantom{0}0.031400(97)	& 	-1.037458(47)\phantom{0} 	& 	 \phantom{0}0.00043(11) 	\\[1pt]  
50 & 4 & 42 & PBC & 		\phantom{0}-1.079276(66) & 	-1.053756(64) 	& 	0.02552(13) 		& 	-1.078908(69)\phantom{0} 	& 	 \phantom{0}0.00037(14) 	\\
50 & 4 & 26 & PBC &	\phantom{0}-1.115640(20) & -1.113874(16) &  \phantom{0}0.001766(36) & -1.1159016(39) & \phantom{0}-0.000262(24) \\
\botrule
\end{tabular}
\end{table*}
\endgroup

\begin{figure}
\centering
\includegraphics[scale=1]{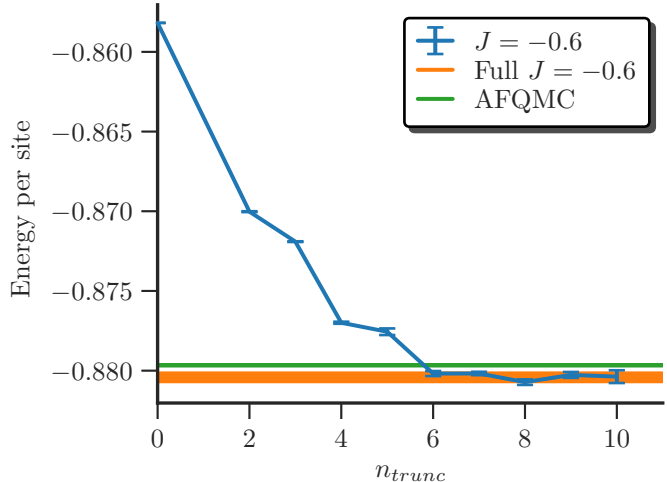}
\caption{\label{fig:50-exlevel}(Color online) The energy per site versus the excitation level truncation $n_{trunc}$ in the half-filled, 50-site, U/t = 4, k = 0 Hubbard model. AFQMC reference\protect\cite{rep-qmc} and non-truncated i-ST-FCIQMC results are also shown. Similar to the 18-site system at half-filling (Fig. 6.), the energies are well-converged at 6-fold excitations.}
\end{figure}

\section{Discussion, Conclusion and Outlook}
\label{sec:discussion}
We have used a projective solution based on the restricted Hartree-Fock determinant to obtain an optimized Gutzwiller correlation parameter. For low to intermediate interaction strength, this method generally recovers over $80\%$ of the ground-state energy. Based on this mean-field solution we derived a similarity transformed Hubbard Hamiltonian, generated by the Gutzwiller Ansatz, in a momentum-space basis. We solved for the exact ground- and excited states energy of this non-Hermitian operator with the FCIQMC method. We have shown that the right eigenvector of the non-Hermitian Hamiltonian has a dramatically more compact form, due to suppression of energetically unfavorable double occupancies, via the Gutzwiller Ansatz. This increased compactness of the right eigenvectors allowed us to solve the Hubbard model for system sizes, which were previously unreachable with the i-FCIQMC method. We benchmarked our results with highly accurate AFQMC reference results and find extremely good agreement at and off half-filling up to interaction strengths of $U/t = 4$. We hope this combination of a similarity transformation based on a correlated Ansatz for the ground-state wavefunction and subsequent beyond mean-field solution with FCIQMC can aid the ongoing search for the phase diagram of the two-dimensional Hubbard model in the thermodynamic limit.\\
An important extension  of the present work will be to compute
observables other than the energy. 
To compute the expectation values of operators $\hat O$ which do not commute with the Hamiltonian
we need additionally to obtain the left eigenvector of the non-Hermitian $\bar H$ with the Ansatz $\bra{\Psi_L} = \bra{\Phi} \e^{-\hat\tau}$
\begin{equation}
\bra{\Psi} \hat H = \bra{\Phi}\e^{-\hat\tau} H = E \bra\Phi \e^{-\hat\tau}.
\end{equation}
The expectation value of the similarity transformed operator 
$\bar O = \e^{-\hat\tau}\hat O \e^{\hat \tau}$ with $\ket{\Phi_{R/L}}$ yields the desired 
\begin{equation}
\frac{\braopket{\Phi_L}{\bar O}{\Phi_R}}{\bracket{\Phi_L}{\Phi_R}} = \frac{\braopket{\Psi}{\e^{\hat\tau} \e^{-\hat\tau} \hat O \e^{\hat \tau}\e^{-\hat\tau}}{\Psi}}{\braopket{\Psi}{\e^{\hat\tau}\e^{-\hat\tau}}{\Psi}} = \expect{\hat O}.
\end{equation}
As already observed in \ref{subsec:18-site}, applying $\bar H$ with $-J$ yields the left eigenvector $\ket{\Phi_L} = \e^{\hat\tau}\ket{\Psi}$. To perform this in the FCIQMC we only need to run two independent simulations in parallel, as is already done in replica-sampling of reduced density matrices\cite{fciqmc-rdm}, where the two runs use an opposite sign of the correlation parameter $J$. Observables, $\hat O$, which commute with the chosen Gutzwiller correlator $\commutator{\hat \tau}{\hat O} = 0$, such as the double occupancy $\expect{n_\u n_\d}$, can be calculated by the 2-body RDM obtained with the left and right eigenvector 
\begin{equation}
\Gamma_{pq,rs} = \braopket{\Phi_L}{c_p^\dagger c_q^\dagger c_s c_r}{\Phi_R}, 
\end{equation} 
with normalized $\bracket{\Phi_L}{\Phi_R} = 1$ and $p,q,r$ and $s$ denoting spin-orbital labels in the momentum space. Non-commuting observable, $\commutator{\hat\tau}{\hat O}$, have to be similarity transformed $\bar O = \e^{-\hat\tau}\hat O\e^{\hat\tau}$ and might require higher order density matrices. 

Simultaneous calculation of the left eigenvectors $\ket{\Phi^i_L}$ also allows us to obtain the correct excited state wave-functions, in addition to the already-correct excited state energy via the shift energy $E_{S,i}$ mentioned in \ref{subsubsec:14-electrons} and App. \ref{app:6-site}, in the following manner: For $m$ excited states we run $2m$ simulations in parallel, where every odd numbered calculation solves for a right eigenstate $\ket{\Phi^i_R}$, which is orthogonalized against all $\ket{\Phi_L^j}$ with $E_j < E_i$. And vice versa, every even numbered simulation solves for a left eigenvector $\ket{\Phi_L^i}$, orthogonalised against each $\ket{\Phi_R^j}$ with $E_j < E_i$. In this \emph{shoelace}-manner $m$ left and right excited state eigenvectors are obtained based on the bi-orthogonal property of left and right eigenvectors of non-Hermitian Operators $\bracket{\Phi^i_L}{\Phi^j_R} = 0$ for $ i\ne j$. Results on observables other than the energy and correct left and right eigenvectors of excited states will be reported in future work. \\
To perform accurate thermodynamic-limit extrapolations, we also need to reduce the finite size errors of the kinetic term in \ref{eq:real-space-start}. This can be done by twist averaged boundary conditions\cite{twist-average-1,twist-average-2,zhang,twist-average-3}, which is readily applicable for the similarity transformed Hamiltonian in FCIQMC, and will be reported in future work.\\

\begin{acknowledgments}
The authors thank Dr. Olle Gunnarsson for providing the 18-site exact diagonalization results, Dr. Pablo R. Lopez for the VMC correlation parameter optimization and Kai Guther for helpful discussions. 
\end{acknowledgments}

\appendix
\section{\label{app:optimize-j}Optimization of $J$}
As mention in \ref{subsec:analytic_results}, similar to the optimization of coupled cluster amplitudes\cite{size-consistency}, we want to solve for the single parameter $J$ of the Ansatz\eqnref{eq:jastrow-ansatz} by projection. Projecting the Ansatz on $\bra{\Phi_{HF}}$ would yield us the energy $E_J^{HF}$
\begin{equation}
\label{eq:energy-proj}
\bra{\Phi_{HF}}\underbrace{\e^{-\hat\tau}\hat H \e^{\hat \tau}}_{\bar H}\ket{\Phi_{HF}} = E_J.
\end{equation}
And projecting onto $\bra{\Phi_{HF}}\hat \tau^\dagger$ 
\begin{equation}
\label{eq:coeff-proj}
\bra{\Phi_{HF}}\hat\tau^\dagger \bar H \ket{\Phi_{HF}} = E_J^{HF} \braopket{\Phi_{HF}}{\hat \tau^\dagger}{\Phi_{HF}}, 
\end{equation}
where $\expect{\hat\tau^\dagger}_{HF}\neq 0$ only for $\mbf k = 0$ terms in the momentum space representation of $\hat \tau:$ 
\begin{equation}
\hat \tau = \frac{J}{M} \sum_{\mbf{p,q,k},\s} c_{\mbf{p-k},\s}^\dagger c_{\mbf{q+k},\bar \s}^\dagger c_{\mbf q,\bar\s} c_{\mbf p,\s}. 
\end{equation} 
Combining Eq.~(\ref{eq:energy-proj}) and (\ref{eq:coeff-proj}) yields 
\begin{equation}
\label{eq:opt-j-combined}
\expect{\left(\hat \tau - \expect{\hat\tau}\right)^\dagger\bar H}_{HF}  = 0, 
\end{equation}
where the diagonal, $\mbf k = 0$, terms cancel. (In the language of the coupled cluster approach an equivalent expression to Eq.~(\ref{eq:opt-j-combined}) is $\braket{\hat \tau^\dagger H}_c = 0$, where $\expect{\dots}_c$ denotes a cumulant expression over linked diagrams \cite{fulde-book} only.)
To optimize $J$ based on a single determinant $\ket{\Phi_{HF}}$ we need to solve Eq.~(\ref{eq:opt-j-combined}), which can also be seen as a projection of the eigenvalue equation $(\bar H - E)\ket{\Phi_{HF}} = 0$ on the single basis of the correlation factor $\hat \tau$. The remaining contributing contractions ($\mbf k \neq 0$) of (\ref{eq:opt-j-combined}) of $\bar H$ are
\begin{align}
\label{eq:opt-j-explicit}
&\expect{\hat\tau^\dagger \bar H}_c  = \frac{1}{M^2}\sum_{\mbf{pqk},\s} n_{\mbf p,\s} n_{\mbf q,\bar\s}(1-n_{\mbf{p-k},\s})(1-n_{\mbf{q+k},\bar\s})  \nonumber \\
 &\times \Bigg\lbrace \underbrace{\omega_2(J,\mbf{p,k})}_{\text{2-body}} + 2t\frac{\cosh J - 1}{M}  \\
 &\times\bigg[\underbrace{N_{\bar\s}\left(\epsilon_{\mbf p}+\epsilon_{\mbf{p-k}}\right)}_{\text{3-body RPA}} - \underbrace{\sum_{\mbf s} \left(\epsilon_{\mbf{p+q-s}}+\epsilon_{\mbf{p-q-k+s}}\right)n_{\mbf s,\bar\s}}_{\text{3-body exchange}}\bigg] \Bigg\rbrace.\nonumber
\end{align}
Equation (\ref{eq:opt-j-explicit}) can be evaluated directly, or since $\hat \tau \ket{\Phi_{HF}} = c_{HF}\ket{\Phi_{HF}} + \sum_i c_i \ket{D_i}$ corresponds to all the double excitation on top of the Hartree-Fock determinant, it is the sum of all the double excitation matrix elements with the Hartree-Fock determinant. The diagonal contribution again cancels with the $\expect{\hat\tau}$ term in (\ref{eq:opt-j-combined}). 
The specific optimal $J$ values for the lattice sizes, fillings and $U/t$ values used in this study are listed in Table~\ref{tab:opt-j-explicit}.

\begingroup
\begin{table}
\centering
\caption{\small $J_{opt}$ obtained by solving Eq.~(\ref{eq:optimize-j}) for the specific lattice sizes, fillings and $U/t$ values used in this paper. $J_{ex}$ is the value, which sets set the $J$-dependent Hartree-Fock energy, $E_J^{HF}$, to the exact energy, if available, or to the AFQMC reference energies for larger systems.}
\label{tab:opt-j-explicit}
\begin{tabular}{cccccccc}
\toprule
$M$ & $U/t$ & $n_{el}$ & $J_{opt}$ & $J_{ex}$ & $e_{ex}$ & $e_{J}$ & $e_{J}/e_{ex} [\%]$ \\ 
\hline 
6 & 4 & 6 & 	  	  	-0.67769 & 	-0.74282 & 	-0.61145 &  	-0.56306 & 92.1 \\ 
18 & 2 & 18 & 	  	-0.27053 & 	-0.28536 & 	-1.32141 & 	-1.31697 & 99.7\\ 
18 & 4 & 18 & 	  	-0.52345 & 	-0.57472 & 	-0.95847 & 	-0.92697 & 96.7 \\ 
18 & 4 & 14 & 	  	-0.55794 & 	-0.62474 & 	-1.13644  & -1.09786 & 96.7 \\ 
36 & 2 & 36 & 	  	-0.30485 &	-0.45423 & 	-1.15158 &	-1.09840 & 95.4\\
36\footnote[1]{anti-periodic BC along y-axis} & 2 & 36 & 	-0.28683 & 	-0.31783 & 	-1.20831 &	-1.19904 & 99.3 \\
36 & 4 & 36 & 		-0.58521 &	-0.79141 &	-0.85736 &	-0.71675 & 83.6\\
36\footnotemark[1] & 4 & 36 &	-0.55295 &	-0.65181 &	-0.87306 &	-0.81145 & 92.9 \\
36\footnotemark[1]  & 4 & 24 & 	-0.53570 &   &  & 					-1.13399 & - \\
36\footnote[2]{open-shell $\mbf k = 0$ reference} &4&24& 	-0.52372 & 	-0.57014 & 	-1.18530 & 	-1.16457 & 98.3 \\
50 & 1 & 50 & 		-0.14290 & 	-0.15357 & 	-1.43718 &  -1.43561 & 99.9 \\ 
50 & 2 & 50 & 		-0.28298 & 	-0.30852 & 	-1.22278 &  -1.21523 & 99.4 \\ 
50 & 3 & 50 & 		-0.41788 & 	-0.46639 & 	-1.03460 & 	-1.01278 & 97.9 \\ 
50 & 4 & 50 & 		-0.54600 & 	-0.63177 & 	-0.87966 & 	-0.82601 & 93.9\\ 
50 & 4 & 48 & 		-0.54945 & 	-0.62810 & 	-0.93720 & 	-0.88954 & 94.9 \\ 
50 & 4 & 46 & 		-0.55208 & 	-0.62227 &	-0.99114  & 	-0.95008 & 95.9 \\ 
50 & 4 & 44 & 		-0.54772 & 	-0.61530 & 	-1.03788 & 	-1.00016 & 96.4 \\ 
50 & 4 & 42 & 		-0.54324 & 	-0.60263 & 	-1.08002 &  	-1.04765 & 97.0 \\ 
50 & 4 & 26 & 		-0.51076 & 	-0.56162 & 	-1.11564 &	-1.09946 & 98.6 \\ 
\botrule
\end{tabular}
\end{table}
\endgroup


\section{\label{app0-analytic-j}Analytic optimization of $J$ in the thermodynamic limit at half-filling}
For a infinite system at half-filling, we define
\begin{align}
T_{0}({\bf k}) &= \frac{1}{M}\sum_{{\bf q}}\Theta(\epsilon_{F}-\epsilon_{{\bf q}})\Theta(\epsilon_{{\bf q}+{\bf k}}-\epsilon_{F}),\label{eq:t0}\\
T_{1}({\bf k}) &=  \frac{1}{M}\sum_{{\bf p}}\Theta(\epsilon_{F}-\epsilon_{{\bf p}})\Theta(\epsilon_{{\bf p}-{\bf k}}-\epsilon_{F})\sum_{{\bf d}}e^{i({\bf p}-{\bf k})\cdot{\bf d}},\label{eq:t1}\\
T_{2}({\bf k}) &= \frac{1}{M}\sum_{{\bf p}}\Theta(\epsilon_{F}-\epsilon_{{\bf p}})\Theta(\epsilon_{{\bf p}-{\bf k}}-\epsilon_{F})\sum_{{\bf d}}e^{i{\bf p}\cdot{\bf d}}.\label{eq:t2}
\end{align}
The 2-body contributions of Eq.~(\ref{eq:optimize-j}) can be expressed as
\begin{widetext}
\begin{equation}
\frac{U}{2}\frac{1}{M}\sum_{{\bf k}}T_{0}^{2}({\bf k})-\frac{t}{M}\left((e^{J}-1)\sum_{{\bf k}}T_{0}({\bf k})T_{1}({\bf k})+(e^{-J}-1)\sum_{{\bf k}}T_{0}({\bf k})T_{2}({\bf k})\right)=0.
\end{equation}
\end{widetext}
In the thermodynamic limit ($M\rightarrow\infty$) the summation in
the expression of the $T_{m}$ factors~(\ref{eq:t0}-\ref{eq:t2}) become integrals
\[
\frac{1}{M}\sum_{{\bf q}}\longrightarrow\frac{1}{(2\pi)^{d}}\int d^{d}q.
\]
For an un-polarized system at half filling, the factor $\Theta(\epsilon_{F}-\epsilon_{{\bf q}})$ leads to a square region in the $k_x-k_y$ plane and $T_{m}({\bf k})$ integrals can be easily calculated after a rotation of coordinates
\begin{equation}
\label{eq:rotation}
k_{x}' = \frac{1}{\sqrt{2}}(k_{x}-k_{y}), \quad k_{y}' = \frac{1}{\sqrt{2}}(k_{x}+k_{y}).
\end{equation}
With this rotation, $T_{0}$ is found to be symmetric with respect
to $k_{x}'\rightarrow-k'_{x}$ and $k_{y}'\rightarrow-k'_{y}$, so
it reduces to a function of $|k_{x}|$ and $|k_{y}|$
\begin{align}
T_{0}({\bf k}) &= \frac{1}{(2\pi)^{2}}\int\Theta(\epsilon_{F}-\epsilon_{{\bf q}})\Theta(\epsilon_{{\bf q}+{\bf k}}-\epsilon_{F})d^{2}q\\
 &= \frac{1}{(2\pi)^{2}}\Bigg(\int_{-\pi/\sqrt{2}}^{-\pi/\sqrt{2}+|k_{x}'|}\int_{-\pi/\sqrt{2}+|k_{y}'|}^{\pi/\sqrt{2}}dq'_{x}dq'_{y}\\
 &+\int_{-\pi/\sqrt{2}+|k'_{x}|}^{\pi/\sqrt{2}}\int_{-\pi/\sqrt{2}}^{-\pi/\sqrt{2}+|k'_{y}|}dq'_{x}dq'_{y}\Bigg)\\
 &= \frac{1}{(2\pi)^{2}}\left(\sqrt{2}\pi(|k_{x}'|+|k_{y}'|)-2|k_{x}'k_{y}'|\right)\\
 &= \frac{1}{(2\pi)^{2}}\left(\pi(|k_{x}-k_{y}|+|k_{x}+k_{y}|)-|k_{x}^{2}-k_{y}^{2}|\right).
\end{align}
With the coordinate rotation~(\ref{eq:rotation}), the integrand of $T_{1}$ can be factorized as
\begin{align*}
\sum_{\mbf d} e^{i({\bf p}-{\bf k})\cdot{\bf d}} &= \cos(p_{x}-k_{x})+\cos(p_{y}-k_{y})\\
 &= \cos\left(\frac{1}{\sqrt{2}}(p_{x}'-k_{x}')+\frac{1}{\sqrt{2}}(p_{y}'-k_{y}')\right)\\
 &+\cos\left(\frac{1}{\sqrt{2}}(p_{x}'-k_{x}')-\frac{1}{\sqrt{2}}(p_{y}'-k_{y}')\right)\\
 &= 2\cos\left(\frac{1}{\sqrt{2}}(p_{x}'-k_{x}')\right)\cos\left(\frac{1}{\sqrt{2}}(p_{y}'-k_{y}')\right),
\end{align*}
and $T_{1}$ can also be found as a function of $|k_{x}|$ and $|k_{y}|$
\begin{equation}
T_{1}({\bf k}) =  \frac{16}{(2\pi)^{2}}\left[\cos\left(\frac{k_{x}-k_{y}}{2}\right)\cos\left(\frac{k_{x}+k_{y}}{2}\right)-1\right].
\end{equation}
In a similar way $T_{2}$ can be calculated as 
\begin{align}
T_{2}({\bf k})&= \frac{16}{(2\pi)^{2}}\left[1-\cos\left(\frac{k_{x}-k_{y}}{2}\right)\cos\left(\frac{k_{x}+k_{y}}{2}\right)\right]\nonumber\\
 &= -T_{1}({\bf k}).
\end{align}
The exchange part of the three body contribution in~(\ref{eq:optimize-j}) to the correlation energy can be calculated as (using here again the rotation\eqnref{eq:rotation} for ${\bf p}$)
\begin{align}
 & \frac{1}{M^{2}}\sum_{{\bf pqk}\sigma}\left(n_{{\bf p},\sigma}n_{{\bf q}+{\bf k},\bar{\sigma}}n_{{\bf q},\bar{\sigma}}\sum_{{\bf d}}\cos({\bf p}\cdot{\bf d})e^{i{\bf k}\cdot{\bf d}}\right)\nonumber\\
 &= \frac{1}{M}\sum_{{\bf pk}\sigma}\left(n_{{\bf p},\sigma}(\frac{1}{2}-T_{0}({\bf k}))\sum_{{\bf d}}\cos({\bf p}\cdot{\bf d})e^{i{\bf k}\cdot{\bf d}}\right)\nonumber\\
 &= \frac{2M}{(2\pi)^{4}}\int\int_{-\pi/\sqrt{2}}^{\pi/\sqrt{2}}dp'_{x}dp'_{y}\int\int_{-\pi}^{\pi}dk{}_{x}dk{}_{y}\times \nonumber\\
 &\left(\frac{1}{2}-\frac{1}{(2\pi)^{2}}\left(\pi(|k_{x}-k_{y}|+|k_{x}+k_{y}|)-|k_{x}^{2}-k_{y}^{2}|\right)\right)\nonumber\\
 &\left(\cos(\frac{p'_{x}+p'_{y}}{\sqrt{2}})\cos(k_{x})+\cos(\frac{p'_{x}-p'_{y}}{\sqrt{2}})\cos(k_{y})\right)\nonumber\\
 &= \frac{32M}{\pi^{6}}.\label{eq:exchange-tdl}
\end{align}
The final results are
\begin{align}
T_{0}({\bf k}) &= \frac{1}{(2\pi)^{2}}\left(\pi(|k_{x}-k_{y}|+|k_{x}+k_{y}|)-|k_{x}^{2}-k_{y}^{2}|\right),\\
T_{1}({\bf k}) &= \frac{16}{(2\pi)^{2}}\left[\cos\left(\frac{k_{x}-k_{y}}{2}\right)\cos\left(\frac{k_{x}+k_{y}}{2}\right)-1\right],\\
T_{2}({\bf k}) &= -T_{1}({\bf k}),
\end{align}
and the summations can also be calculate as integrals
\begin{align}
\frac{1}{M}\sum_{{\bf k}}T_{0}^{2}({\bf k}) &= \frac{5}{72},\\
\frac{1}{M}\sum_{{\bf k}}T_{0}({\bf k})T_{1}({\bf k}) &=  -\frac{16+\pi^{4}}{\pi^{6}}.
\end{align}
$J_{opt}^{TDL}$ can be obtained by solving 
\begin{equation}
\frac{5U}{144}+t\frac{16+\pi^{4}}{\pi^{6}}\left((e^{J}-e^{-J}\right)=0,
\end{equation}
which, for small $U/t$, can be approximated as
\begin{equation}
J_{opt}^{TDL}=\mbox{argsinh}(-\frac{5U}{288t}\times\frac{\pi^{6}}{16+\pi^{4}})\approx-0.14717\frac{U}{t}.
\end{equation}
At half-filling Hartree-Fock energy of the original Hubbard Hamiltonian \eqref{eq:original-k-space}, with $\mbf k = 0$ in the two-body term, 
\begin{equation}
E_{HF}^{J=0} = \expect{-t\sum_{\mbf k,\s} \epsilon_{\mbf k} n_{\mbf k,\s}}_{HF} + \frac{U}{2} \expect{\frac{1}{M} \sum_{\mbf{p,q},\s} n_{\mbf p,\s} n_{\mbf q,\bar\s}}_{HF}
\end{equation}
results to 
\begin{equation}
E_{HF}^{J=0} = M \left(-t\frac{64}{(2\pi)^2} + \frac{U}{4}\right)
\end{equation}
in the thermodynamic limit (TDL). The additional contributions arising due to the similarity transformation 
\begin{equation}
E_{HF}^J = \expect{-2t\frac{\cosh(J-1)}{M} \sum_{\mbf{p,q},\s} \epsilon_{\mbf p} n_{\mbf p,\s} n_{\mbf q,\bar\s}}_{HF} - \expect{2t\frac{\cosh(J-1)}{M^2} \sum_{\mbf{p,q,k},\s} \epsilon_{\mbf p+k} n_{\mbf p,\s} n_{\mbf{q+k},\bar\s} n_{\mbf q,\bar s}}_{HF}
\end{equation}
can be estimated, with $\cosh(J-1) \approx J^2$ for small $J$ and Eq.~\eqnref{eq:exchange-tdl}, as 
\begin{equation}
E_{HF}^J \approx -tJ^2 M \left(\frac{16}{(2\pi)^2} + \frac{64}{\pi^6}\right).
\end{equation} 
Hence, the energy per site in the TDL for an un-polarized system at half-filling is given by
\begin{align}
E^{TDL}_{opt}&=-t\frac{64}{(2\pi)^{2}}+\frac{U}{4}-tJ^{2}\left(\frac{16}{(2\pi)^{2}}+\frac{64}{\pi^{6}}\right).
\end{align}

\section{\label{app:6-site} Excited states}
As discussed in \ref{subsubsec:14-electrons} the right eigenvectors of a non-Hermitian operator $\bar H \ket{\Phi_i^R} = E_i \ket{\Phi_i^R}$ are in general not orthogonal to each other. And hence the way excited states are obtained with the FCIQMC method\cite{excited-fciqmc} should in general not be applicable to excited states of a non-Hermitian operator, since they are sampled by orthogonalizing the $n$-th excited state to all lower energy states $m < n$. But it turns out that we are still able to use the dynamically adapted shift energy $E_i^S$ of Eq.~(\ref{eq:excited-states-projector}) as a valid estimator for the excited state energies. In Fig.~\ref{fig:6-site-excited} the difference to the exact energy, obtained by the projected $e_p$ and shift $e_s$ energy estimator, for the first 10 states of the 1D 6 $e^-$ in 6 site, periodic, $U/t = 4$, $\mbf k = 0$ Hubbard model with a correlation parameter $J = -0.1$ are shown. Also shown is the difference of the sum of the overlap of the $i$-th excited states to all lower lying states $j$ with $E_j < E_i$, for the exact right eigenvectors obtained by exact diagonalization and the sampled eigenvectors within FCIQMC 
\begin{equation}
\label{eq:overlap-difference}
\Delta O_i = \sum_{j} \vert \bracket{\Phi_i^{ex}}{\Phi_j^{ex}} - \bracket{\Phi_i^{qmc}}{\Phi_j^{qmc}} \vert \quad \forall  j: E_j < E_i.
\end{equation}
As mentioned $\bracket{\Phi_i^R}{\Phi_j^R} \neq 0$ is possible for non-Hermitian operators, and is the case for states $3,4$ and $5$ shown in Fig.~\ref{fig:6-site-excited}, indicated by a large value of $\Delta O_i$, since within FCIQMC the incorrect $\bracket{\Phi_i^{qmc}}{\Phi_j^{qmc}} \mustbe 0$ is tried to be enforced. The partially incorrect wave-function form is additionally marked by an large error in the projected energy $e_p$ compared to the exact result.\\
But as the $i$-th excited state is only orthogonalised to all the lower lying excited states to converge to the next higher energy governed by the dynamics (\ref{eq:first-order-proj}) and the spectrum of the Hamiltonian\eqnref{eq:hamil-real}, which is unchanged by the similarity transformation (\ref{eq:similarity-tranformation}), the shift energy remains a good energy estimator. This can clearly be seen in Fig.~\ref{fig:6-site-excited}, as the shift energy is a good estimate of all the targeted eigenstates.\\
The only exception in Fig.~\ref{fig:6-site-excited}, which could be misleading, is state number $7$, which appears to have a large error in $\Delta O_i$, but the projected energy is still a good estimator for the energy. This comes from the fact that state $6$ and $7$ are actually degenerate and thus the exact eigenvectors $\ket{\Phi_6^{ex}}$ and $\ket{\Phi_7^{ex}}$ obtained by \emph{Lapack}\cite{lapack} are an arbitrary linear combination and could be chosen to be both orthogonal to the states $i < 6$.
\begin{figure}
\centering
\includegraphics[scale=1]{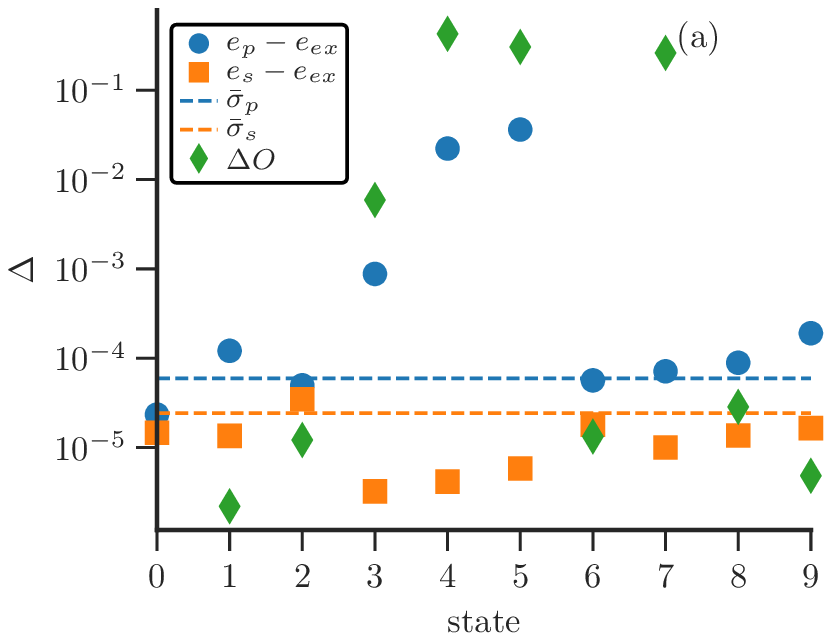}
\caption{\label{fig:6-site-excited}(Color online) Error of the first 10 eigenstate energies obtained by the projected energy $e_p$ and shift energy $e_s$ for the 6 $e^-$ in 6 site 1D periodic Hubbard model at $U/t = 4$ and $\mbf k = 0$ compared to exact diagonalization results. The horizontal dashed lines indicate the averaged statistical errors. The green pluses show the difference of the exact overlaps to the overlaps obtained within FCIQMC, see Eq.~(\ref{eq:overlap-difference}). A correlation parameter of $J = -0.1$, initiator threshold of $n_{init} = 1.2$ and a maximum walker number of $N_w = 10^5$ were used. (a) Exact overlap is ill-defined due to degeneracy of states $6$ and $7$.}
\end{figure}
The $n$-th excited state in FCIQMC is obtained\cite{excited-fciqmc} by 
\begin{equation}
\label{eq:excited-recursion}
\ket{\Phi_n(t+ \delta t)} = \hat P_n(t + \delta t) \left[\mbf 1 - \delta t\left(\hat H - E_n^S \right) \right] \ket{\Phi_n(t)}
\end{equation}
with 
\begin{equation}
\label{eq:gram-schmidt}
\hat P_n(t) = \mbf 1 - \sum_{m<n} \frac{\ket{\Phi_m(t)}\bra{\Phi_m(t)}}{\bracket{\Phi_m(t)}{\Phi_m(t)}},\quad E_m < E_n,
\end{equation}
being the Gram-Schmidt projector, which removes all contributions of lower lying states $\ket{\Phi_m}$ and thus orthogonalises $\ket{\Phi_n}$ to each state with $E_m < E_n$. For the set of right eigenvector of a non-Hermitian Hamiltonian the assumption of them being orthogonal to each other does not hold in general. So this method of obtaining the excited states of $\bar H$ should in principle not work. But the results above indicate, that the shift energy still provides a correct energy estimate.\\ 
To see why the shift energy is a valid estimate for the exact excited states energy, let's look at the right eigenvalue equation for a general (Hermitian or non-Hermitian) Hamiltonian $\hat H$ for the $i$-th excited state
\begin{equation}
\label{eq:exc-ev-equation}
\hat H \ket{\Psi_i} = E_i \ket{\Psi_i}, 
\end{equation}
where $\ket{\Psi_i}$ is the $i$-th right eigenvector of $\hat H$. 
We now want to show that there exists a vector $\ket{\Phi_i}$, which is a eigenvector of the composite operator $\hat P_i \hat H$ with the same eigenvalue $E_i$
\begin{equation}
\label{eq:exc-proj-ev-equation}
\hat P_i \hat H \ket{\Phi_i} = E_i \ket{\Phi_i},
\end{equation}
where $\hat P_i$ is the Gram-Schmidt projector~(\ref{eq:gram-schmidt}) and $\ket{\Phi_0} = \ket{\Psi_0}$, which creates an orthonormal basis out of the linear-independent, but not necessarily orthonormal set $\lbrace\ket{\Psi_i}\rbrace$. We assume all states to be normalized. Multiplying Eq.~(\ref{eq:exc-ev-equation}) with $\hat P_i$ from the left, we obtain
\begin{equation}
\label{eq:phi-definition}
\hat P_i \hat H \ket{\Psi_i} = E_i \hat P_i \ket{\Psi_i} = E_i \ket{\Phi_i} \ra \ket{\Phi_i} = \hat P_i \ket{\Psi_i}.
\end{equation}
And we assume $\ket{\Phi_i}$ to be the desired eigenvector of $\hat P_i \hat H$. To show that we plug (\ref{eq:phi-definition}) into Eq.~(\ref{eq:exc-proj-ev-equation})
\begin{align}
\label{eq:induction}
&\hat P_i \hat H \ket{\Phi_i} = \hat P_i \hat H \hat P_i \ket{\Psi_i} = \hat P_i \hat H \left(\ket{\Psi_i} - \sum_{j<i} \bracket{\Phi_j}{\Psi_i}\ket{\Phi_j}\right) \nonumber \\
&= E_i\hat P_i \ket{\Psi_i} - \sum_{j<i} b_{ji} \hat P_i \hat H \ket{\Phi_j} = E_i\ket{\Phi_i} - \sum_{j<i} b_{ji} \hat P_i \hat H \ket{\Phi_j}, 
\end{align}
with $b_{ij} = \bracket{\Phi_j}{\Psi_i}$. We can express $\ket{\Phi_j}$ in Eq.~(\ref{eq:induction}) and all subsequent appearances of $\ket{\Phi_k}$ with $k < i$ as $\ket{\Phi_k} = \hat P_k \ket{\Psi_k}$ until we reach $\ket{\Psi_0}$. So the remaining thing to show is that $\hat P_i \ket{\Psi_j} = 0$ for $i > j$. \\
For $i > j$
\begin{equation}
\label{eq:zero-proj-phi}
\hat P_i \ket{\Phi_j} = \ket{\Phi_j} - \sum_{k<i} \underbrace{\bracket{\Phi_k}{\Phi_j}}_{\delta_{jk}}\ket{\Phi_k} = 0, \quad \text{with} \quad i > j
\end{equation}
is easy to show since $\lbrace \ket{\Phi_j} \rbrace$ is a orthonormal basis. We prove $\hat P_i \ket{\Psi_j} = 0, j < i$ by induction. For $i = 1$ we have 
\begin{equation}
\hat P_1 \ket{\Psi_0} = \ket{\Psi_0} - \bracket{\Psi_0}{\Psi_0}\ket{\Psi_0} = 0. 
\end{equation}
Let's assume $\hat P_i \ket{\Psi_j} = 0$ for $i < j$, performing the induction step $i \ra i + 1$ yields
\begin{align}
\underline{j < i}: \hat P_{i+1} \ket{\Psi_j} &= \underbrace{\hat P_i \ket{\Psi_j}}_{=0} - \bracket{\Phi_i}{\Psi_j}\ket{\Phi_i}\nonumber \\
 &= -\braopket{\Psi_i}{\underbrace{\hat P_i^\dagger}_{=\hat P_i}}{\Psi_j} \ket{\Phi_i} = 0 \\
\underline{j = i}: \hat P_{i+1}\ket{\Psi_i} &= \underbrace{\hat P_i \ket{\Psi_i}}_{=\ket{\Phi_i}} - \bracket{\Phi_i}{\Psi_i}\ket{\Phi_i} \nonumber \\ 
&= \ket{\Phi_i} - \braopket{\Psi_i}{\underbrace{\hat P_i}_{=\hat P_i^2}}{\Psi_i}\ket{\Phi_i} \nonumber \\
&= \ket{\Phi_i} - \bracket{\Phi_i}{\Phi_i} \ket{\Phi_i} = 0, 
\end{align}
where we used the Hermiticity $\hat P_i^\dagger = \hat P_i$ and idempotency $\hat P_i^2 = \hat P_i$ of the projection operator.
With $\hat P_i \ket{\Psi_j} = 0$ Eq.~(\ref{eq:induction}) gives the desired 
\begin{equation}
\label{eq:phi-ev-eq}
\hat P_i \hat H \ket{\Phi_i} = E_i \ket{\Phi_i}. 
\end{equation}
And this eigenvector $\ket{\Phi_i}$ of the composite operator $\hat P_i \hat H$ is the stationary vector we sample in FCIQMC. Since it has the same eigenvalue $E_i$, we obtain the correct excited state energy estimate from the shift energy $E_i^S$ in the propagator (\ref{eq:excited-recursion}). Since the same argument holds for the long-time limit of the projection 
\begin{equation}
\hat Q_i(E_i^S)\ket{\Psi_i} = \left(\mbf 1 - \Delta t\left(\hat H - E_i^S\right)\right) \ket{\Psi_i} = \ket{\Psi_i}, 
\end{equation}
with stationary $\ket{\Psi_i}$ for $E_i^S = E_i$. There is an eigenvector $\ket{\Phi_i}$ of the composite operator 
\begin{equation}
\hat P_i \hat Q_i(E_i^S) \ket{\Phi_i} = \ket{\Phi_i} 
\end{equation}
for $E_i^S = E_i$ with 
\begin{equation}
\ket{\Phi_i} = P_i \ket{\Psi_i}, \quad \text{since} \quad \hat P_i \hat Q_i(E_i^S) \ket{\Psi_i} = \hat P_i \ket{\Psi_i}. 
\end{equation}
This $\ket{\Phi_i}$ is sampled by the walkers in a FCIQMC simulation and the shift energy $E_i^S(t)$ is adapted to keep the walker population fixed. The projected energy is in general not a good energy estimate, since 
\begin{align}
\label{eq:exc-proje}
E_i^P &= \frac{\braopket{D_i}{\hat H}{\Phi_i}}{\bracket{D_i}{\Phi_i}} = \frac{\braopket{D_I}{\hat H\hat P_i}{\Psi_i}}{\braopket{D_I}{\hat P_i}{\Psi_i}} \nonumber \\
&= \frac{\braopket{D_I}{\hat H\left(\mbf 1 - \sum_{j<i} \ket{\Phi_j}\bra{\Phi_j}\right)}{\Psi_i}}{\bracket{D_I}{\Psi_i} - \sum_{j<i}\bracket{D_I}{\Phi_j}\bracket{\Phi_j}{\Psi_i}} \nonumber \\
&= \frac{E_i c_{I,i} - \sum_{j<i} b_{ij} \braopket{D_I}{\hat H}{\Phi_j}}{c_{I,i} - \sum_{j<i} d_{I,j} b_{ij}} \nonumber \\
&= \frac{E_i c_{I,i} - \sum_{j<i} b_{ij} d_{I,j} E_j^P}{c_{I,i} - \sum_{j<i} d_{I,j} b_{ij}} \\ 
 \quad \text{with} \quad c_{I,i} &= \bracket{D_I}{\Psi_i}, b_{ij} = \bracket{\Phi_j}{\Psi_i}, d_{I,j} = \bracket{D_I}{\Phi_j}
\end{align}
and $\ket{D_I}$ being the reference determinant of state $i$. 
With Eq.~(\ref{eq:exc-proje}) and knowledge of the exact eigenfunctions $\lbrace \ket{\Psi_i}\rbrace$ the excited state energy could be calculated as 
\begin{equation}
\label{eq:exc-proje-corr}
E_i = \left[E_i^P\left(c_{I,i} - \sum_{j<i} d_{I,j} b_{ij}  \right) + \sum_{j<i} b_{ij}d_{I,j} E_j^P\right]c_{I,i}^{-1}.
\end{equation}
For states where $\bracket{D_I}{\Phi_i} \approx c_{I,i}$ and $b_{ij} \approx 0$ the projected energy remains a good estimator for the exact $E_i$. But especially in cases where the exact right eigenvectors are not orthogonal to all lower lying ones, as demonstrated in Fig.~\ref{fig:6-site-excited}, the projected energy should not be trusted. Another correction for the projected energy would be 
\begin{align}
&E_i^P = \frac{\braopket{D_I}{\hat H}{\Phi_i}}{\bracket{D_I}{\Phi_i}} \\
&\bracket{D_I}{\Phi_i} E_i^P = \braopket{D_I}{\hat H \hat P_i}{\Psi_i} \nonumber \\
&= \bracket{D_I}{\Psi_I} E_i - \sum_{j<i} \bracket{\Phi_j}{\Psi_i}\underbrace{\braopket{D_I}{\hat H}{\Phi_j}}_{=\bracket{D_I}{\Phi_j}E_j^P} \\
&\ra E_i = \frac{\bracket{D_I}{\Phi_i} E_i^P + \sum_{j<i}\bracket{\Phi_j}{\Psi_i}\bracket{D_I}{\Phi_j} E_j^P}{\bracket{D_I}{\Psi_i}} \\
&\text{with} \quad \bracket{D_I}{\Psi_i} = \bracket{D_I}{\Phi_i} + \sum_{j<i}\bracket{\Phi_j}{\Psi_i}\bracket{D_I}{\Phi_j} \\
&\ra E_i = \frac{\bracket{D_I}{\Phi_i} E_i^P + \sum_{j<i}\bracket{\Phi_j}{\Psi_i}\bracket{D_I}{\Phi_j} E_j^P}{\bracket{D_I}{\Phi_i} + \sum_{j<i}\bracket{\Phi_j}{\Psi_i}\bracket{D_I}{\Phi_j}}.
\end{align}
Where we can estimate the overlap $\bracket{\Phi_j}{\Psi_i}$ from the orthogonalisation procedure.\\
Actually for the correct projected energy one needs to calculate 
\begin{equation}
\label{eq:correct-proje}
\bar E_i^P = \frac{\braopket{D_I}{\hat P_i \hat H}{\Phi_i}}{\bracket{D_I}{\Phi_i}} = E_i, \quad \text{since} \quad \hat P_i \hat H \ket{\Phi_i} = E_i \ket{\Phi_i}. 
\end{equation}
Unfortunately the numerator of Eq.~(\ref{eq:correct-proje}) takes the following form 
\begin{equation}
\braopket{D_I}{\hat P_i \hat H}{\Phi_i} = \braopket{D_I}{\hat H}{\Phi_i} - \sum_{j<i} \bracket{D_I}{\Phi_j}\braopket{\Phi_j}{\hat H}{\Phi_i}.
\end{equation} 
To calculate $\braopket{\Phi_j}{\hat H}{\Phi_i}$ we would need the transition (reduced) density matrices (t-(R)DM) between all states $j < i$. And for the similarity transformed momentum-space Hubbard Hamiltonian even up to the 3-body t-RDM. So we have to rely on the shift energy to yield the correct excited state energy in the ST-FCIQMC method or apply the mentioned \emph{shoelace} technique in Sec.~\ref{sec:discussion}. 


\begin{thebibliography}{115}%
\makeatletter
\providecommand \@ifxundefined [1]{%
 \@ifx{#1\undefined}
}%
\providecommand \@ifnum [1]{%
 \ifnum #1\expandafter \@firstoftwo
 \else \expandafter \@secondoftwo
 \fi
}%
\providecommand \@ifx [1]{%
 \ifx #1\expandafter \@firstoftwo
 \else \expandafter \@secondoftwo
 \fi
}%
\providecommand \natexlab [1]{#1}%
\providecommand \enquote  [1]{``#1''}%
\providecommand \bibnamefont  [1]{#1}%
\providecommand \bibfnamefont [1]{#1}%
\providecommand \citenamefont [1]{#1}%
\providecommand \href@noop [0]{\@secondoftwo}%
\providecommand \href [0]{\begingroup \@sanitize@url \@href}%
\providecommand \@href[1]{\@@startlink{#1}\@@href}%
\providecommand \@@href[1]{\endgroup#1\@@endlink}%
\providecommand \@sanitize@url [0]{\catcode `\\12\catcode `\$12\catcode
  `\&12\catcode `\#12\catcode `\^12\catcode `\_12\catcode `\%12\relax}%
\providecommand \@@startlink[1]{}%
\providecommand \@@endlink[0]{}%
\providecommand \url  [0]{\begingroup\@sanitize@url \@url }%
\providecommand \@url [1]{\endgroup\@href {#1}{\urlprefix }}%
\providecommand \urlprefix  [0]{URL }%
\providecommand \Eprint [0]{\href }%
\providecommand \doibase [0]{http://dx.doi.org/}%
\providecommand \selectlanguage [0]{\@gobble}%
\providecommand \bibinfo  [0]{\@secondoftwo}%
\providecommand \bibfield  [0]{\@secondoftwo}%
\providecommand \translation [1]{[#1]}%
\providecommand \BibitemOpen [0]{}%
\providecommand \bibitemStop [0]{}%
\providecommand \bibitemNoStop [0]{.\EOS\space}%
\providecommand \EOS [0]{\spacefactor3000\relax}%
\providecommand \BibitemShut  [1]{\csname bibitem#1\endcsname}%
\let\auto@bib@innerbib\@empty
\bibitem [{\citenamefont {Hubbard}(1963)}]{hubbard-model-1}%
  \BibitemOpen
  \bibfield  {author} {\bibinfo {author} {\bibfnamefont {J.}~\bibnamefont
  {Hubbard}},\ }\href {\doibase 10.1098/rspa.1963.0204} {\bibfield  {journal}
  {\bibinfo  {journal} {Proceedings of the Royal Society of London A:
  Mathematical, Physical and Engineering Sciences}\ }\textbf {\bibinfo {volume}
  {276}},\ \bibinfo {pages} {238} (\bibinfo {year} {1963})}\BibitemShut
  {NoStop}%
\bibitem [{\citenamefont {Gutzwiller}(1963)}]{hubbard-model-2}%
  \BibitemOpen
  \bibfield  {author} {\bibinfo {author} {\bibfnamefont {M.~C.}\ \bibnamefont
  {Gutzwiller}},\ }\href {\doibase 10.1103/PhysRevLett.10.159} {\bibfield
  {journal} {\bibinfo  {journal} {Phys. Rev. Lett.}\ }\textbf {\bibinfo
  {volume} {10}},\ \bibinfo {pages} {159} (\bibinfo {year} {1963})}\BibitemShut
  {NoStop}%
\bibitem [{\citenamefont {Kanamori}(1963)}]{hubbard-model-3}%
  \BibitemOpen
  \bibfield  {author} {\bibinfo {author} {\bibfnamefont {J.}~\bibnamefont
  {Kanamori}},\ }\href {\doibase 10.1143/PTP.30.275} {\bibfield  {journal}
  {\bibinfo  {journal} {Progress of Theoretical Physics}\ }\textbf {\bibinfo
  {volume} {30}},\ \bibinfo {pages} {275} (\bibinfo {year} {1963})}\BibitemShut
  {NoStop}%
\bibitem [{\citenamefont {Zhang}\ and\ \citenamefont {Rice}(1988)}]{cuprates}%
  \BibitemOpen
  \bibfield  {author} {\bibinfo {author} {\bibfnamefont {F.~C.}\ \bibnamefont
  {Zhang}}\ and\ \bibinfo {author} {\bibfnamefont {T.~M.}\ \bibnamefont
  {Rice}},\ }\href {\doibase 10.1103/PhysRevB.37.3759} {\bibfield  {journal}
  {\bibinfo  {journal} {Phys. Rev. B}\ }\textbf {\bibinfo {volume} {37}},\
  \bibinfo {pages} {3759} (\bibinfo {year} {1988})}\BibitemShut {NoStop}%
\bibitem [{\citenamefont {Dagotto}(1994)}]{high-tc}%
  \BibitemOpen
  \bibfield  {author} {\bibinfo {author} {\bibfnamefont {E.}~\bibnamefont
  {Dagotto}},\ }\href {\doibase 10.1103/RevModPhys.66.763} {\bibfield
  {journal} {\bibinfo  {journal} {Rev. Mod. Phys.}\ }\textbf {\bibinfo {volume}
  {66}},\ \bibinfo {pages} {763} (\bibinfo {year} {1994})}\BibitemShut
  {NoStop}%
\bibitem [{\citenamefont {Scalapino}(2007)}]{scalapino}%
  \BibitemOpen
  \bibfield  {author} {\bibinfo {author} {\bibfnamefont {D.~J.}\ \bibnamefont
  {Scalapino}},\ }\enquote {\bibinfo {title} {Numerical studies of the 2d
  hubbard model},}\ in\ \href {\doibase 10.1007/978-0-387-68734-6_13} {\emph
  {\bibinfo {booktitle} {Handbook of High-Temperature Superconductivity: Theory
  and Experiment}}},\ \bibinfo {editor} {edited by\ \bibinfo {editor}
  {\bibfnamefont {J.~R.}\ \bibnamefont {Schrieffer}}\ and\ \bibinfo {editor}
  {\bibfnamefont {J.~S.}\ \bibnamefont {Brooks}}}\ (\bibinfo  {publisher}
  {Springer New York},\ \bibinfo {address} {New York, NY},\ \bibinfo {year}
  {2007})\ pp.\ \bibinfo {pages} {495--526}\BibitemShut {NoStop}%
\bibitem [{\citenamefont {White}\ and\ \citenamefont
  {Scalapino}(2000)}]{dmrg-1}%
  \BibitemOpen
  \bibfield  {author} {\bibinfo {author} {\bibfnamefont {S.~R.}\ \bibnamefont
  {White}}\ and\ \bibinfo {author} {\bibfnamefont {D.~J.}\ \bibnamefont
  {Scalapino}},\ }\href {\doibase 10.1103/PhysRevB.61.6320} {\bibfield
  {journal} {\bibinfo  {journal} {Phys. Rev. B}\ }\textbf {\bibinfo {volume}
  {61}},\ \bibinfo {pages} {6320} (\bibinfo {year} {2000})}\BibitemShut
  {NoStop}%
\bibitem [{\citenamefont {Scalapino}\ and\ \citenamefont
  {White}(2001)}]{dmrg-2}%
  \BibitemOpen
  \bibfield  {author} {\bibinfo {author} {\bibfnamefont {D.}~\bibnamefont
  {Scalapino}}\ and\ \bibinfo {author} {\bibfnamefont {S.}~\bibnamefont
  {White}},\ }\href {\doibase 10.1023/A:1004147703543} {\bibfield  {journal}
  {\bibinfo  {journal} {Foundations of Physics}\ }\textbf {\bibinfo {volume}
  {31}},\ \bibinfo {pages} {27} (\bibinfo {year} {2001})}\BibitemShut {NoStop}%
\bibitem [{\citenamefont {White}\ and\ \citenamefont
  {Scalapino}(2003)}]{dmrg-3}%
  \BibitemOpen
  \bibfield  {author} {\bibinfo {author} {\bibfnamefont {S.~R.}\ \bibnamefont
  {White}}\ and\ \bibinfo {author} {\bibfnamefont {D.~J.}\ \bibnamefont
  {Scalapino}},\ }\href {\doibase 10.1103/PhysRevLett.91.136403} {\bibfield
  {journal} {\bibinfo  {journal} {Phys. Rev. Lett.}\ }\textbf {\bibinfo
  {volume} {91}},\ \bibinfo {pages} {136403} (\bibinfo {year}
  {2003})}\BibitemShut {NoStop}%
\bibitem [{\citenamefont {Tocchio}\ \emph {et~al.}(2008)\citenamefont
  {Tocchio}, \citenamefont {Becca}, \citenamefont {Parola},\ and\ \citenamefont
  {Sorella}}]{vmc-1}%
  \BibitemOpen
  \bibfield  {author} {\bibinfo {author} {\bibfnamefont {L.~F.}\ \bibnamefont
  {Tocchio}}, \bibinfo {author} {\bibfnamefont {F.}~\bibnamefont {Becca}},
  \bibinfo {author} {\bibfnamefont {A.}~\bibnamefont {Parola}}, \ and\ \bibinfo
  {author} {\bibfnamefont {S.}~\bibnamefont {Sorella}},\ }\href {\doibase
  10.1103/PhysRevB.78.041101} {\bibfield  {journal} {\bibinfo  {journal} {Phys.
  Rev. B}\ }\textbf {\bibinfo {volume} {78}},\ \bibinfo {pages} {041101}
  (\bibinfo {year} {2008})}\BibitemShut {NoStop}%
\bibitem [{\citenamefont {Yokoyama}\ and\ \citenamefont {Shiba}(1987)}]{vmc-2}%
  \BibitemOpen
  \bibfield  {author} {\bibinfo {author} {\bibfnamefont {H.}~\bibnamefont
  {Yokoyama}}\ and\ \bibinfo {author} {\bibfnamefont {H.}~\bibnamefont
  {Shiba}},\ }\href {\doibase 10.1143/JPSJ.56.1490} {\bibfield  {journal}
  {\bibinfo  {journal} {Journal of the Physical Society of Japan}\ }\textbf
  {\bibinfo {volume} {56}},\ \bibinfo {pages} {1490} (\bibinfo {year}
  {1987})}\BibitemShut {NoStop}%
\bibitem [{\citenamefont {Eichenberger}\ and\ \citenamefont
  {Baeriswyl}(2007)}]{vmc-3}%
  \BibitemOpen
  \bibfield  {author} {\bibinfo {author} {\bibfnamefont {D.}~\bibnamefont
  {Eichenberger}}\ and\ \bibinfo {author} {\bibfnamefont {D.}~\bibnamefont
  {Baeriswyl}},\ }\href {\doibase 10.1103/PhysRevB.76.180504} {\bibfield
  {journal} {\bibinfo  {journal} {Phys. Rev. B}\ }\textbf {\bibinfo {volume}
  {76}},\ \bibinfo {pages} {180504} (\bibinfo {year} {2007})}\BibitemShut
  {NoStop}%
\bibitem [{\citenamefont {Yamaji}\ \emph {et~al.}(1998)\citenamefont {Yamaji},
  \citenamefont {Yanagisawa}, \citenamefont {Nakanishi},\ and\ \citenamefont
  {Koike}}]{vmc-4}%
  \BibitemOpen
  \bibfield  {author} {\bibinfo {author} {\bibfnamefont {K.}~\bibnamefont
  {Yamaji}}, \bibinfo {author} {\bibfnamefont {T.}~\bibnamefont {Yanagisawa}},
  \bibinfo {author} {\bibfnamefont {T.}~\bibnamefont {Nakanishi}}, \ and\
  \bibinfo {author} {\bibfnamefont {S.}~\bibnamefont {Koike}},\ }\href
  {\doibase https://doi.org/10.1016/S0921-4534(98)00283-4} {\bibfield
  {journal} {\bibinfo  {journal} {Physica C: Superconductivity}\ }\textbf
  {\bibinfo {volume} {304}},\ \bibinfo {pages} {225 } (\bibinfo {year}
  {1998})}\BibitemShut {NoStop}%
\bibitem [{\citenamefont {Giamarchi}\ and\ \citenamefont
  {Lhuillier}(1991)}]{vmc-5}%
  \BibitemOpen
  \bibfield  {author} {\bibinfo {author} {\bibfnamefont {T.}~\bibnamefont
  {Giamarchi}}\ and\ \bibinfo {author} {\bibfnamefont {C.}~\bibnamefont
  {Lhuillier}},\ }\href {\doibase 10.1103/PhysRevB.43.12943} {\bibfield
  {journal} {\bibinfo  {journal} {Phys. Rev. B}\ }\textbf {\bibinfo {volume}
  {43}},\ \bibinfo {pages} {12943} (\bibinfo {year} {1991})}\BibitemShut
  {NoStop}%
\bibitem [{\citenamefont {Becca}\ \emph {et~al.}(2000)\citenamefont {Becca},
  \citenamefont {Capone},\ and\ \citenamefont {Sorella}}]{gf-qmc-1}%
  \BibitemOpen
  \bibfield  {author} {\bibinfo {author} {\bibfnamefont {F.}~\bibnamefont
  {Becca}}, \bibinfo {author} {\bibfnamefont {M.}~\bibnamefont {Capone}}, \
  and\ \bibinfo {author} {\bibfnamefont {S.}~\bibnamefont {Sorella}},\ }\href
  {\doibase 10.1103/PhysRevB.62.12700} {\bibfield  {journal} {\bibinfo
  {journal} {Phys. Rev. B}\ }\textbf {\bibinfo {volume} {62}},\ \bibinfo
  {pages} {12700} (\bibinfo {year} {2000})}\BibitemShut {NoStop}%
\bibitem [{\citenamefont {Cosentini}\ \emph {et~al.}(1998)\citenamefont
  {Cosentini}, \citenamefont {Capone}, \citenamefont {Guidoni},\ and\
  \citenamefont {Bachelet}}]{gf-qmc-2}%
  \BibitemOpen
  \bibfield  {author} {\bibinfo {author} {\bibfnamefont {A.~C.}\ \bibnamefont
  {Cosentini}}, \bibinfo {author} {\bibfnamefont {M.}~\bibnamefont {Capone}},
  \bibinfo {author} {\bibfnamefont {L.}~\bibnamefont {Guidoni}}, \ and\
  \bibinfo {author} {\bibfnamefont {G.~B.}\ \bibnamefont {Bachelet}},\ }\href
  {\doibase 10.1103/PhysRevB.58.R14685} {\bibfield  {journal} {\bibinfo
  {journal} {Phys. Rev. B}\ }\textbf {\bibinfo {volume} {58}},\ \bibinfo
  {pages} {R14685} (\bibinfo {year} {1998})}\BibitemShut {NoStop}%
\bibitem [{\citenamefont {van Bemmel}\ \emph {et~al.}(1994)\citenamefont {van
  Bemmel}, \citenamefont {ten Haaf}, \citenamefont {van Saarloos},
  \citenamefont {van Leeuwen},\ and\ \citenamefont {An}}]{gf-qmc-3}%
  \BibitemOpen
  \bibfield  {author} {\bibinfo {author} {\bibfnamefont {H.~J.~M.}\
  \bibnamefont {van Bemmel}}, \bibinfo {author} {\bibfnamefont {D.~F.~B.}\
  \bibnamefont {ten Haaf}}, \bibinfo {author} {\bibfnamefont {W.}~\bibnamefont
  {van Saarloos}}, \bibinfo {author} {\bibfnamefont {J.~M.~J.}\ \bibnamefont
  {van Leeuwen}}, \ and\ \bibinfo {author} {\bibfnamefont {G.}~\bibnamefont
  {An}},\ }\href {\doibase 10.1103/PhysRevLett.72.2442} {\bibfield  {journal}
  {\bibinfo  {journal} {Phys. Rev. Lett.}\ }\textbf {\bibinfo {volume} {72}},\
  \bibinfo {pages} {2442} (\bibinfo {year} {1994})}\BibitemShut {NoStop}%
\bibitem [{\citenamefont {Zhang}\ \emph
  {et~al.}(1997{\natexlab{a}})\citenamefont {Zhang}, \citenamefont {Carlson},\
  and\ \citenamefont {Gubernatis}}]{afqmc-1}%
  \BibitemOpen
  \bibfield  {author} {\bibinfo {author} {\bibfnamefont {S.}~\bibnamefont
  {Zhang}}, \bibinfo {author} {\bibfnamefont {J.}~\bibnamefont {Carlson}}, \
  and\ \bibinfo {author} {\bibfnamefont {J.~E.}\ \bibnamefont {Gubernatis}},\
  }\href {\doibase 10.1103/PhysRevB.55.7464} {\bibfield  {journal} {\bibinfo
  {journal} {Phys. Rev. B}\ }\textbf {\bibinfo {volume} {55}},\ \bibinfo
  {pages} {7464} (\bibinfo {year} {1997}{\natexlab{a}})}\BibitemShut {NoStop}%
\bibitem [{\citenamefont {Chang}\ and\ \citenamefont {Zhang}(2008)}]{afqmc-2}%
  \BibitemOpen
  \bibfield  {author} {\bibinfo {author} {\bibfnamefont {C.-C.}\ \bibnamefont
  {Chang}}\ and\ \bibinfo {author} {\bibfnamefont {S.}~\bibnamefont {Zhang}},\
  }\href {\doibase 10.1103/PhysRevB.78.165101} {\bibfield  {journal} {\bibinfo
  {journal} {Phys. Rev. B}\ }\textbf {\bibinfo {volume} {78}},\ \bibinfo
  {pages} {165101} (\bibinfo {year} {2008})}\BibitemShut {NoStop}%
\bibitem [{\citenamefont {Chang}\ and\ \citenamefont {Zhang}(2010)}]{afqmc-3}%
  \BibitemOpen
  \bibfield  {author} {\bibinfo {author} {\bibfnamefont {C.-C.}\ \bibnamefont
  {Chang}}\ and\ \bibinfo {author} {\bibfnamefont {S.}~\bibnamefont {Zhang}},\
  }\href {\doibase 10.1103/PhysRevLett.104.116402} {\bibfield  {journal}
  {\bibinfo  {journal} {Phys. Rev. Lett.}\ }\textbf {\bibinfo {volume} {104}},\
  \bibinfo {pages} {116402} (\bibinfo {year} {2010})}\BibitemShut {NoStop}%
\bibitem [{\citenamefont {Varney}\ \emph {et~al.}(2009)\citenamefont {Varney},
  \citenamefont {Lee}, \citenamefont {Bai}, \citenamefont {Chiesa},
  \citenamefont {Jarrell},\ and\ \citenamefont {Scalettar}}]{dqmc}%
  \BibitemOpen
  \bibfield  {author} {\bibinfo {author} {\bibfnamefont {C.~N.}\ \bibnamefont
  {Varney}}, \bibinfo {author} {\bibfnamefont {C.-R.}\ \bibnamefont {Lee}},
  \bibinfo {author} {\bibfnamefont {Z.~J.}\ \bibnamefont {Bai}}, \bibinfo
  {author} {\bibfnamefont {S.}~\bibnamefont {Chiesa}}, \bibinfo {author}
  {\bibfnamefont {M.}~\bibnamefont {Jarrell}}, \ and\ \bibinfo {author}
  {\bibfnamefont {R.~T.}\ \bibnamefont {Scalettar}},\ }\href {\doibase
  10.1103/PhysRevB.80.075116} {\bibfield  {journal} {\bibinfo  {journal} {Phys.
  Rev. B}\ }\textbf {\bibinfo {volume} {80}},\ \bibinfo {pages} {075116}
  (\bibinfo {year} {2009})}\BibitemShut {NoStop}%
\bibitem [{\citenamefont {Deng}\ \emph {et~al.}(2015)\citenamefont {Deng},
  \citenamefont {Kozik}, \citenamefont {Prokof'ev},\ and\ \citenamefont
  {Svistunov}}]{Svistunov2015}%
  \BibitemOpen
  \bibfield  {author} {\bibinfo {author} {\bibfnamefont {Y.}~\bibnamefont
  {Deng}}, \bibinfo {author} {\bibfnamefont {E.}~\bibnamefont {Kozik}},
  \bibinfo {author} {\bibfnamefont {N.~V.}\ \bibnamefont {Prokof'ev}}, \ and\
  \bibinfo {author} {\bibfnamefont {B.~V.}\ \bibnamefont {Svistunov}},\ }\href
  {http://stacks.iop.org/0295-5075/110/i=5/a=57001} {\bibfield  {journal}
  {\bibinfo  {journal} {EPL (Europhysics Letters)}\ }\textbf {\bibinfo {volume}
  {110}},\ \bibinfo {pages} {57001} (\bibinfo {year} {2015})}\BibitemShut
  {NoStop}%
\bibitem [{\citenamefont {Hettler}\ \emph {et~al.}(1998)\citenamefont
  {Hettler}, \citenamefont {Tahvildar-Zadeh}, \citenamefont {Jarrell},
  \citenamefont {Pruschke},\ and\ \citenamefont {Krishnamurthy}}]{dca-1}%
  \BibitemOpen
  \bibfield  {author} {\bibinfo {author} {\bibfnamefont {M.~H.}\ \bibnamefont
  {Hettler}}, \bibinfo {author} {\bibfnamefont {A.~N.}\ \bibnamefont
  {Tahvildar-Zadeh}}, \bibinfo {author} {\bibfnamefont {M.}~\bibnamefont
  {Jarrell}}, \bibinfo {author} {\bibfnamefont {T.}~\bibnamefont {Pruschke}}, \
  and\ \bibinfo {author} {\bibfnamefont {H.~R.}\ \bibnamefont
  {Krishnamurthy}},\ }\href {\doibase 10.1103/PhysRevB.58.R7475} {\bibfield
  {journal} {\bibinfo  {journal} {Phys. Rev. B}\ }\textbf {\bibinfo {volume}
  {58}},\ \bibinfo {pages} {R7475} (\bibinfo {year} {1998})}\BibitemShut
  {NoStop}%
\bibitem [{\citenamefont {Maier}\ \emph {et~al.}(2010)\citenamefont {Maier},
  \citenamefont {Alvarez}, \citenamefont {Summers},\ and\ \citenamefont
  {Schulthess}}]{PhysRevLett.104.247001}%
  \BibitemOpen
  \bibfield  {author} {\bibinfo {author} {\bibfnamefont {T.~A.}\ \bibnamefont
  {Maier}}, \bibinfo {author} {\bibfnamefont {G.}~\bibnamefont {Alvarez}},
  \bibinfo {author} {\bibfnamefont {M.}~\bibnamefont {Summers}}, \ and\
  \bibinfo {author} {\bibfnamefont {T.~C.}\ \bibnamefont {Schulthess}},\ }\href
  {\doibase 10.1103/PhysRevLett.104.247001} {\bibfield  {journal} {\bibinfo
  {journal} {Phys. Rev. Lett.}\ }\textbf {\bibinfo {volume} {104}},\ \bibinfo
  {pages} {247001} (\bibinfo {year} {2010})}\BibitemShut {NoStop}%
\bibitem [{\citenamefont {Chen}\ \emph {et~al.}(2013)\citenamefont {Chen},
  \citenamefont {Meng}, \citenamefont {Yang}, \citenamefont {Pruschke},
  \citenamefont {Moreno},\ and\ \citenamefont {Jarrell}}]{PhysRevB.88.245110}%
  \BibitemOpen
  \bibfield  {author} {\bibinfo {author} {\bibfnamefont {K.-S.}\ \bibnamefont
  {Chen}}, \bibinfo {author} {\bibfnamefont {Z.~Y.}\ \bibnamefont {Meng}},
  \bibinfo {author} {\bibfnamefont {S.-X.}\ \bibnamefont {Yang}}, \bibinfo
  {author} {\bibfnamefont {T.}~\bibnamefont {Pruschke}}, \bibinfo {author}
  {\bibfnamefont {J.}~\bibnamefont {Moreno}}, \ and\ \bibinfo {author}
  {\bibfnamefont {M.}~\bibnamefont {Jarrell}},\ }\href {\doibase
  10.1103/PhysRevB.88.245110} {\bibfield  {journal} {\bibinfo  {journal} {Phys.
  Rev. B}\ }\textbf {\bibinfo {volume} {88}},\ \bibinfo {pages} {245110}
  (\bibinfo {year} {2013})}\BibitemShut {NoStop}%
\bibitem [{\citenamefont {Potthoff}\ \emph {et~al.}(2003)\citenamefont
  {Potthoff}, \citenamefont {Aichhorn},\ and\ \citenamefont {Dahnken}}]{vca-1}%
  \BibitemOpen
  \bibfield  {author} {\bibinfo {author} {\bibfnamefont {M.}~\bibnamefont
  {Potthoff}}, \bibinfo {author} {\bibfnamefont {M.}~\bibnamefont {Aichhorn}},
  \ and\ \bibinfo {author} {\bibfnamefont {C.}~\bibnamefont {Dahnken}},\ }\href
  {\doibase 10.1103/PhysRevLett.91.206402} {\bibfield  {journal} {\bibinfo
  {journal} {Phys. Rev. Lett.}\ }\textbf {\bibinfo {volume} {91}},\ \bibinfo
  {pages} {206402} (\bibinfo {year} {2003})}\BibitemShut {NoStop}%
\bibitem [{\citenamefont {Dahnken}\ \emph {et~al.}(2004)\citenamefont
  {Dahnken}, \citenamefont {Aichhorn}, \citenamefont {Hanke}, \citenamefont
  {Arrigoni},\ and\ \citenamefont {Potthoff}}]{vca-2}%
  \BibitemOpen
  \bibfield  {author} {\bibinfo {author} {\bibfnamefont {C.}~\bibnamefont
  {Dahnken}}, \bibinfo {author} {\bibfnamefont {M.}~\bibnamefont {Aichhorn}},
  \bibinfo {author} {\bibfnamefont {W.}~\bibnamefont {Hanke}}, \bibinfo
  {author} {\bibfnamefont {E.}~\bibnamefont {Arrigoni}}, \ and\ \bibinfo
  {author} {\bibfnamefont {M.}~\bibnamefont {Potthoff}},\ }\href {\doibase
  10.1103/PhysRevB.70.245110} {\bibfield  {journal} {\bibinfo  {journal} {Phys.
  Rev. B}\ }\textbf {\bibinfo {volume} {70}},\ \bibinfo {pages} {245110}
  (\bibinfo {year} {2004})}\BibitemShut {NoStop}%
\bibitem [{\citenamefont {Georges}\ \emph {et~al.}(1996)\citenamefont
  {Georges}, \citenamefont {Kotliar}, \citenamefont {Krauth},\ and\
  \citenamefont {Rozenberg}}]{RevModPhys.68.13}%
  \BibitemOpen
  \bibfield  {author} {\bibinfo {author} {\bibfnamefont {A.}~\bibnamefont
  {Georges}}, \bibinfo {author} {\bibfnamefont {G.}~\bibnamefont {Kotliar}},
  \bibinfo {author} {\bibfnamefont {W.}~\bibnamefont {Krauth}}, \ and\ \bibinfo
  {author} {\bibfnamefont {M.~J.}\ \bibnamefont {Rozenberg}},\ }\href {\doibase
  10.1103/RevModPhys.68.13} {\bibfield  {journal} {\bibinfo  {journal} {Rev.
  Mod. Phys.}\ }\textbf {\bibinfo {volume} {68}},\ \bibinfo {pages} {13}
  (\bibinfo {year} {1996})}\BibitemShut {NoStop}%
\bibitem [{\citenamefont {Lichtenstein}\ and\ \citenamefont
  {Katsnelson}(2000)}]{dmft-1}%
  \BibitemOpen
  \bibfield  {author} {\bibinfo {author} {\bibfnamefont {A.~I.}\ \bibnamefont
  {Lichtenstein}}\ and\ \bibinfo {author} {\bibfnamefont {M.~I.}\ \bibnamefont
  {Katsnelson}},\ }\href {\doibase 10.1103/PhysRevB.62.R9283} {\bibfield
  {journal} {\bibinfo  {journal} {Phys. Rev. B}\ }\textbf {\bibinfo {volume}
  {62}},\ \bibinfo {pages} {R9283} (\bibinfo {year} {2000})}\BibitemShut
  {NoStop}%
\bibitem [{\citenamefont {Kotliar}\ \emph {et~al.}(2001)\citenamefont
  {Kotliar}, \citenamefont {Savrasov}, \citenamefont {P\'alsson},\ and\
  \citenamefont {Biroli}}]{dmft-2}%
  \BibitemOpen
  \bibfield  {author} {\bibinfo {author} {\bibfnamefont {G.}~\bibnamefont
  {Kotliar}}, \bibinfo {author} {\bibfnamefont {S.~Y.}\ \bibnamefont
  {Savrasov}}, \bibinfo {author} {\bibfnamefont {G.}~\bibnamefont {P\'alsson}},
  \ and\ \bibinfo {author} {\bibfnamefont {G.}~\bibnamefont {Biroli}},\ }\href
  {\doibase 10.1103/PhysRevLett.87.186401} {\bibfield  {journal} {\bibinfo
  {journal} {Phys. Rev. Lett.}\ }\textbf {\bibinfo {volume} {87}},\ \bibinfo
  {pages} {186401} (\bibinfo {year} {2001})}\BibitemShut {NoStop}%
\bibitem [{\citenamefont {Gull}\ \emph {et~al.}(2013)\citenamefont {Gull},
  \citenamefont {Parcollet},\ and\ \citenamefont
  {Millis}}]{PhysRevLett.110.216405}%
  \BibitemOpen
  \bibfield  {author} {\bibinfo {author} {\bibfnamefont {E.}~\bibnamefont
  {Gull}}, \bibinfo {author} {\bibfnamefont {O.}~\bibnamefont {Parcollet}}, \
  and\ \bibinfo {author} {\bibfnamefont {A.~J.}\ \bibnamefont {Millis}},\
  }\href {\doibase 10.1103/PhysRevLett.110.216405} {\bibfield  {journal}
  {\bibinfo  {journal} {Phys. Rev. Lett.}\ }\textbf {\bibinfo {volume} {110}},\
  \bibinfo {pages} {216405} (\bibinfo {year} {2013})}\BibitemShut {NoStop}%
\bibitem [{\citenamefont {Gull}\ \emph {et~al.}(2010)\citenamefont {Gull},
  \citenamefont {Ferrero}, \citenamefont {Parcollet}, \citenamefont {Georges},\
  and\ \citenamefont {Millis}}]{PhysRevB.82.155101}%
  \BibitemOpen
  \bibfield  {author} {\bibinfo {author} {\bibfnamefont {E.}~\bibnamefont
  {Gull}}, \bibinfo {author} {\bibfnamefont {M.}~\bibnamefont {Ferrero}},
  \bibinfo {author} {\bibfnamefont {O.}~\bibnamefont {Parcollet}}, \bibinfo
  {author} {\bibfnamefont {A.}~\bibnamefont {Georges}}, \ and\ \bibinfo
  {author} {\bibfnamefont {A.~J.}\ \bibnamefont {Millis}},\ }\href {\doibase
  10.1103/PhysRevB.82.155101} {\bibfield  {journal} {\bibinfo  {journal} {Phys.
  Rev. B}\ }\textbf {\bibinfo {volume} {82}},\ \bibinfo {pages} {155101}
  (\bibinfo {year} {2010})}\BibitemShut {NoStop}%
\bibitem [{\citenamefont {Corboz}(2016)}]{PhysRevB.93.045116}%
  \BibitemOpen
  \bibfield  {author} {\bibinfo {author} {\bibfnamefont {P.}~\bibnamefont
  {Corboz}},\ }\href {\doibase 10.1103/PhysRevB.93.045116} {\bibfield
  {journal} {\bibinfo  {journal} {Phys. Rev. B}\ }\textbf {\bibinfo {volume}
  {93}},\ \bibinfo {pages} {045116} (\bibinfo {year} {2016})}\BibitemShut
  {NoStop}%
\bibitem [{\citenamefont {Verstraete}\ \emph {et~al.}(2008)\citenamefont
  {Verstraete}, \citenamefont {Murg},\ and\ \citenamefont {Cirac}}]{peps}%
  \BibitemOpen
  \bibfield  {author} {\bibinfo {author} {\bibfnamefont {F.}~\bibnamefont
  {Verstraete}}, \bibinfo {author} {\bibfnamefont {V.}~\bibnamefont {Murg}}, \
  and\ \bibinfo {author} {\bibfnamefont {J.}~\bibnamefont {Cirac}},\ }\href
  {\doibase 10.1080/14789940801912366} {\bibfield  {journal} {\bibinfo
  {journal} {Advances in Physics}\ }\textbf {\bibinfo {volume} {57}},\ \bibinfo
  {pages} {143} (\bibinfo {year} {2008})},\ \Eprint
  {http://arxiv.org/abs/https://doi.org/10.1080/14789940801912366}
  {https://doi.org/10.1080/14789940801912366} \BibitemShut {NoStop}%
\bibitem [{\citenamefont {LeBlanc}\ \emph {et~al.}(2015)\citenamefont
  {LeBlanc}, \citenamefont {Antipov}, \citenamefont {Becca}, \citenamefont
  {Bulik}, \citenamefont {Chan}, \citenamefont {Chung}, \citenamefont {Deng},
  \citenamefont {Ferrero}, \citenamefont {Henderson}, \citenamefont
  {Jim\'enez-Hoyos}, \citenamefont {Kozik}, \citenamefont {Liu}, \citenamefont
  {Millis}, \citenamefont {Prokof'ev}, \citenamefont {Qin}, \citenamefont
  {Scuseria}, \citenamefont {Shi}, \citenamefont {Svistunov}, \citenamefont
  {Tocchio}, \citenamefont {Tupitsyn}, \citenamefont {White}, \citenamefont
  {Zhang}, \citenamefont {Zheng}, \citenamefont {Zhu},\ and\ \citenamefont
  {Gull}}]{leblanc}%
  \BibitemOpen
  \bibfield  {author} {\bibinfo {author} {\bibfnamefont {J.~P.~F.}\
  \bibnamefont {LeBlanc}}, \bibinfo {author} {\bibfnamefont {A.~E.}\
  \bibnamefont {Antipov}}, \bibinfo {author} {\bibfnamefont {F.}~\bibnamefont
  {Becca}}, \bibinfo {author} {\bibfnamefont {I.~W.}\ \bibnamefont {Bulik}},
  \bibinfo {author} {\bibfnamefont {G.~K.-L.}\ \bibnamefont {Chan}}, \bibinfo
  {author} {\bibfnamefont {C.-M.}\ \bibnamefont {Chung}}, \bibinfo {author}
  {\bibfnamefont {Y.}~\bibnamefont {Deng}}, \bibinfo {author} {\bibfnamefont
  {M.}~\bibnamefont {Ferrero}}, \bibinfo {author} {\bibfnamefont {T.~M.}\
  \bibnamefont {Henderson}}, \bibinfo {author} {\bibfnamefont {C.~A.}\
  \bibnamefont {Jim\'enez-Hoyos}}, \bibinfo {author} {\bibfnamefont
  {E.}~\bibnamefont {Kozik}}, \bibinfo {author} {\bibfnamefont {X.-W.}\
  \bibnamefont {Liu}}, \bibinfo {author} {\bibfnamefont {A.~J.}\ \bibnamefont
  {Millis}}, \bibinfo {author} {\bibfnamefont {N.~V.}\ \bibnamefont
  {Prokof'ev}}, \bibinfo {author} {\bibfnamefont {M.}~\bibnamefont {Qin}},
  \bibinfo {author} {\bibfnamefont {G.~E.}\ \bibnamefont {Scuseria}}, \bibinfo
  {author} {\bibfnamefont {H.}~\bibnamefont {Shi}}, \bibinfo {author}
  {\bibfnamefont {B.~V.}\ \bibnamefont {Svistunov}}, \bibinfo {author}
  {\bibfnamefont {L.~F.}\ \bibnamefont {Tocchio}}, \bibinfo {author}
  {\bibfnamefont {I.~S.}\ \bibnamefont {Tupitsyn}}, \bibinfo {author}
  {\bibfnamefont {S.~R.}\ \bibnamefont {White}}, \bibinfo {author}
  {\bibfnamefont {S.}~\bibnamefont {Zhang}}, \bibinfo {author} {\bibfnamefont
  {B.-X.}\ \bibnamefont {Zheng}}, \bibinfo {author} {\bibfnamefont
  {Z.}~\bibnamefont {Zhu}}, \ and\ \bibinfo {author} {\bibfnamefont
  {E.}~\bibnamefont {Gull}} (\bibinfo {collaboration} {Simons Collaboration on
  the Many-Electron Problem}),\ }\href {\doibase 10.1103/PhysRevX.5.041041}
  {\bibfield  {journal} {\bibinfo  {journal} {Phys. Rev. X}\ }\textbf {\bibinfo
  {volume} {5}},\ \bibinfo {pages} {041041} (\bibinfo {year}
  {2015})}\BibitemShut {NoStop}%
\bibitem [{\citenamefont {Qin}\ \emph {et~al.}(2016)\citenamefont {Qin},
  \citenamefont {Shi},\ and\ \citenamefont {Zhang}}]{zhang}%
  \BibitemOpen
  \bibfield  {author} {\bibinfo {author} {\bibfnamefont {M.}~\bibnamefont
  {Qin}}, \bibinfo {author} {\bibfnamefont {H.}~\bibnamefont {Shi}}, \ and\
  \bibinfo {author} {\bibfnamefont {S.}~\bibnamefont {Zhang}},\ }\href
  {\doibase 10.1103/PhysRevB.94.085103} {\bibfield  {journal} {\bibinfo
  {journal} {Phys. Rev. B}\ }\textbf {\bibinfo {volume} {94}},\ \bibinfo
  {pages} {085103} (\bibinfo {year} {2016})}\BibitemShut {NoStop}%
\bibitem [{\citenamefont {Zheng}\ \emph {et~al.}(2017)\citenamefont {Zheng},
  \citenamefont {Chung}, \citenamefont {Corboz}, \citenamefont {Ehlers},
  \citenamefont {Qin}, \citenamefont {Noack}, \citenamefont {Shi},
  \citenamefont {White}, \citenamefont {Zhang},\ and\ \citenamefont
  {Chan}}]{chan}%
  \BibitemOpen
  \bibfield  {author} {\bibinfo {author} {\bibfnamefont {B.-X.}\ \bibnamefont
  {Zheng}}, \bibinfo {author} {\bibfnamefont {C.-M.}\ \bibnamefont {Chung}},
  \bibinfo {author} {\bibfnamefont {P.}~\bibnamefont {Corboz}}, \bibinfo
  {author} {\bibfnamefont {G.}~\bibnamefont {Ehlers}}, \bibinfo {author}
  {\bibfnamefont {M.-P.}\ \bibnamefont {Qin}}, \bibinfo {author} {\bibfnamefont
  {R.~M.}\ \bibnamefont {Noack}}, \bibinfo {author} {\bibfnamefont
  {H.}~\bibnamefont {Shi}}, \bibinfo {author} {\bibfnamefont {S.~R.}\
  \bibnamefont {White}}, \bibinfo {author} {\bibfnamefont {S.}~\bibnamefont
  {Zhang}}, \ and\ \bibinfo {author} {\bibfnamefont {G.~K.-L.}\ \bibnamefont
  {Chan}},\ }\href {\doibase 10.1126/science.aam7127} {\bibfield  {journal}
  {\bibinfo  {journal} {Science}\ }\textbf {\bibinfo {volume} {358}},\ \bibinfo
  {pages} {1155} (\bibinfo {year} {2017})},\ \Eprint
  {http://arxiv.org/abs/http://science.sciencemag.org/content/358/6367/1155.full.pdf}
  {http://science.sciencemag.org/content/358/6367/1155.full.pdf} \BibitemShut
  {NoStop}%
\bibitem [{\citenamefont {Brinkman}\ and\ \citenamefont
  {Rice}(1970)}]{Rice1970}%
  \BibitemOpen
  \bibfield  {author} {\bibinfo {author} {\bibfnamefont {W.~F.}\ \bibnamefont
  {Brinkman}}\ and\ \bibinfo {author} {\bibfnamefont {T.~M.}\ \bibnamefont
  {Rice}},\ }\href {\doibase 10.1103/PhysRevB.2.4302} {\bibfield  {journal}
  {\bibinfo  {journal} {Phys. Rev. B}\ }\textbf {\bibinfo {volume} {2}},\
  \bibinfo {pages} {4302} (\bibinfo {year} {1970})}\BibitemShut {NoStop}%
\bibitem [{\citenamefont {Gutzwiller}(1965)}]{ga}%
  \BibitemOpen
  \bibfield  {author} {\bibinfo {author} {\bibfnamefont {M.~C.}\ \bibnamefont
  {Gutzwiller}},\ }\href {\doibase 10.1103/PhysRev.137.A1726} {\bibfield
  {journal} {\bibinfo  {journal} {Phys. Rev.}\ }\textbf {\bibinfo {volume}
  {137}},\ \bibinfo {pages} {A1726} (\bibinfo {year} {1965})}\BibitemShut
  {NoStop}%
\bibitem [{\citenamefont {Ogawa}\ \emph {et~al.}(1975)\citenamefont {Ogawa},
  \citenamefont {Kanda},\ and\ \citenamefont
  {Matsubara}}]{gutzwiller-ansatz-2}%
  \BibitemOpen
  \bibfield  {author} {\bibinfo {author} {\bibfnamefont {T.}~\bibnamefont
  {Ogawa}}, \bibinfo {author} {\bibfnamefont {K.}~\bibnamefont {Kanda}}, \ and\
  \bibinfo {author} {\bibfnamefont {T.}~\bibnamefont {Matsubara}},\ }\href
  {\doibase 10.1143/PTP.53.614} {\bibfield  {journal} {\bibinfo  {journal}
  {Progress of Theoretical Physics}\ }\textbf {\bibinfo {volume} {53}},\
  \bibinfo {pages} {614} (\bibinfo {year} {1975})}\BibitemShut {NoStop}%
\bibitem [{\citenamefont {Vollhardt}(1984)}]{Vollhardt1984}%
  \BibitemOpen
  \bibfield  {author} {\bibinfo {author} {\bibfnamefont {D.}~\bibnamefont
  {Vollhardt}},\ }\href {\doibase 10.1103/RevModPhys.56.99} {\bibfield
  {journal} {\bibinfo  {journal} {Rev. Mod. Phys.}\ }\textbf {\bibinfo {volume}
  {56}},\ \bibinfo {pages} {99} (\bibinfo {year} {1984})}\BibitemShut {NoStop}%
\bibitem [{\citenamefont {Zhang}\ \emph {et~al.}(1988)\citenamefont {Zhang},
  \citenamefont {Gros}, \citenamefont {Rice},\ and\ \citenamefont
  {Shiba}}]{Shiba1988}%
  \BibitemOpen
  \bibfield  {author} {\bibinfo {author} {\bibfnamefont {F.~C.}\ \bibnamefont
  {Zhang}}, \bibinfo {author} {\bibfnamefont {C.}~\bibnamefont {Gros}},
  \bibinfo {author} {\bibfnamefont {T.~M.}\ \bibnamefont {Rice}}, \ and\
  \bibinfo {author} {\bibfnamefont {H.}~\bibnamefont {Shiba}},\ }\href
  {http://stacks.iop.org/0953-2048/1/i=1/a=009} {\bibfield  {journal} {\bibinfo
   {journal} {Superconductor Science and Technology}\ }\textbf {\bibinfo
  {volume} {1}},\ \bibinfo {pages} {36} (\bibinfo {year} {1988})}\BibitemShut
  {NoStop}%
\bibitem [{\citenamefont {Metzner}\ and\ \citenamefont
  {Vollhardt}(1989)}]{PhysRevLett.62.324}%
  \BibitemOpen
  \bibfield  {author} {\bibinfo {author} {\bibfnamefont {W.}~\bibnamefont
  {Metzner}}\ and\ \bibinfo {author} {\bibfnamefont {D.}~\bibnamefont
  {Vollhardt}},\ }\href {\doibase 10.1103/PhysRevLett.62.324} {\bibfield
  {journal} {\bibinfo  {journal} {Phys. Rev. Lett.}\ }\textbf {\bibinfo
  {volume} {62}},\ \bibinfo {pages} {324} (\bibinfo {year} {1989})}\BibitemShut
  {NoStop}%
\bibitem [{\citenamefont {Gros}\ \emph
  {et~al.}(1987{\natexlab{a}})\citenamefont {Gros}, \citenamefont {Joynt},\
  and\ \citenamefont {Rice}}]{rice}%
  \BibitemOpen
  \bibfield  {author} {\bibinfo {author} {\bibfnamefont {C.}~\bibnamefont
  {Gros}}, \bibinfo {author} {\bibfnamefont {R.}~\bibnamefont {Joynt}}, \ and\
  \bibinfo {author} {\bibfnamefont {T.~M.}\ \bibnamefont {Rice}},\ }\href
  {\doibase 10.1103/PhysRevB.36.381} {\bibfield  {journal} {\bibinfo  {journal}
  {Phys. Rev. B}\ }\textbf {\bibinfo {volume} {36}},\ \bibinfo {pages} {381}
  (\bibinfo {year} {1987}{\natexlab{a}})}\BibitemShut {NoStop}%
\bibitem [{\citenamefont {Horsch}\ and\ \citenamefont
  {Kaplan}(1983)}]{Horsch1983}%
  \BibitemOpen
  \bibfield  {author} {\bibinfo {author} {\bibfnamefont {P.}~\bibnamefont
  {Horsch}}\ and\ \bibinfo {author} {\bibfnamefont {T.~A.}\ \bibnamefont
  {Kaplan}},\ }\href {http://stacks.iop.org/0022-3719/16/i=35/a=002} {\bibfield
   {journal} {\bibinfo  {journal} {Journal of Physics C: Solid State Physics}\
  }\textbf {\bibinfo {volume} {16}},\ \bibinfo {pages} {L1203} (\bibinfo {year}
  {1983})}\BibitemShut {NoStop}%
\bibitem [{\citenamefont {Kaplan}\ \emph {et~al.}(1982)\citenamefont {Kaplan},
  \citenamefont {Horsch},\ and\ \citenamefont {Fulde}}]{fulde-1}%
  \BibitemOpen
  \bibfield  {author} {\bibinfo {author} {\bibfnamefont {T.~A.}\ \bibnamefont
  {Kaplan}}, \bibinfo {author} {\bibfnamefont {P.}~\bibnamefont {Horsch}}, \
  and\ \bibinfo {author} {\bibfnamefont {P.}~\bibnamefont {Fulde}},\ }\href
  {\doibase 10.1103/PhysRevLett.49.889} {\bibfield  {journal} {\bibinfo
  {journal} {Phys. Rev. Lett.}\ }\textbf {\bibinfo {volume} {49}},\ \bibinfo
  {pages} {889} (\bibinfo {year} {1982})}\BibitemShut {NoStop}%
\bibitem [{\citenamefont {Metzner}\ and\ \citenamefont
  {Vollhardt}(1987)}]{vollhardt-1}%
  \BibitemOpen
  \bibfield  {author} {\bibinfo {author} {\bibfnamefont {W.}~\bibnamefont
  {Metzner}}\ and\ \bibinfo {author} {\bibfnamefont {D.}~\bibnamefont
  {Vollhardt}},\ }\href {\doibase 10.1103/PhysRevLett.59.121} {\bibfield
  {journal} {\bibinfo  {journal} {Phys. Rev. Lett.}\ }\textbf {\bibinfo
  {volume} {59}},\ \bibinfo {pages} {121} (\bibinfo {year} {1987})}\BibitemShut
  {NoStop}%
\bibitem [{\citenamefont {Gebhard}\ and\ \citenamefont
  {Vollhardt}(1987)}]{vollhardt-2}%
  \BibitemOpen
  \bibfield  {author} {\bibinfo {author} {\bibfnamefont {F.}~\bibnamefont
  {Gebhard}}\ and\ \bibinfo {author} {\bibfnamefont {D.}~\bibnamefont
  {Vollhardt}},\ }\href {\doibase 10.1103/PhysRevLett.59.1472} {\bibfield
  {journal} {\bibinfo  {journal} {Phys. Rev. Lett.}\ }\textbf {\bibinfo
  {volume} {59}},\ \bibinfo {pages} {1472} (\bibinfo {year}
  {1987})}\BibitemShut {NoStop}%
\bibitem [{\citenamefont {Jastrow}(1955)}]{Jastrow1955}%
  \BibitemOpen
  \bibfield  {author} {\bibinfo {author} {\bibfnamefont {R.}~\bibnamefont
  {Jastrow}},\ }\href {\doibase 10.1103/PhysRev.98.1479} {\bibfield  {journal}
  {\bibinfo  {journal} {Phys. Rev.}\ }\textbf {\bibinfo {volume} {98}},\
  \bibinfo {pages} {1479} (\bibinfo {year} {1955})}\BibitemShut {NoStop}%
\bibitem [{\citenamefont {Capello}\ \emph {et~al.}(2005)\citenamefont
  {Capello}, \citenamefont {Becca}, \citenamefont {Fabrizio}, \citenamefont
  {Sorella},\ and\ \citenamefont {Tosatti}}]{density-density}%
  \BibitemOpen
  \bibfield  {author} {\bibinfo {author} {\bibfnamefont {M.}~\bibnamefont
  {Capello}}, \bibinfo {author} {\bibfnamefont {F.}~\bibnamefont {Becca}},
  \bibinfo {author} {\bibfnamefont {M.}~\bibnamefont {Fabrizio}}, \bibinfo
  {author} {\bibfnamefont {S.}~\bibnamefont {Sorella}}, \ and\ \bibinfo
  {author} {\bibfnamefont {E.}~\bibnamefont {Tosatti}},\ }\href {\doibase
  10.1103/PhysRevLett.94.026406} {\bibfield  {journal} {\bibinfo  {journal}
  {Phys. Rev. Lett.}\ }\textbf {\bibinfo {volume} {94}},\ \bibinfo {pages}
  {026406} (\bibinfo {year} {2005})}\BibitemShut {NoStop}%
\bibitem [{\citenamefont {Liu}\ \emph {et~al.}(2005)\citenamefont {Liu},
  \citenamefont {Schmalian},\ and\ \citenamefont {Trivedi}}]{holon-doublon-2}%
  \BibitemOpen
  \bibfield  {author} {\bibinfo {author} {\bibfnamefont {J.}~\bibnamefont
  {Liu}}, \bibinfo {author} {\bibfnamefont {J.}~\bibnamefont {Schmalian}}, \
  and\ \bibinfo {author} {\bibfnamefont {N.}~\bibnamefont {Trivedi}},\ }\href
  {\doibase 10.1103/PhysRevLett.94.127003} {\bibfield  {journal} {\bibinfo
  {journal} {Phys. Rev. Lett.}\ }\textbf {\bibinfo {volume} {94}},\ \bibinfo
  {pages} {127003} (\bibinfo {year} {2005})}\BibitemShut {NoStop}%
\bibitem [{\citenamefont {Watanabe}\ \emph {et~al.}(2006)\citenamefont
  {Watanabe}, \citenamefont {Yokoyama}, \citenamefont {Tanaka},\ and\
  \citenamefont {Inoue}}]{holon-doublon-3}%
  \BibitemOpen
  \bibfield  {author} {\bibinfo {author} {\bibfnamefont {T.}~\bibnamefont
  {Watanabe}}, \bibinfo {author} {\bibfnamefont {H.}~\bibnamefont {Yokoyama}},
  \bibinfo {author} {\bibfnamefont {Y.}~\bibnamefont {Tanaka}}, \ and\ \bibinfo
  {author} {\bibfnamefont {J.-i.}\ \bibnamefont {Inoue}},\ }\href {\doibase
  10.1143/JPSJ.75.074707} {\bibfield  {journal} {\bibinfo  {journal} {Journal
  of the Physical Society of Japan}\ }\textbf {\bibinfo {volume} {75}},\
  \bibinfo {pages} {074707} (\bibinfo {year} {2006})}\BibitemShut {NoStop}%
\bibitem [{\citenamefont {Li}\ and\ \citenamefont
  {d'Ambrumenil}(1993)}]{gutzwiller-hf-spin-density}%
  \BibitemOpen
  \bibfield  {author} {\bibinfo {author} {\bibfnamefont {Y.~M.}\ \bibnamefont
  {Li}}\ and\ \bibinfo {author} {\bibfnamefont {N.}~\bibnamefont
  {d'Ambrumenil}},\ }\href {\doibase 10.1063/1.352555} {\bibfield  {journal}
  {\bibinfo  {journal} {Journal of Applied Physics}\ }\textbf {\bibinfo
  {volume} {73}},\ \bibinfo {pages} {6537} (\bibinfo {year}
  {1993})}\BibitemShut {NoStop}%
\bibitem [{\citenamefont {Anderson}(1987)}]{Anderson1987}%
  \BibitemOpen
  \bibfield  {author} {\bibinfo {author} {\bibfnamefont {P.~W.}\ \bibnamefont
  {Anderson}},\ }\href {\doibase 10.1126/science.235.4793.1196} {\bibfield
  {journal} {\bibinfo  {journal} {Science}\ }\textbf {\bibinfo {volume}
  {235}},\ \bibinfo {pages} {1196} (\bibinfo {year} {1987})},\ \Eprint
  {http://arxiv.org/abs/http://science.sciencemag.org/content/235/4793/1196.full.pdf}
  {http://science.sciencemag.org/content/235/4793/1196.full.pdf} \BibitemShut
  {NoStop}%
\bibitem [{\citenamefont {Edegger}\ \emph {et~al.}(2005)\citenamefont
  {Edegger}, \citenamefont {Fukushima}, \citenamefont {Gros},\ and\
  \citenamefont {Muthukumar}}]{gutzwiller-bcs}%
  \BibitemOpen
  \bibfield  {author} {\bibinfo {author} {\bibfnamefont {B.}~\bibnamefont
  {Edegger}}, \bibinfo {author} {\bibfnamefont {N.}~\bibnamefont {Fukushima}},
  \bibinfo {author} {\bibfnamefont {C.}~\bibnamefont {Gros}}, \ and\ \bibinfo
  {author} {\bibfnamefont {V.~N.}\ \bibnamefont {Muthukumar}},\ }\href
  {\doibase 10.1103/PhysRevB.72.134504} {\bibfield  {journal} {\bibinfo
  {journal} {Phys. Rev. B}\ }\textbf {\bibinfo {volume} {72}},\ \bibinfo
  {pages} {134504} (\bibinfo {year} {2005})}\BibitemShut {NoStop}%
\bibitem [{\citenamefont {Anderson}\ and\ \citenamefont
  {Ong}(2006)}]{gutzwiller-bcs-2}%
  \BibitemOpen
  \bibfield  {author} {\bibinfo {author} {\bibfnamefont {P.}~\bibnamefont
  {Anderson}}\ and\ \bibinfo {author} {\bibfnamefont {N.}~\bibnamefont {Ong}},\
  }\href {\doibase https://doi.org/10.1016/j.jpcs.2005.10.132} {\bibfield
  {journal} {\bibinfo  {journal} {Journal of Physics and Chemistry of Solids}\
  }\textbf {\bibinfo {volume} {67}},\ \bibinfo {pages} {1 } (\bibinfo {year}
  {2006})},\ \bibinfo {note} {spectroscopies in Novel Superconductors
  2004}\BibitemShut {NoStop}%
\bibitem [{\citenamefont {Yokoyama}\ and\ \citenamefont
  {Shiba}(1988)}]{gutzwiller-bcs-3}%
  \BibitemOpen
  \bibfield  {author} {\bibinfo {author} {\bibfnamefont {H.}~\bibnamefont
  {Yokoyama}}\ and\ \bibinfo {author} {\bibfnamefont {H.}~\bibnamefont
  {Shiba}},\ }\href {\doibase 10.1143/JPSJ.57.2482} {\bibfield  {journal}
  {\bibinfo  {journal} {Journal of the Physical Society of Japan}\ }\textbf
  {\bibinfo {volume} {57}},\ \bibinfo {pages} {2482} (\bibinfo {year}
  {1988})}\BibitemShut {NoStop}%
\bibitem [{\citenamefont {Paramekanti}\ \emph {et~al.}(2001)\citenamefont
  {Paramekanti}, \citenamefont {Randeria},\ and\ \citenamefont
  {Trivedi}}]{gutzwiller-bcs-4}%
  \BibitemOpen
  \bibfield  {author} {\bibinfo {author} {\bibfnamefont {A.}~\bibnamefont
  {Paramekanti}}, \bibinfo {author} {\bibfnamefont {M.}~\bibnamefont
  {Randeria}}, \ and\ \bibinfo {author} {\bibfnamefont {N.}~\bibnamefont
  {Trivedi}},\ }\href {\doibase 10.1103/PhysRevLett.87.217002} {\bibfield
  {journal} {\bibinfo  {journal} {Phys. Rev. Lett.}\ }\textbf {\bibinfo
  {volume} {87}},\ \bibinfo {pages} {217002} (\bibinfo {year}
  {2001})}\BibitemShut {NoStop}%
\bibitem [{\citenamefont {Gros}(1989)}]{Gros1989}%
  \BibitemOpen
  \bibfield  {author} {\bibinfo {author} {\bibfnamefont {C.}~\bibnamefont
  {Gros}},\ }\href {\doibase https://doi.org/10.1016/0003-4916(89)90077-8}
  {\bibfield  {journal} {\bibinfo  {journal} {Annals of Physics}\ }\textbf
  {\bibinfo {volume} {189}},\ \bibinfo {pages} {53 } (\bibinfo {year}
  {1989})}\BibitemShut {NoStop}%
\bibitem [{\citenamefont {Gros}(1988)}]{Gros1988}%
  \BibitemOpen
  \bibfield  {author} {\bibinfo {author} {\bibfnamefont {C.}~\bibnamefont
  {Gros}},\ }\href {\doibase 10.1103/PhysRevB.38.931} {\bibfield  {journal}
  {\bibinfo  {journal} {Phys. Rev. B}\ }\textbf {\bibinfo {volume} {38}},\
  \bibinfo {pages} {931} (\bibinfo {year} {1988})}\BibitemShut {NoStop}%
\bibitem [{\citenamefont {Kaczmarczyk}\ \emph {et~al.}(2013)\citenamefont
  {Kaczmarczyk}, \citenamefont {Spa\l{}ek}, \citenamefont {Schickling},\ and\
  \citenamefont {B\"unemann}}]{PhysRevB.88.115127}%
  \BibitemOpen
  \bibfield  {author} {\bibinfo {author} {\bibfnamefont {J.}~\bibnamefont
  {Kaczmarczyk}}, \bibinfo {author} {\bibfnamefont {J.}~\bibnamefont
  {Spa\l{}ek}}, \bibinfo {author} {\bibfnamefont {T.}~\bibnamefont
  {Schickling}}, \ and\ \bibinfo {author} {\bibfnamefont {J.}~\bibnamefont
  {B\"unemann}},\ }\href {\doibase 10.1103/PhysRevB.88.115127} {\bibfield
  {journal} {\bibinfo  {journal} {Phys. Rev. B}\ }\textbf {\bibinfo {volume}
  {88}},\ \bibinfo {pages} {115127} (\bibinfo {year} {2013})}\BibitemShut
  {NoStop}%
\bibitem [{\citenamefont {Baeriswyl}(2018)}]{Baeriswyl2018}%
  \BibitemOpen
  \bibfield  {author} {\bibinfo {author} {\bibfnamefont {D.}~\bibnamefont
  {Baeriswyl}},\ }\href@noop {} {\enquote {\bibinfo {title} {Superconductivity
  from repulsion: Variational results for the 2d hubbard model in the limit of
  weak interaction},}\ } (\bibinfo {year} {2018}),\ \Eprint
  {http://arxiv.org/abs/arXiv:1809.04916} {arXiv:1809.04916} \BibitemShut
  {NoStop}%
\bibitem [{\citenamefont {Lee}\ and\ \citenamefont {Feng}(1988)}]{bcs-af-1}%
  \BibitemOpen
  \bibfield  {author} {\bibinfo {author} {\bibfnamefont {T.~K.}\ \bibnamefont
  {Lee}}\ and\ \bibinfo {author} {\bibfnamefont {S.}~\bibnamefont {Feng}},\
  }\href {\doibase 10.1103/PhysRevB.38.11809} {\bibfield  {journal} {\bibinfo
  {journal} {Phys. Rev. B}\ }\textbf {\bibinfo {volume} {38}},\ \bibinfo
  {pages} {11809} (\bibinfo {year} {1988})}\BibitemShut {NoStop}%
\bibitem [{\citenamefont {Huang}\ \emph {et~al.}(2005)\citenamefont {Huang},
  \citenamefont {Li},\ and\ \citenamefont {Zhang}}]{charge-order}%
  \BibitemOpen
  \bibfield  {author} {\bibinfo {author} {\bibfnamefont {H.-X.}\ \bibnamefont
  {Huang}}, \bibinfo {author} {\bibfnamefont {Y.-Q.}\ \bibnamefont {Li}}, \
  and\ \bibinfo {author} {\bibfnamefont {F.-C.}\ \bibnamefont {Zhang}},\ }\href
  {\doibase 10.1103/PhysRevB.71.184514} {\bibfield  {journal} {\bibinfo
  {journal} {Phys. Rev. B}\ }\textbf {\bibinfo {volume} {71}},\ \bibinfo
  {pages} {184514} (\bibinfo {year} {2005})}\BibitemShut {NoStop}%
\bibitem [{\citenamefont {B\"unemann}\ \emph {et~al.}(2012)\citenamefont
  {B\"unemann}, \citenamefont {Schickling},\ and\ \citenamefont
  {Gebhard}}]{Gebhard2012}%
  \BibitemOpen
  \bibfield  {author} {\bibinfo {author} {\bibfnamefont {J.}~\bibnamefont
  {B\"unemann}}, \bibinfo {author} {\bibfnamefont {T.}~\bibnamefont
  {Schickling}}, \ and\ \bibinfo {author} {\bibfnamefont {F.}~\bibnamefont
  {Gebhard}},\ }\href {http://stacks.iop.org/0295-5075/98/i=2/a=27006}
  {\bibfield  {journal} {\bibinfo  {journal} {EPL (Europhysics Letters)}\
  }\textbf {\bibinfo {volume} {98}},\ \bibinfo {pages} {27006} (\bibinfo {year}
  {2012})}\BibitemShut {NoStop}%
\bibitem [{\citenamefont {Lanat\`a}\ \emph {et~al.}(2017)\citenamefont
  {Lanat\`a}, \citenamefont {Lee}, \citenamefont {Yao},\ and\ \citenamefont
  {Dobrosavljevi\'{c}}}]{PhysRevB.96.195126}%
  \BibitemOpen
  \bibfield  {author} {\bibinfo {author} {\bibfnamefont {N.}~\bibnamefont
  {Lanat\`a}}, \bibinfo {author} {\bibfnamefont {T.-H.}\ \bibnamefont {Lee}},
  \bibinfo {author} {\bibfnamefont {Y.-X.}\ \bibnamefont {Yao}}, \ and\
  \bibinfo {author} {\bibfnamefont {V.}~\bibnamefont {Dobrosavljevi\'{c}}},\
  }\href {\doibase 10.1103/PhysRevB.96.195126} {\bibfield  {journal} {\bibinfo
  {journal} {Phys. Rev. B}\ }\textbf {\bibinfo {volume} {96}},\ \bibinfo
  {pages} {195126} (\bibinfo {year} {2017})}\BibitemShut {NoStop}%
\bibitem [{\citenamefont {Fabrizio}(2017)}]{PhysRevB.95.075156}%
  \BibitemOpen
  \bibfield  {author} {\bibinfo {author} {\bibfnamefont {M.}~\bibnamefont
  {Fabrizio}},\ }\href {\doibase 10.1103/PhysRevB.95.075156} {\bibfield
  {journal} {\bibinfo  {journal} {Phys. Rev. B}\ }\textbf {\bibinfo {volume}
  {95}},\ \bibinfo {pages} {075156} (\bibinfo {year} {2017})}\BibitemShut
  {NoStop}%
\bibitem [{\citenamefont {Wysoki\'{n}ski}\ and\ \citenamefont
  {Fabrizio}(2017)}]{PhysRevB.95.161106}%
  \BibitemOpen
  \bibfield  {author} {\bibinfo {author} {\bibfnamefont {M.~M.}\ \bibnamefont
  {Wysoki\'{n}ski}}\ and\ \bibinfo {author} {\bibfnamefont {M.}~\bibnamefont
  {Fabrizio}},\ }\href {\doibase 10.1103/PhysRevB.95.161106} {\bibfield
  {journal} {\bibinfo  {journal} {Phys. Rev. B}\ }\textbf {\bibinfo {volume}
  {95}},\ \bibinfo {pages} {161106} (\bibinfo {year} {2017})}\BibitemShut
  {NoStop}%
\bibitem [{\citenamefont {Boys}\ and\ \citenamefont
  {Handy}(1969{\natexlab{a}})}]{transcorrelated-method-1}%
  \BibitemOpen
  \bibfield  {author} {\bibinfo {author} {\bibfnamefont {S.}~\bibnamefont
  {Boys}}\ and\ \bibinfo {author} {\bibfnamefont {N.}~\bibnamefont {Handy}},\
  }\href {\doibase 10.1098/rspa.1969.0038} {\bibfield  {journal} {\bibinfo
  {journal} {Proceedings of the Royal Society of London A: Mathematical,
  Physical and Engineering Sciences}\ }\textbf {\bibinfo {volume} {309}},\
  \bibinfo {pages} {209} (\bibinfo {year} {1969}{\natexlab{a}})}\BibitemShut
  {NoStop}%
\bibitem [{\citenamefont {Boys}\ and\ \citenamefont
  {Handy}(1969{\natexlab{b}})}]{transcorrelated-method-3}%
  \BibitemOpen
  \bibfield  {author} {\bibinfo {author} {\bibfnamefont {S.}~\bibnamefont
  {Boys}}\ and\ \bibinfo {author} {\bibfnamefont {N.}~\bibnamefont {Handy}},\
  }\href {\doibase 10.1098/rspa.1969.0062} {\bibfield  {journal} {\bibinfo
  {journal} {Proceedings of the Royal Society of London A: Mathematical,
  Physical and Engineering Sciences}\ }\textbf {\bibinfo {volume} {310}},\
  \bibinfo {pages} {63} (\bibinfo {year} {1969}{\natexlab{b}})}\BibitemShut
  {NoStop}%
\bibitem [{\citenamefont {Hirschfelder}(1963)}]{transcorrelated-method-0}%
  \BibitemOpen
  \bibfield  {author} {\bibinfo {author} {\bibfnamefont {J.~O.}\ \bibnamefont
  {Hirschfelder}},\ }\href {\doibase 10.1063/1.1734157} {\bibfield  {journal}
  {\bibinfo  {journal} {The Journal of Chemical Physics}\ }\textbf {\bibinfo
  {volume} {39}},\ \bibinfo {pages} {3145} (\bibinfo {year}
  {1963})}\BibitemShut {NoStop}%
\bibitem [{\citenamefont {Tsuneyuki}(2008)}]{tsuneyuki}%
  \BibitemOpen
  \bibfield  {author} {\bibinfo {author} {\bibfnamefont {S.}~\bibnamefont
  {Tsuneyuki}},\ }\href {\doibase 10.1143/PTPS.176.134} {\bibfield  {journal}
  {\bibinfo  {journal} {Progress of Theoretical Physics Supplement}\ }\textbf
  {\bibinfo {volume} {176}},\ \bibinfo {pages} {134} (\bibinfo {year}
  {2008})}\BibitemShut {NoStop}%
\bibitem [{\citenamefont {Wahlen-Strothman}\ \emph {et~al.}(2015)\citenamefont
  {Wahlen-Strothman}, \citenamefont {Jim\'enez-Hoyos}, \citenamefont
  {Henderson},\ and\ \citenamefont {Scuseria}}]{scuseria-1}%
  \BibitemOpen
  \bibfield  {author} {\bibinfo {author} {\bibfnamefont {J.~M.}\ \bibnamefont
  {Wahlen-Strothman}}, \bibinfo {author} {\bibfnamefont {C.~A.}\ \bibnamefont
  {Jim\'enez-Hoyos}}, \bibinfo {author} {\bibfnamefont {T.~M.}\ \bibnamefont
  {Henderson}}, \ and\ \bibinfo {author} {\bibfnamefont {G.~E.}\ \bibnamefont
  {Scuseria}},\ }\href {\doibase 10.1103/PhysRevB.91.041114} {\bibfield
  {journal} {\bibinfo  {journal} {Phys. Rev. B}\ }\textbf {\bibinfo {volume}
  {91}},\ \bibinfo {pages} {041114} (\bibinfo {year} {2015})}\BibitemShut
  {NoStop}%
\bibitem [{\citenamefont {Neuscamman}\ \emph {et~al.}(2011)\citenamefont
  {Neuscamman}, \citenamefont {Changlani}, \citenamefont {Kinder},\ and\
  \citenamefont {Chan}}]{non-stochastic}%
  \BibitemOpen
  \bibfield  {author} {\bibinfo {author} {\bibfnamefont {E.}~\bibnamefont
  {Neuscamman}}, \bibinfo {author} {\bibfnamefont {H.}~\bibnamefont
  {Changlani}}, \bibinfo {author} {\bibfnamefont {J.}~\bibnamefont {Kinder}}, \
  and\ \bibinfo {author} {\bibfnamefont {G.~K.-L.}\ \bibnamefont {Chan}},\
  }\href {\doibase 10.1103/PhysRevB.84.205132} {\bibfield  {journal} {\bibinfo
  {journal} {Phys. Rev. B}\ }\textbf {\bibinfo {volume} {84}},\ \bibinfo
  {pages} {205132} (\bibinfo {year} {2011})}\BibitemShut {NoStop}%
\bibitem [{\citenamefont {Cleland}\ \emph {et~al.}(2010)\citenamefont
  {Cleland}, \citenamefont {Booth},\ and\ \citenamefont
  {Alavi}}]{initiator-fciqmc}%
  \BibitemOpen
  \bibfield  {author} {\bibinfo {author} {\bibfnamefont {D.}~\bibnamefont
  {Cleland}}, \bibinfo {author} {\bibfnamefont {G.~H.}\ \bibnamefont {Booth}},
  \ and\ \bibinfo {author} {\bibfnamefont {A.}~\bibnamefont {Alavi}},\ }\href
  {\doibase 10.1063/1.3302277} {\bibfield  {journal} {\bibinfo  {journal} {The
  Journal of Chemical Physics}\ }\textbf {\bibinfo {volume} {132}},\ \bibinfo
  {pages} {041103} (\bibinfo {year} {2010})}\BibitemShut {NoStop}%
\bibitem [{\citenamefont {B\"unemann}\ \emph {et~al.}(1998)\citenamefont
  {B\"unemann}, \citenamefont {Weber},\ and\ \citenamefont
  {Gebhard}}]{gutzwiller-correlator}%
  \BibitemOpen
  \bibfield  {author} {\bibinfo {author} {\bibfnamefont {J.}~\bibnamefont
  {B\"unemann}}, \bibinfo {author} {\bibfnamefont {W.}~\bibnamefont {Weber}}, \
  and\ \bibinfo {author} {\bibfnamefont {F.}~\bibnamefont {Gebhard}},\ }\href
  {\doibase 10.1103/PhysRevB.57.6896} {\bibfield  {journal} {\bibinfo
  {journal} {Phys. Rev. B}\ }\textbf {\bibinfo {volume} {57}},\ \bibinfo
  {pages} {6896} (\bibinfo {year} {1998})}\BibitemShut {NoStop}%
\bibitem [{\citenamefont {Gebhard}\ and\ \citenamefont
  {Vollhardt}(1988)}]{correl-functions}%
  \BibitemOpen
  \bibfield  {author} {\bibinfo {author} {\bibfnamefont {F.}~\bibnamefont
  {Gebhard}}\ and\ \bibinfo {author} {\bibfnamefont {D.}~\bibnamefont
  {Vollhardt}},\ }\href {\doibase 10.1103/PhysRevB.38.6911} {\bibfield
  {journal} {\bibinfo  {journal} {Phys. Rev. B}\ }\textbf {\bibinfo {volume}
  {38}},\ \bibinfo {pages} {6911} (\bibinfo {year} {1988})}\BibitemShut
  {NoStop}%
\bibitem [{\citenamefont {Boys}\ and\ \citenamefont
  {Handy}(1969{\natexlab{c}})}]{transcorrelated-method-2}%
  \BibitemOpen
  \bibfield  {author} {\bibinfo {author} {\bibfnamefont {S.}~\bibnamefont
  {Boys}}\ and\ \bibinfo {author} {\bibfnamefont {N.}~\bibnamefont {Handy}},\
  }\href {\doibase 10.1098/rspa.1969.0061} {\bibfield  {journal} {\bibinfo
  {journal} {Proceedings of the Royal Society of London A: Mathematical,
  Physical and Engineering Sciences}\ }\textbf {\bibinfo {volume} {310}},\
  \bibinfo {pages} {43} (\bibinfo {year} {1969}{\natexlab{c}})}\BibitemShut
  {NoStop}%
\bibitem [{\citenamefont {Yanai}\ and\ \citenamefont
  {Shiozaki}(2012)}]{canonical-transcorr}%
  \BibitemOpen
  \bibfield  {author} {\bibinfo {author} {\bibfnamefont {T.}~\bibnamefont
  {Yanai}}\ and\ \bibinfo {author} {\bibfnamefont {T.}~\bibnamefont
  {Shiozaki}},\ }\href {\doibase 10.1063/1.3688225} {\bibfield  {journal}
  {\bibinfo  {journal} {The Journal of Chemical Physics}\ }\textbf {\bibinfo
  {volume} {136}},\ \bibinfo {pages} {084107} (\bibinfo {year}
  {2012})}\BibitemShut {NoStop}%
\bibitem [{\citenamefont {Sharma}\ \emph {et~al.}(2014)\citenamefont {Sharma},
  \citenamefont {Yanai}, \citenamefont {Booth}, \citenamefont {Umrigar},\ and\
  \citenamefont {Chan}}]{fciqmc-transcorr}%
  \BibitemOpen
  \bibfield  {author} {\bibinfo {author} {\bibfnamefont {S.}~\bibnamefont
  {Sharma}}, \bibinfo {author} {\bibfnamefont {T.}~\bibnamefont {Yanai}},
  \bibinfo {author} {\bibfnamefont {G.~H.}\ \bibnamefont {Booth}}, \bibinfo
  {author} {\bibfnamefont {C.~J.}\ \bibnamefont {Umrigar}}, \ and\ \bibinfo
  {author} {\bibfnamefont {G.~K.-L.}\ \bibnamefont {Chan}},\ }\href {\doibase
  10.1063/1.4867383} {\bibfield  {journal} {\bibinfo  {journal} {The Journal of
  Chemical Physics}\ }\textbf {\bibinfo {volume} {140}},\ \bibinfo {pages}
  {104112} (\bibinfo {year} {2014})}\BibitemShut {NoStop}%
\bibitem [{\citenamefont {Luo}(2012)}]{luo-ueg}%
  \BibitemOpen
  \bibfield  {author} {\bibinfo {author} {\bibfnamefont {H.}~\bibnamefont
  {Luo}},\ }\href {\doibase 10.1063/1.4727852} {\bibfield  {journal} {\bibinfo
  {journal} {The Journal of Chemical Physics}\ }\textbf {\bibinfo {volume}
  {136}},\ \bibinfo {pages} {224111} (\bibinfo {year} {2012})}\BibitemShut
  {NoStop}%
\bibitem [{\citenamefont {Luo}\ and\ \citenamefont {Alavi}(2018)}]{ali-ueg}%
  \BibitemOpen
  \bibfield  {author} {\bibinfo {author} {\bibfnamefont {H.}~\bibnamefont
  {Luo}}\ and\ \bibinfo {author} {\bibfnamefont {A.}~\bibnamefont {Alavi}},\
  }\href {\doibase 10.1021/acs.jctc.7b01257} {\bibfield  {journal} {\bibinfo
  {journal} {Journal of Chemical Theory and Computation}\ }\textbf {\bibinfo
  {volume} {14}},\ \bibinfo {pages} {1403} (\bibinfo {year} {2018})},\ \bibinfo
  {note} {pMID: 29431996}\BibitemShut {NoStop}%
\bibitem [{\citenamefont {Ten-no}(2000{\natexlab{a}})}]{tenno1}%
  \BibitemOpen
  \bibfield  {author} {\bibinfo {author} {\bibfnamefont {S.}~\bibnamefont
  {Ten-no}},\ }\href {\doibase https://doi.org/10.1016/S0009-2614(00)01066-6}
  {\bibfield  {journal} {\bibinfo  {journal} {Chemical Physics Letters}\
  }\textbf {\bibinfo {volume} {330}},\ \bibinfo {pages} {169 } (\bibinfo {year}
  {2000}{\natexlab{a}})}\BibitemShut {NoStop}%
\bibitem [{\citenamefont {Ten-no}(2000{\natexlab{b}})}]{tenno2}%
  \BibitemOpen
  \bibfield  {author} {\bibinfo {author} {\bibfnamefont {S.}~\bibnamefont
  {Ten-no}},\ }\href {\doibase https://doi.org/10.1016/S0009-2614(00)01067-8}
  {\bibfield  {journal} {\bibinfo  {journal} {Chemical Physics Letters}\
  }\textbf {\bibinfo {volume} {330}},\ \bibinfo {pages} {175 } (\bibinfo {year}
  {2000}{\natexlab{b}})}\BibitemShut {NoStop}%
\bibitem [{\citenamefont {Wahlen-Strothman}\ and\ \citenamefont
  {Scuseria}(2016)}]{scuseria-2}%
  \BibitemOpen
  \bibfield  {author} {\bibinfo {author} {\bibfnamefont {J.~M.}\ \bibnamefont
  {Wahlen-Strothman}}\ and\ \bibinfo {author} {\bibfnamefont {G.~E.}\
  \bibnamefont {Scuseria}},\ }\href
  {http://stacks.iop.org/0953-8984/28/i=48/a=485502} {\bibfield  {journal}
  {\bibinfo  {journal} {Journal of Physics: Condensed Matter}\ }\textbf
  {\bibinfo {volume} {28}},\ \bibinfo {pages} {485502} (\bibinfo {year}
  {2016})}\BibitemShut {NoStop}%
\bibitem [{\citenamefont {Booth}\ \emph {et~al.}(2009)\citenamefont {Booth},
  \citenamefont {Thom},\ and\ \citenamefont {Alavi}}]{original-fciqmc}%
  \BibitemOpen
  \bibfield  {author} {\bibinfo {author} {\bibfnamefont {G.~H.}\ \bibnamefont
  {Booth}}, \bibinfo {author} {\bibfnamefont {A.~J.~W.}\ \bibnamefont {Thom}},
  \ and\ \bibinfo {author} {\bibfnamefont {A.}~\bibnamefont {Alavi}},\ }\href
  {\doibase 10.1063/1.3193710} {\bibfield  {journal} {\bibinfo  {journal} {The
  Journal of Chemical Physics}\ }\textbf {\bibinfo {volume} {131}},\ \bibinfo
  {pages} {054106} (\bibinfo {year} {2009})}\BibitemShut {NoStop}%
\bibitem [{\citenamefont {Ohtsuka}\ and\ \citenamefont
  {Ten-no}(2012)}]{tenno-ohtsuka}%
  \BibitemOpen
  \bibfield  {author} {\bibinfo {author} {\bibfnamefont {Y.}~\bibnamefont
  {Ohtsuka}}\ and\ \bibinfo {author} {\bibfnamefont {S.}~\bibnamefont
  {Ten-no}},\ }\href {\doibase 10.1063/1.4730647} {\bibfield  {journal}
  {\bibinfo  {journal} {AIP Conference Proceedings}\ }\textbf {\bibinfo
  {volume} {1456}},\ \bibinfo {pages} {97} (\bibinfo {year}
  {2012})}\BibitemShut {NoStop}%
\bibitem [{\citenamefont {Golub}\ and\ \citenamefont {Loan}(1989)}]{Golub}%
  \BibitemOpen
  \bibfield  {author} {\bibinfo {author} {\bibfnamefont {G.}~\bibnamefont
  {Golub}}\ and\ \bibinfo {author} {\bibfnamefont {C.~V.}\ \bibnamefont
  {Loan}},\ }\href@noop {} {\emph {\bibinfo {title} {Matrix Computations}}},\
  \bibinfo {edition} {2nd}\ ed.\ (\bibinfo  {publisher} {John Hopkins Univ.
  Press},\ \bibinfo {address} {Baltimore, MD},\ \bibinfo {year}
  {1989})\BibitemShut {NoStop}%
\bibitem [{\citenamefont {Yanai}\ \emph {et~al.}(2010)\citenamefont {Yanai},
  \citenamefont {Kurashige}, \citenamefont {Neuscamman},\ and\ \citenamefont
  {Chan}}]{yanai-chan}%
  \BibitemOpen
  \bibfield  {author} {\bibinfo {author} {\bibfnamefont {T.}~\bibnamefont
  {Yanai}}, \bibinfo {author} {\bibfnamefont {Y.}~\bibnamefont {Kurashige}},
  \bibinfo {author} {\bibfnamefont {E.}~\bibnamefont {Neuscamman}}, \ and\
  \bibinfo {author} {\bibfnamefont {G.~K.-L.}\ \bibnamefont {Chan}},\ }\href
  {\doibase 10.1063/1.3275806} {\bibfield  {journal} {\bibinfo  {journal} {The
  Journal of Chemical Physics}\ }\textbf {\bibinfo {volume} {132}},\ \bibinfo
  {pages} {024105} (\bibinfo {year} {2010})}\BibitemShut {NoStop}%
\bibitem [{\citenamefont {Yanai}\ and\ \citenamefont
  {Chan}(2007)}]{yanai-chan-2}%
  \BibitemOpen
  \bibfield  {author} {\bibinfo {author} {\bibfnamefont {T.}~\bibnamefont
  {Yanai}}\ and\ \bibinfo {author} {\bibfnamefont {G.~K.-L.}\ \bibnamefont
  {Chan}},\ }\href {\doibase 10.1063/1.2761870} {\bibfield  {journal} {\bibinfo
   {journal} {The Journal of Chemical Physics}\ }\textbf {\bibinfo {volume}
  {127}},\ \bibinfo {pages} {104107} (\bibinfo {year} {2007})}\BibitemShut
  {NoStop}%
\bibitem [{\citenamefont {Fazekas}(1988)}]{fazekas-1}%
  \BibitemOpen
  \bibfield  {author} {\bibinfo {author} {\bibfnamefont {P.}~\bibnamefont
  {Fazekas}},\ }\href {\doibase https://doi.org/10.1016/0921-4534(88)90282-1}
  {\bibfield  {journal} {\bibinfo  {journal} {Physica C: Superconductivity}\
  }\textbf {\bibinfo {volume} {153-155}},\ \bibinfo {pages} {1283 } (\bibinfo
  {year} {1988})},\ \bibinfo {note} {proceedings of the International
  Conference on High Temperature Superconductors and Materials and Mechanisms
  of Superconductivity Part II}\BibitemShut {NoStop}%
\bibitem [{\citenamefont {Dobrautz}\ \emph {et~al.}()\citenamefont {Dobrautz},
  \citenamefont {Luo}, \citenamefont {Rios},\ and\ \citenamefont
  {Alavi}}]{to-be-published-1}%
  \BibitemOpen
  \bibfield  {author} {\bibinfo {author} {\bibfnamefont {W.}~\bibnamefont
  {Dobrautz}}, \bibinfo {author} {\bibfnamefont {H.}~\bibnamefont {Luo}},
  \bibinfo {author} {\bibfnamefont {P.}~\bibnamefont {Rios}}, \ and\ \bibinfo
  {author} {\bibfnamefont {A.}~\bibnamefont {Alavi}},\ }\href@noop {} {\enquote
  {\bibinfo {title} {Combining vmc and similarity tranformed fciqmc for the
  real-space hubbard model},}\ }\bibinfo {note} {To be published}\BibitemShut
  {NoStop}%
\bibitem [{\citenamefont {Hirsch}(1985)}]{afqmc-sign}%
  \BibitemOpen
  \bibfield  {author} {\bibinfo {author} {\bibfnamefont {J.~E.}\ \bibnamefont
  {Hirsch}},\ }\href {\doibase 10.1103/PhysRevB.31.4403} {\bibfield  {journal}
  {\bibinfo  {journal} {Phys. Rev. B}\ }\textbf {\bibinfo {volume} {31}},\
  \bibinfo {pages} {4403} (\bibinfo {year} {1985})}\BibitemShut {NoStop}%
\bibitem [{\citenamefont {Gros}\ \emph
  {et~al.}(1987{\natexlab{b}})\citenamefont {Gros}, \citenamefont {Joynt},\
  and\ \citenamefont {Rice}}]{rice-2}%
  \BibitemOpen
  \bibfield  {author} {\bibinfo {author} {\bibfnamefont {C.}~\bibnamefont
  {Gros}}, \bibinfo {author} {\bibfnamefont {R.}~\bibnamefont {Joynt}}, \ and\
  \bibinfo {author} {\bibfnamefont {T.~M.}\ \bibnamefont {Rice}},\ }\href
  {\doibase 10.1007/BF01471072} {\bibfield  {journal} {\bibinfo  {journal}
  {Zeitschrift f{\"u}r Physik B Condensed Matter}\ }\textbf {\bibinfo {volume}
  {68}},\ \bibinfo {pages} {425} (\bibinfo {year}
  {1987}{\natexlab{b}})}\BibitemShut {NoStop}%
\bibitem [{\citenamefont {Shi}\ and\ \citenamefont {Zhang}(2013)}]{6x6_nel24}%
  \BibitemOpen
  \bibfield  {author} {\bibinfo {author} {\bibfnamefont {H.}~\bibnamefont
  {Shi}}\ and\ \bibinfo {author} {\bibfnamefont {S.}~\bibnamefont {Zhang}},\
  }\href {\doibase 10.1103/PhysRevB.88.125132} {\bibfield  {journal} {\bibinfo
  {journal} {Phys. Rev. B}\ }\textbf {\bibinfo {volume} {88}},\ \bibinfo
  {pages} {125132} (\bibinfo {year} {2013})}\BibitemShut {NoStop}%
\bibitem [{\citenamefont {Rios}(2018)}]{pablo-priv-comm}%
  \BibitemOpen
  \bibfield  {author} {\bibinfo {author} {\bibfnamefont {P.~L.}\ \bibnamefont
  {Rios}},\ }\href@noop {} {}\bibinfo {howpublished} {private communication}
  (\bibinfo {year} {2018})\BibitemShut {NoStop}%
\bibitem [{\citenamefont {Trivedi}\ and\ \citenamefont
  {Ceperley}(1990)}]{spectral-width}%
  \BibitemOpen
  \bibfield  {author} {\bibinfo {author} {\bibfnamefont {N.}~\bibnamefont
  {Trivedi}}\ and\ \bibinfo {author} {\bibfnamefont {D.~M.}\ \bibnamefont
  {Ceperley}},\ }\href {\doibase 10.1103/PhysRevB.41.4552} {\bibfield
  {journal} {\bibinfo  {journal} {Phys. Rev. B}\ }\textbf {\bibinfo {volume}
  {41}},\ \bibinfo {pages} {4552} (\bibinfo {year} {1990})}\BibitemShut
  {NoStop}%
\bibitem [{\citenamefont {Booth}\ and\ \citenamefont {Alavi~et.
  al.}(2013)}]{neci}%
  \BibitemOpen
  \bibfield  {author} {\bibinfo {author} {\bibfnamefont {G.~H.}\ \bibnamefont
  {Booth}}\ and\ \bibinfo {author} {\bibfnamefont {A.}~\bibnamefont {Alavi~et.
  al.}},\ }\href@noop {} {\enquote {\bibinfo {title} {Standalone neci codebase
  designed for fciqmc and other stochastic quantum chemistry methods.}}\
  }\bibinfo {howpublished} {\url{https://github.com/ghb24/NECI_STABLE}}
  (\bibinfo {year} {2013})\BibitemShut {NoStop}%
\bibitem [{\citenamefont {Gunnarsson}(2018)}]{olle-priv-comm}%
  \BibitemOpen
  \bibfield  {author} {\bibinfo {author} {\bibfnamefont {O.}~\bibnamefont
  {Gunnarsson}},\ }\href@noop {} {}\bibinfo {howpublished} {private
  communication} (\bibinfo {year} {2018})\BibitemShut {NoStop}%
\bibitem [{\citenamefont {Sherrill}\ and\ \citenamefont
  {Schaefer}(1999)}]{truncated-ci}%
  \BibitemOpen
  \bibfield  {author} {\bibinfo {author} {\bibfnamefont {C.~D.}\ \bibnamefont
  {Sherrill}}\ and\ \bibinfo {author} {\bibfnamefont {H.~F.}\ \bibnamefont
  {Schaefer}}\ }(\bibinfo  {publisher} {Academic Press},\ \bibinfo {year}
  {1999})\ pp.\ \bibinfo {pages} {143 -- 269}\BibitemShut {NoStop}%
\bibitem [{\citenamefont {Blunt}\ \emph {et~al.}(2015)\citenamefont {Blunt},
  \citenamefont {Smart}, \citenamefont {Booth},\ and\ \citenamefont
  {Alavi}}]{excited-fciqmc}%
  \BibitemOpen
  \bibfield  {author} {\bibinfo {author} {\bibfnamefont {N.~S.}\ \bibnamefont
  {Blunt}}, \bibinfo {author} {\bibfnamefont {S.~D.}\ \bibnamefont {Smart}},
  \bibinfo {author} {\bibfnamefont {G.~H.}\ \bibnamefont {Booth}}, \ and\
  \bibinfo {author} {\bibfnamefont {A.}~\bibnamefont {Alavi}},\ }\href
  {\doibase 10.1063/1.4932595} {\bibfield  {journal} {\bibinfo  {journal} {The
  Journal of Chemical Physics}\ }\textbf {\bibinfo {volume} {143}},\ \bibinfo
  {pages} {134117} (\bibinfo {year} {2015})}\BibitemShut {NoStop}%
\bibitem [{\citenamefont {White}\ \emph {et~al.}(1989)\citenamefont {White},
  \citenamefont {Scalapino}, \citenamefont {Sugar}, \citenamefont {Loh},
  \citenamefont {Gubernatis},\ and\ \citenamefont {Scalettar}}]{afqmc-white}%
  \BibitemOpen
  \bibfield  {author} {\bibinfo {author} {\bibfnamefont {S.~R.}\ \bibnamefont
  {White}}, \bibinfo {author} {\bibfnamefont {D.~J.}\ \bibnamefont
  {Scalapino}}, \bibinfo {author} {\bibfnamefont {R.~L.}\ \bibnamefont
  {Sugar}}, \bibinfo {author} {\bibfnamefont {E.~Y.}\ \bibnamefont {Loh}},
  \bibinfo {author} {\bibfnamefont {J.~E.}\ \bibnamefont {Gubernatis}}, \ and\
  \bibinfo {author} {\bibfnamefont {R.~T.}\ \bibnamefont {Scalettar}},\ }\href
  {\doibase 10.1103/PhysRevB.40.506} {\bibfield  {journal} {\bibinfo  {journal}
  {Phys. Rev. B}\ }\textbf {\bibinfo {volume} {40}},\ \bibinfo {pages} {506}
  (\bibinfo {year} {1989})}\BibitemShut {NoStop}%
\bibitem [{\citenamefont {Sorella}\ \emph {et~al.}(1989)\citenamefont
  {Sorella}, \citenamefont {Baroni}, \citenamefont {Car},\ and\ \citenamefont
  {Parrinello}}]{afqmc-sorella}%
  \BibitemOpen
  \bibfield  {author} {\bibinfo {author} {\bibfnamefont {S.}~\bibnamefont
  {Sorella}}, \bibinfo {author} {\bibfnamefont {S.}~\bibnamefont {Baroni}},
  \bibinfo {author} {\bibfnamefont {R.}~\bibnamefont {Car}}, \ and\ \bibinfo
  {author} {\bibfnamefont {M.}~\bibnamefont {Parrinello}},\ }\href
  {http://stacks.iop.org/0295-5075/8/i=7/a=014} {\bibfield  {journal} {\bibinfo
   {journal} {EPL (Europhysics Letters)}\ }\textbf {\bibinfo {volume} {8}},\
  \bibinfo {pages} {663} (\bibinfo {year} {1989})}\BibitemShut {NoStop}%
\bibitem [{\citenamefont {Zhang}\ \emph
  {et~al.}(1997{\natexlab{b}})\citenamefont {Zhang}, \citenamefont {Carlson},\
  and\ \citenamefont {Gubernatis}}]{cp-afqmc-1}%
  \BibitemOpen
  \bibfield  {author} {\bibinfo {author} {\bibfnamefont {S.}~\bibnamefont
  {Zhang}}, \bibinfo {author} {\bibfnamefont {J.}~\bibnamefont {Carlson}}, \
  and\ \bibinfo {author} {\bibfnamefont {J.~E.}\ \bibnamefont {Gubernatis}},\
  }\href {\doibase 10.1103/PhysRevLett.78.4486} {\bibfield  {journal} {\bibinfo
   {journal} {Phys. Rev. Lett.}\ }\textbf {\bibinfo {volume} {78}},\ \bibinfo
  {pages} {4486} (\bibinfo {year} {1997}{\natexlab{b}})}\BibitemShut {NoStop}%
\bibitem [{\citenamefont {Carlson}\ \emph {et~al.}(1999)\citenamefont
  {Carlson}, \citenamefont {Gubernatis}, \citenamefont {Ortiz},\ and\
  \citenamefont {Zhang}}]{cp-afqmc-2}%
  \BibitemOpen
  \bibfield  {author} {\bibinfo {author} {\bibfnamefont {J.}~\bibnamefont
  {Carlson}}, \bibinfo {author} {\bibfnamefont {J.~E.}\ \bibnamefont
  {Gubernatis}}, \bibinfo {author} {\bibfnamefont {G.}~\bibnamefont {Ortiz}}, \
  and\ \bibinfo {author} {\bibfnamefont {S.}~\bibnamefont {Zhang}},\ }\href
  {\doibase 10.1103/PhysRevB.59.12788} {\bibfield  {journal} {\bibinfo
  {journal} {Phys. Rev. B}\ }\textbf {\bibinfo {volume} {59}},\ \bibinfo
  {pages} {12788} (\bibinfo {year} {1999})}\BibitemShut {NoStop}%
\bibitem [{\citenamefont {Sorella}(2011)}]{sorella-lafqmc}%
  \BibitemOpen
  \bibfield  {author} {\bibinfo {author} {\bibfnamefont {S.}~\bibnamefont
  {Sorella}},\ }\href {\doibase 10.1103/PhysRevB.84.241110} {\bibfield
  {journal} {\bibinfo  {journal} {Phys. Rev. B}\ }\textbf {\bibinfo {volume}
  {84}},\ \bibinfo {pages} {241110} (\bibinfo {year} {2011})}\BibitemShut
  {NoStop}%
\bibitem [{\citenamefont {Sorella}(2016)}]{sorella-priv-comm}%
  \BibitemOpen
  \bibfield  {author} {\bibinfo {author} {\bibfnamefont {S.}~\bibnamefont
  {Sorella}},\ }\href@noop {} {}\bibinfo {howpublished} {private communication}
  (\bibinfo {year} {2016})\BibitemShut {NoStop}%
\bibitem [{\citenamefont {Carleo}\ \emph {et~al.}(2010)\citenamefont {Carleo},
  \citenamefont {Becca}, \citenamefont {Moroni},\ and\ \citenamefont
  {Baroni}}]{rep-qmc}%
  \BibitemOpen
  \bibfield  {author} {\bibinfo {author} {\bibfnamefont {G.}~\bibnamefont
  {Carleo}}, \bibinfo {author} {\bibfnamefont {F.}~\bibnamefont {Becca}},
  \bibinfo {author} {\bibfnamefont {S.}~\bibnamefont {Moroni}}, \ and\ \bibinfo
  {author} {\bibfnamefont {S.}~\bibnamefont {Baroni}},\ }\href {\doibase
  10.1103/PhysRevE.82.046710} {\bibfield  {journal} {\bibinfo  {journal} {Phys.
  Rev. E}\ }\textbf {\bibinfo {volume} {82}},\ \bibinfo {pages} {046710}
  (\bibinfo {year} {2010})}\BibitemShut {NoStop}%
\bibitem [{\citenamefont {Overy}\ \emph {et~al.}(2014)\citenamefont {Overy},
  \citenamefont {Booth}, \citenamefont {Blunt}, \citenamefont {Shepherd},
  \citenamefont {Cleland},\ and\ \citenamefont {Alavi}}]{fciqmc-rdm}%
  \BibitemOpen
  \bibfield  {author} {\bibinfo {author} {\bibfnamefont {C.}~\bibnamefont
  {Overy}}, \bibinfo {author} {\bibfnamefont {G.~H.}\ \bibnamefont {Booth}},
  \bibinfo {author} {\bibfnamefont {N.~S.}\ \bibnamefont {Blunt}}, \bibinfo
  {author} {\bibfnamefont {J.~J.}\ \bibnamefont {Shepherd}}, \bibinfo {author}
  {\bibfnamefont {D.}~\bibnamefont {Cleland}}, \ and\ \bibinfo {author}
  {\bibfnamefont {A.}~\bibnamefont {Alavi}},\ }\href {\doibase
  10.1063/1.4904313} {\bibfield  {journal} {\bibinfo  {journal} {The Journal of
  Chemical Physics}\ }\textbf {\bibinfo {volume} {141}},\ \bibinfo {pages}
  {244117} (\bibinfo {year} {2014})}\BibitemShut {NoStop}%
\bibitem [{\citenamefont {Lin}\ \emph {et~al.}(2001)\citenamefont {Lin},
  \citenamefont {Zong},\ and\ \citenamefont {Ceperley}}]{twist-average-1}%
  \BibitemOpen
  \bibfield  {author} {\bibinfo {author} {\bibfnamefont {C.}~\bibnamefont
  {Lin}}, \bibinfo {author} {\bibfnamefont {F.~H.}\ \bibnamefont {Zong}}, \
  and\ \bibinfo {author} {\bibfnamefont {D.~M.}\ \bibnamefont {Ceperley}},\
  }\href {\doibase 10.1103/PhysRevE.64.016702} {\bibfield  {journal} {\bibinfo
  {journal} {Phys. Rev. E}\ }\textbf {\bibinfo {volume} {64}},\ \bibinfo
  {pages} {016702} (\bibinfo {year} {2001})}\BibitemShut {NoStop}%
\bibitem [{\citenamefont {Poilblanc}(1991)}]{twist-average-2}%
  \BibitemOpen
  \bibfield  {author} {\bibinfo {author} {\bibfnamefont {D.}~\bibnamefont
  {Poilblanc}},\ }\href {\doibase 10.1103/PhysRevB.44.9562} {\bibfield
  {journal} {\bibinfo  {journal} {Phys. Rev. B}\ }\textbf {\bibinfo {volume}
  {44}},\ \bibinfo {pages} {9562} (\bibinfo {year} {1991})}\BibitemShut
  {NoStop}%
\bibitem [{\citenamefont {Sorella}(2015)}]{twist-average-3}%
  \BibitemOpen
  \bibfield  {author} {\bibinfo {author} {\bibfnamefont {S.}~\bibnamefont
  {Sorella}},\ }\href {\doibase 10.1103/PhysRevB.91.241116} {\bibfield
  {journal} {\bibinfo  {journal} {Phys. Rev. B}\ }\textbf {\bibinfo {volume}
  {91}},\ \bibinfo {pages} {241116} (\bibinfo {year} {2015})}\BibitemShut
  {NoStop}%
\bibitem [{\citenamefont {Taylor}(1994)}]{size-consistency}%
  \BibitemOpen
  \bibfield  {author} {\bibinfo {author} {\bibfnamefont {P.}~\bibnamefont
  {Taylor}},\ }\enquote {\bibinfo {title} {Lecture notes in quantum chemistry
  ii. lecture notes in chemistry},}\ \ (\bibinfo  {publisher} {Springer},\
  \bibinfo {address} {Berlin, Heidelberg},\ \bibinfo {year} {1994})\ Chap.\
  \bibinfo {chapter} {Coupled-cluster Methods in Quantum Chemistry}\BibitemShut
  {NoStop}%
\bibitem [{\citenamefont {Fulde}(2012)}]{fulde-book}%
  \BibitemOpen
  \bibfield  {author} {\bibinfo {author} {\bibfnamefont {P.}~\bibnamefont
  {Fulde}},\ }\href
  {https://www.amazon.com/Electron-Correlations-Molecules-Springer-Solid-State-ebook/dp/B000SBCA36?SubscriptionId=AKIAIOBINVZYXZQZ2U3A&tag=chimbori05-20&linkCode=xm2&camp=2025&creative=165953&creativeASIN=B000SBCA36}
  {\emph {\bibinfo {title} {Electron Correlations in Molecules and Solids
  (Springer Series in Solid-State Sciences)}}}\ (\bibinfo  {publisher}
  {Springer},\ \bibinfo {year} {2012})\BibitemShut {NoStop}%
\bibitem [{\citenamefont {Anderson}\ \emph {et~al.}(1999)\citenamefont
  {Anderson}, \citenamefont {Bai}, \citenamefont {Bischof}, \citenamefont
  {Blackford}, \citenamefont {Demmel}, \citenamefont {Dongarra}, \citenamefont
  {Du~Croz}, \citenamefont {Greenbaum}, \citenamefont {Hammarling},
  \citenamefont {McKenney},\ and\ \citenamefont {Sorensen}}]{lapack}%
  \BibitemOpen
  \bibfield  {author} {\bibinfo {author} {\bibfnamefont {E.}~\bibnamefont
  {Anderson}}, \bibinfo {author} {\bibfnamefont {Z.}~\bibnamefont {Bai}},
  \bibinfo {author} {\bibfnamefont {C.}~\bibnamefont {Bischof}}, \bibinfo
  {author} {\bibfnamefont {S.}~\bibnamefont {Blackford}}, \bibinfo {author}
  {\bibfnamefont {J.}~\bibnamefont {Demmel}}, \bibinfo {author} {\bibfnamefont
  {J.}~\bibnamefont {Dongarra}}, \bibinfo {author} {\bibfnamefont
  {J.}~\bibnamefont {Du~Croz}}, \bibinfo {author} {\bibfnamefont
  {A.}~\bibnamefont {Greenbaum}}, \bibinfo {author} {\bibfnamefont
  {S.}~\bibnamefont {Hammarling}}, \bibinfo {author} {\bibfnamefont
  {A.}~\bibnamefont {McKenney}}, \ and\ \bibinfo {author} {\bibfnamefont
  {D.}~\bibnamefont {Sorensen}},\ }\href@noop {} {\emph {\bibinfo {title}
  {{LAPACK} Users' Guide}}},\ \bibinfo {edition} {3rd}\ ed.\ (\bibinfo
  {publisher} {Society for Industrial and Applied Mathematics},\ \bibinfo
  {address} {Philadelphia, PA},\ \bibinfo {year} {1999})\BibitemShut {NoStop}%
\end{thebibliography}
%
\end{document}